\documentclass{aa}
\usepackage{graphicx}
\usepackage[varg]{txfonts}
\usepackage{longtable}
\usepackage{lscape}
\usepackage{natbib}
\usepackage{url}
\bibpunct{(}{)}{;}{a}{}{,} 
\newcommand\kms{{\rm\,km\,s^{-1}}}

\newcommand\teff{T_{\rm eff}}

\begin{document}

\title{
Chemical evolution of the Galactic bulge as traced by \\
microlensed dwarf and subgiant stars\thanks{Based on observations 
made with the European Southern Observatory
telescopes, Program IDs 082.B-0453 and 083.B-0265.
}
}
\subtitle{
II. Ages, metallicities, detailed elemental abundances, and\\
connections to the Galactic thick disc
}
\titlerunning{Chemical evolution of the Galactic bulge as traced by
microlensed dwarf and subgiant stars. II.}

\author{
T. Bensby\inst{1}
\and
S. Feltzing\inst{2}
\and
J.A. Johnson\inst{3}
\and
A. Gould\inst{3}
\and
D. Ad\'en\inst{2}
\and
M. Asplund\inst{4}
\and
J. Melend\'ez\inst{5}
\and
A. Gal-Yam\inst{6}
\and\\
S. Lucatello\inst{7}
\and
H. Sana\inst{1, 8}
\and
T. Sumi\inst{9}
\and
N. Miyake\inst{9}
\and
D. Suzuki\inst{9}
\and
C. Han\inst{10}
\and
I. Bond\inst{11}
\and
A. Udalski\inst{12}
 }

\institute{European Southern Observatory, Alonso de Cordova 3107, 
Vitacura, Casilla 19001, Santiago 19, Chile
\and
Lund Observatory, Box 43, SE-221\,00 Lund, Sweden
\and
Department of Astronomy, Ohio State University, 140 W. 18th Avenue, 
Columbus, OH 43210, USA
\and
Max Planck Institute for Astrophysik, Garching, Germany
\and
Centro de Astrof\'{\i}sica, Universidade do Porto, Rua das Estrelas, 
4150-762 Porto, Portugal
\and
Benoziyo Center for Astrophysics, Weizmann Institute of Science, 
76100 Rehovot, Israel
\and
INAF-Astronomical Observatory of Padova, Vicolo dell'Osservatorio 5, 
35122 Padova, Italy
\and
Universiteit van Amsterdam, Sterrenkundig Instituut 'Anton Pannekoek',
Postbus 94249 -- 1090 GE Amsterdam, Netherlands
\and
Solar-Terrestrial Enivironment Laboratory, Nagoya University, Furo-cho,
Chikusa-ku, Nagoya, 464-8601, Japan
\and
Department of Physics, Chungbuk National University, Cheongju
361-763, Republic of Korea
\and
Institute of Information and Mathematical Sciences, Massey University,
Albany Campus, Private Bag 102-904, North Shore Mail Centre,
Auckland, New Zealand
\and
Warsaw University Observatory, A1. Ujazdowskie 4, 00-478, Warszawa, 
Poland
}


\date{Received 26 November 2009 / Accepted 6 January 2010}
\offprints{Thomas Bensby, \email{tbensby@eso.org}}
 \abstract%
 {
 The Bulge is the least understood major stellar population of the Milky
 Way. Most of what we know about the 
 formation and evolution of the Bulge comes from bright giant stars.  
 The underlying assumption that giants represent all the 
 stars, and accurately trace the chemical evolution 
 of a stellar population, is under debate. In particular, recent
 observations of a few microlensed dwarf stars give a very 
 different picture of the evolution of the Bulge from that given by 
 the giant stars.
 }
 {
 We aim to resolve the apparent discrepancy between Bulge 
 metallicity  distributions derived from
 microlensed dwarf stars and giant stars. Additionally, we aim to put 
 observational constraints on the elemental abundance trends and
 chemical evolution of the Bulge. 
  }
 {
 We perform a detailed elemental abundance analysis of dwarf stars
 in the Galactic bulge, based on high-resolution spectra that were 
 obtained while the stars were optically magnified during gravitational
 microlensing events. The analysis method is the same as for 
 a large sample of F and G dwarf stars in the Solar neighbourhood, 
 enabling a fully differential comparison between the Bulge and
 the local stellar populations in the Galactic disc.
  }
 {
 We present detailed elemental abundances and stellar ages for six
 new dwarf stars in the Galactic bulge. Combining these with
 previous events, here re-analysed with the same methods, 
 we study a homogeneous sample of 15 stars, which constitute the 
 largest sample to date of microlensed dwarf stars in the 
 Galactic bulge.  We find that the stars span the full range of 
 metallicities from 
 $\rm [Fe/H]=-0.72$ to $+0.54$, and an average metallicity of
 $\rm\langle [Fe/H]\rangle=-0.08\pm 0.47$, close to the average  
 metallicity based on giant stars in the Bulge. Furthermore, the stars 
 follow well-defined abundance trends, that for $\rm [Fe/H]<0$  
 are very similar to those of the local Galactic thick disc. This 
 suggests that the Bulge and the thick disc have had, at least 
 partially, comparable chemical histories. At sub-solar metallicities 
 we find the Bulge dwarf stars to have consistently old ages, while 
 at super-solar metallicities 
 we find a wide range of ages. Using the new age and abundance
 results from the microlensed dwarf stars
 we investigate possible formation scenarios for the Bulge.
  }
 {}
   \maketitle

\section{Introduction}

The Galactic bulge is a major stellar component of the Milky Way.
Estimations of its mass range from 10\,\% \citep{oort1977} to 
25\,\% \citep{sofue2009} of the total stellar mass of the Galaxy. 
It is a peanut shaped barred bulge and occupies the inner $\sim$1\,kpc 
of the Galaxy \citep{frogel1988}. Recent results for the Bulge shape 
give a scale length for the bar major axis of $\sim$1.5\,kpc 
\citep{rattenbury2007}. The kinematic properties of the Bulge are 
intermediate between a rotationally supported system and a velocity
dispersion dominated system \citep[e.g.,][]{minniti2008}. The
markedly different stellar populations that inhabit this region of the
Galaxy make it important to discern the formation and evolution as
part of the understanding of the overall formation of the Galaxy. 
In addition, bulges are common features among luminous galaxies, 
and determining the origin of the Bulge and the corresponding 
observational signatures (e.g., boxy isophotes, high [$\alpha$/Fe] 
ratios, no age dispersion) are important steps in decoding 
the formation of bulges in general.

The differences among the Bulge, the thin and thick discs in the
Solar neighbourhood, and the 
Galactic stellar halo have been discovered and detailed through 
extensive photometric and spectroscopic observations of stars. 
Red giants are generally the only stars bright enough for 
high-resolution spectroscopy at the distance of the Bulge, and 
they have been intensively studied in the optical and infrared, 
particularly in a few fields including Baade's window 
\citep[e.g.][]{mcwilliam1994,fulbright2006,zoccali2008,melendez2008,cunha2006,ryde2009b}.  
Chemically, the Bulge is decidedly more metal-rich than the 
stellar halo, with the mean [Fe/H] from the 
\cite{zoccali2008} sample slightly below solar metallicity, and a
possible vertical metallicity gradient of 0.6\,dex per kpc, 
although a comparison with M giants in the inner Bulge may 
indicate a flattening 
of the gradient in the central regions \citep{rich2007}. 
The abundance 
ratios show enhanced [$\alpha$/Fe] ratios that persist to higher 
[Fe/H] than in local thin disc stars, but show good agreement with 
thick disc giants \citep{melendez2008,alvesbrito2010}.
Furthermore, the stars at the Bulge main-sequence turnoff are red, 
and isochrone fitting has shown that the majority of the stars in 
the Bulge are old 
\cite[e.g.,][]{holtzman1993,feltzing2000b,zoccali2003,clarkson2008}.

These observations have fuelled an intense debate on the origin of the
Bulge. The established concordance model of cosmology provides the
framework to understand the formation of the Milky Way by hierarchical
merging, but the formation of the components of the Galaxy requires
important physics, such as star formation and feedback, that is
unresolved and parametrised in current models. However, there are two
basic scenarios by which a bulge is built up in simulations 
\citep[see][and references therein]{rahimi2009}. The first is 
by mergers, 
where subclumps coming together in the early phases of Galactic 
evolution combine to form the Bulge out of both accreted stars and 
stars formed in situ.  The second is secular evolution, where the Bulge 
is created gradually out of the Galactic disc 
\citep[e.g.,][]{kormendy2004}. 
The merger model is favoured by the metallicity gradient, while the 
secular evolution model is favored by cylindrical 
rotation and by agreement (in terms of mean metallicity  
abundance trends, and ages) with the Galactic thick disc 
\citep[e.g.,][]{howard2009}.

As mentioned above, spectroscopic observations of stars in the Bulge 
have usually been confined to giants. This limits our knowledge of 
the Bulge in several ways. First, because much of our knowledge of 
the Solar neighbourhood relies on dwarf stars 
\citep[e.g.][]{edvardsson1993}, any systematic offsets between the 
metallicity scale of giant and dwarf stars is cause for concern. In 
this context, \cite{taylor2005} have shown that there is a lack of 
very metal-rich ($\rm [Fe/H] > +0.2$) giants in the Solar 
neighbourhood and that the mean metallicity of local giants is 
lower than for dwarfs. Recent studies of nearby giants 
\citep{luck2007,takeda2008} confirm the lack of very metal-rich stars. 
\cite{santos2009} suggest that there may be systematic errors in the 
metallicity determinations of metal-rich giants. Secondly, there are 
several pieces of evidence that giant stars may not accurately represent 
all the stars. For the metal-rich cluster NGC\,6791, \cite{kalirai2007}
proposed that the explanation for the large number of low-mass He white
dwarfs was that about 40\,\% of the stars do not become red clump stars,
skipping entire phases of stellar evolution altogether because of
high mass loss. \cite{kilic2007} proposed a similar mechanism to explain 
the non-binary He white dwarfs found in the Solar neighbourhood.
Finally, with turnoff stars, ages can be determined for individual 
stars, and an age-metallicity relationship derived, which is not 
possible for giant stars. This means that a true study of the Bulge 
requires the study of dwarf stars. However, at the distance of the 
Bulge, dwarf stars are too faint for abundance analyses based on 
high-resolution spectra. Turnoff stars in the Bulge have V magnitudes 
around 19 to 20 (compare, e.g., the colour-magnitude diagrams in 
\citealt{feltzing2000b}). However, in the event that a Bulge dwarf star 
is lensed by a foreground object, the magnitude of the star can 
increase by more than 5 magnitudes, in which case a high-resolution 
spectrum can be obtained and the star analysed in a similar manner as 
the dwarf stars in the Solar neighbourhood.

There were several spectroscopic observations of microlensed Bulge 
stars in the 1990s, but the first high-resolution spectrum of a 
dwarf star was presented in \cite{minniti1998}. Complete 
high-resolution spectroscopic abundance analyses have been published 
for eight microlensing events toward the Bulge \citep{johnson2007,johnson2008,cohen2008,cohen2009,bensby2009letter,bensby2009,epstein2009}. 
Initially, it appeared that the microlensed Bulge dwarf stars were 
typically much more metal-rich than the giants in the Bulge, and 
\cite{epstein2009} found, using a Kolmogorov-Smirnow (KS) test,
a very low probability of only 1.6\,\% that these eight 
microlensed dwarf stars in the Bulge were drawn from the same 
metallicity distribution (MDF) as  the sample of Bulge giants from 
\cite{zoccali2008}. \cite{cohen2008} proposed that a similar
mechanism to the one occurring in NGC 6791 was occurring in the Bulge,
preventing the most metal-rich stars from being included in the
giant surveys. Arguments against this idea
presented by \cite{zoccali2008} were based on the luminosity function
along the main sequence and red giant branch, showing no lack
of RGB stars, with respect to the prediction of theoretical models.

In addition, the microlensed Bulge dwarf stars showed good agreement in 
abundance ratios with the thick disc stars in the Solar neighbourhood 
\citep{bensby2009letter,bensby2009}. However, comparisons have been 
hampered because of the small number of microlensed stars that did not 
always cover the [Fe/H] range of interest. 
Also, individual age estimates were provided for stars near the 
turnoff subgiant branch, including some stars that could be younger 
than the canonical old Bulge population \citep{johnson2008,bensby2009}.

\begin{table*}
\centering
\caption{
Summary$^\dagger$ of the, so far, 15 dwarf star microlensing events in the Bulge that have been observed with high-resolution spectrographs. 
They have been sorted according to their metallicities (as given in Table~\ref{tab:parameters}).
\label{tab:events}
}
\setlength{\tabcolsep}{1.4mm}
\tiny
\begin{tabular}{rcccccrrrrrlll}
\hline\hline
\noalign{\smallskip}
\multicolumn{1}{c}{Object}     &
RAJ2000                        &
DEJ2000                        &
 $l$                           &
 $b$                           &
 \multicolumn{1}{c}{$T_{E}$}   &
 $T_{max}$                     &  
 $A_{max}$                     &
 \multicolumn{1}{c}{$T_{obs}$} &  
 Exp.                          &
 $S/N$                         &
 Spec.                         &
 \multicolumn{1}{c}{$R$}       &
 Ref.                          \\
                               &
[hh:mm:ss]                     &
[dd:mm:ss]                     &
 [deg]                         &    
 [deg]                         &  
 [days]                        &
 [HJD]                         &
                               &   
 \multicolumn{1}{c}{[MJD]}     &  
 \multicolumn{1}{c}{[s]}       &
                               &
                               &
                               &
                               \\  
\noalign{\smallskip}
\hline
\noalign{\smallskip}
 OGLE-2009-BLG-076S$\phantom{^{b}}$ & 17:58:31.9 & $-$29:12:17.8 & $+1.21$  &  $-2.56$ & 36.9 &  4916.46   &  70 & 4916.291 &       7200  &  \phantom{1}30      &  UVES  & 45\,000    & B09a \\ 
  MOA-2009-BLG-493S$\phantom{^{b}}$ & 17:55:46.0 & $-$28:48:25.8 & $+1.25$  &  $-1.84$ & 13.2 &  5094.61   & 150 & 5093.980 &       7200  &  \phantom{1}40      &  UVES  & 45\,000    & TW   \\ 
  MOA-2009-BLG-133S$\phantom{^{b}}$ & 18:06:32.8 & $-$31:30:10.7 & $+0.05$  &  $-5.19$ & 26.6 &  4932.10   &  70 & 4932.233 &       7200  &  \phantom{1}35      &  UVES  & 45\,000    & TW   \\ 
  MOA-2009-BLG-475S$\phantom{^{b}}$ & 18:02:27.3 & $-$27:26:49.9 & $+3.16$  &  $-2.45$ & 34.0 &  5084.92   &  62 & 5084.980 &       7200  &  \phantom{1}20      &  UVES  & 45\,000    & TW   \\ 
MACHO-1999-BLG-022S$^{\ddag}$       & 18:05:05.8 & $-$28:34:39.5 & $+2.46$  &  $-3.51$ &265.0 &  1365.50   &  28 & 1366.315 &      12600  &  \phantom{1}80      &  HIRES & 29\,000    & C03  \\ 
 OGLE-2008-BLG-209S$\phantom{^{b}}$ & 18:04:50.0 & $-$29:42:35.3 & $+1.44$  &  $-4.01$ & 19.5 &  4606.09   &  30 & 4606.833 &       5400  &  \phantom{1}30      &  MIKE  & 47\,000    & B09b \\ 
  MOA-2009-BLG-489S$\phantom{^{b}}$ & 17:57:46.5 & $-$28:38:57.8 & $+1.61$  &  $-2.14$ & 58.9 &  5095.52   & 103 & 5094.982 &       7200  &  \phantom{1}65      &  UVES  & 45\,000    & TW   \\ 
  MOA-2009-BLG-456S$\phantom{^{b}}$ & 17:48:56.3 & $-$34:13:32.3 & $-4.16$  &  $-3.34$ & 36.2 &  5090.94   &  77 & 5090.982 &       7200  &  \phantom{1}45      &  UVES  & 45\,000    & TW   \\ 
 OGLE-2007-BLG-514S$\phantom{^{b}}$ & 17:58:03.0 & $-$27:31:08.2 & $+2.62$  &  $-1.63$ & 76.0 &  4386.28   &1200 & 4385.985 &       4800  &  \phantom{1}30      &  MIKE  & 25\,000    & E09  \\ 
  MOA-2009-BLG-259S$\phantom{^{b}}$ & 17:57:57.6 & $-$29:11:39.1 & $+1.15$  &  $-2.45$ & 69.1 &  5016.77   & 220 & 5016.227 &       7920  &  \phantom{1}50      &  UVES  & 45\,000    & TW   \\ 
  MOA-2008-BLG-311S$\phantom{^{b}}$ & 17:56:53.7 & $-$31:23:40.3 & $-0.87$  &  $-3.35$ & 15.7 &  4655.40   & 400 & 4654.952 &       7200  &  \phantom{1}85      &  MIKE  & 29\,000    & C09  \\ 
  MOA-2008-BLG-310S$\phantom{^{b}}$ & 17:54:14.4 & $-$34:46:37.7 & $-4.09$  &  $-4.56$ &  7.1 &  4656.39   & 280 & 4655.957 &       7200  &  \phantom{1}90      &  MIKE  & 41\,000    & C09  \\ 
 OGLE-2007-BLG-349S$\phantom{^{b}}$ & 18:05:23.0 & $-$26:25:27.1 & $+4.38$  &  $-2.52$ &109.4 &  4348.56   & 400 & 4348.237 &       3050  &  \phantom{1}90      &  HIRES & 48\,000    & C08  \\ 
  MOA-2006-BLG-099S$\phantom{^{b}}$ & 17:54:10.2 & $-$35:13:34.9 & $-4.48$  &  $-4.78$ & 30.1 &  3940.35   & 380 & 3940.090 &       2400  &  \phantom{1}30      &  MIKE  & 19\,000    & J08  \\ 
 OGLE-2006-BLG-265S$\phantom{^{b}}$ & 18:07:18.9 & $-$27:47:44.0 & $+3.38$  &  $-3.55$ & 28.6 &  3893.24   & 212 & 3892.581 &        900  &  \phantom{1}45      &  HIRES & 45\,000    & J07  \\ 
\noalign{\smallskip}
\hline
\end{tabular}
\flushleft
\scriptsize
$^\dagger$ Given for each microlensing event is: RA and DE coordinates (J2000) read from the
fits headers of the spectra (the direction where the telescope pointed during observation); 
galactic coordinates ($l$ and $b$); 
duration of the event in days ($T_{E}$); time when maximum magnification occured ($T_{max}$); 
maximum magnification ($A_{max}$); time when event was observed
with high-resolution spectrograph ($T_{obs}$); magnification at the time
of observation ($A_{obs}$); the exposure time (Exp.),
the measured signal-to-noise ration per pixel at $\sim$6400\,{\AA}; the
spectrograph that was used; the spectral resolution; and the
reference where the star first appeared: TW=This Work, B09a=\cite{bensby2009letter},
B09b=\cite{bensby2009}, J07=\cite{johnson2007}, J08=\cite{johnson2008},
C08=\cite{cohen2008}, C09=\cite{cohen2009}, C03=\cite{cavallo2003}. \\
$^\ddag$ MACHO ID: 109.20893.3423.
\end{table*}


We will here present detailed elemental abundance results for six new 
microlensing events toward the Galactic bulge. We also re-analyse the 
events previously studied by \cite{cavallo2003,cohen2009} and 
\cite{epstein2009}. Combining these data with the results from 
\cite{bensby2009letter} and \cite{bensby2009} (which includes a 
re-analysis of the events from 
\citealt{johnson2007,johnson2008,cohen2008}) we now have a sample of 
15 microlensed dwarf stars in the Bulge that have been homogeneously 
analysed using the exact same methods.

\section{Observations and data reduction}

In order to trigger observations of these highly magnified stars
we rely on the OGLE\footnote{OGLE is short for Optical Gravitational 
Lens Experiment, \url{http://ogle.astrouw.edu.pl} \citep{udalski2003}.} 
and MOA\footnote{MOA is short for Microlensing Observations for 
Astrophysics, \url{http://www.phys.canterbury.ac.nz/moa} 
\citep[e.g.,][]{bond2001}.} projects that every night monitor about 
100 million stars toward the Bulge to detect variations in their 
brightnesses. If an object shows a well-defined rise in brightness, a 
microlensing alert is announced. Every year, 600-800 events are 
detected. Based on the photometric data obtained by the MOA and OGLE 
surveys it is possible to model the event, and make predictions of 
the length of the event, peak brightness, and time of peak brightness. 
Stars are identified as likely dwarf stars based on their unlensed 
magnitudes and colour differences relative to the red clump stars. 
This is done in instrumental magnitudes. The majority are low 
magnification events, and only a few have unlensed brightnesses 
of $V=18-20$, characteristic of dwarf stars in the Bulge at a distance 
of $\sim 8$\,kpc.  During a regular Bulge season in the 
Southern hemisphere, typically around 10 high-magnification events of 
dwarf stars in the Bulge are detected\footnote{
With the new OGLE-IV camera that will be in operation in the first
half of 2010 the field-of-view will increase from the 0.3 square degrees 
of OGLE-III to 1.4 square degrees, resulting in a substantial 
increase in the number of detected microlensing events.}. 
To catch these unpredictable events, we have an ongoing Target of 
Opportunity (ToO) program at the ESO Very Large Telescope on Paranal 
in Chile. Observations can then be triggered with only a few hours 
notice.

\begin{figure*}
\resizebox{\hsize}{!}{
\includegraphics[bb=10 40 520 550,clip]{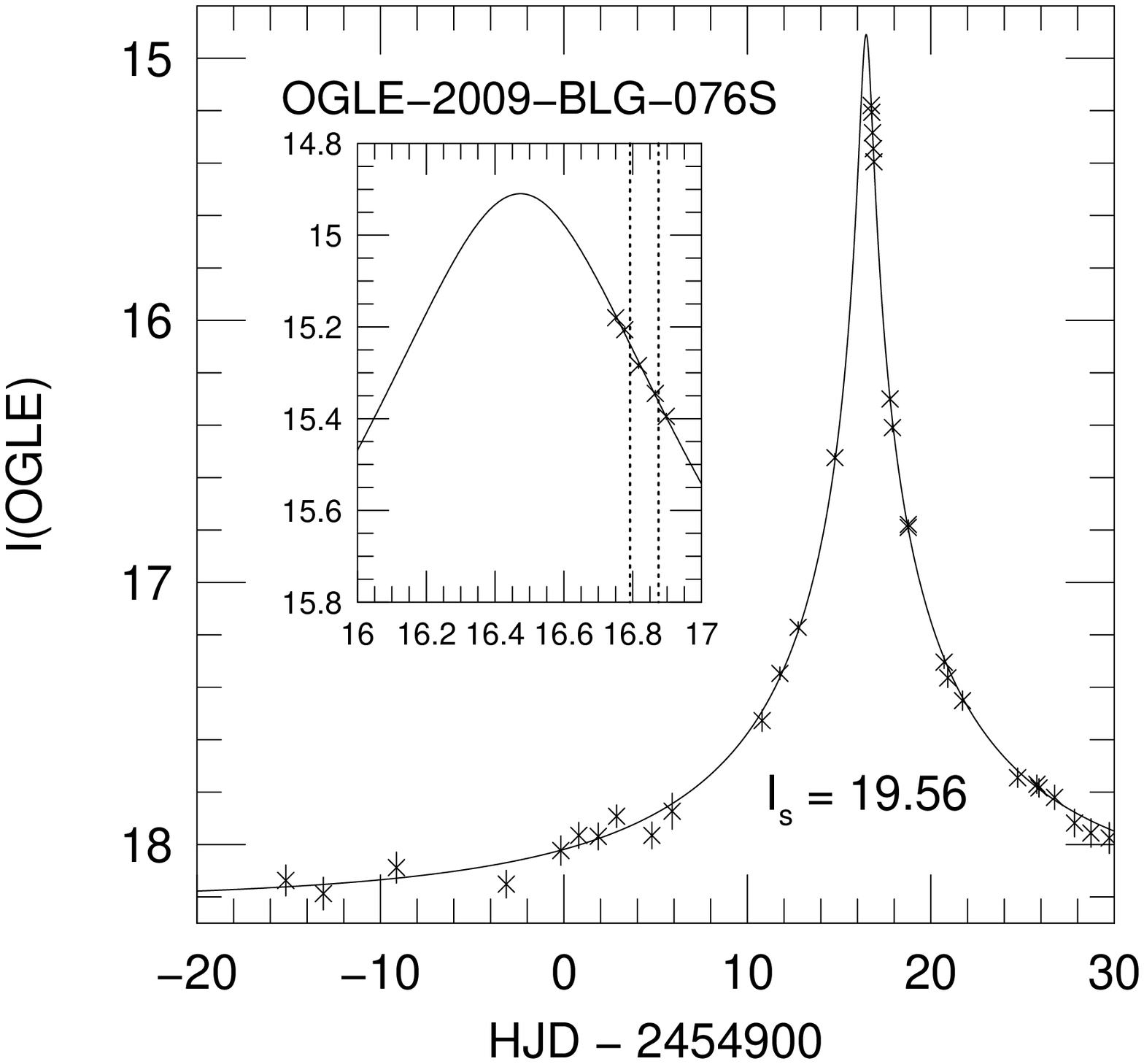}
\includegraphics[bb=60 40 520 550,clip]{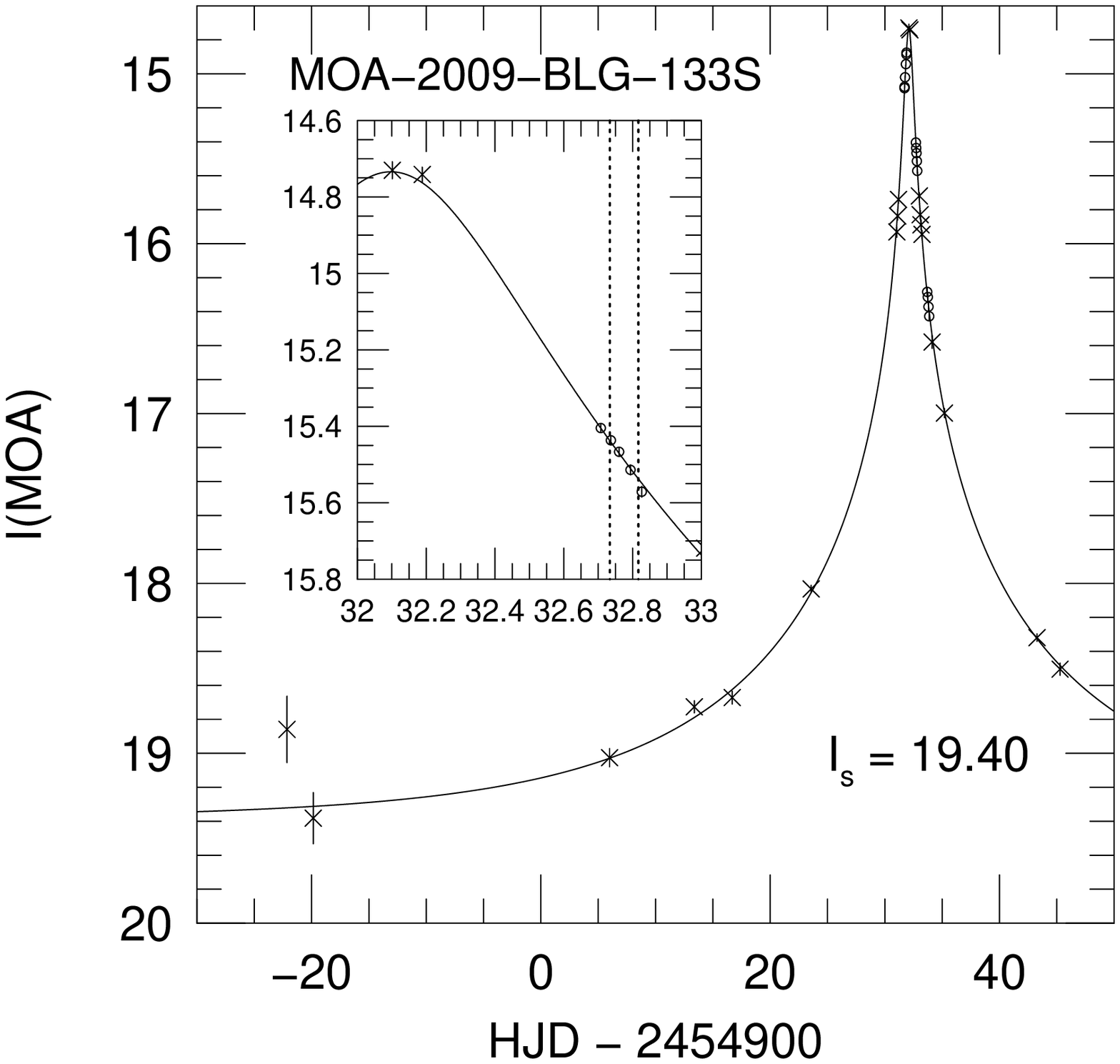}
\includegraphics[bb=60 40 520 550,clip]{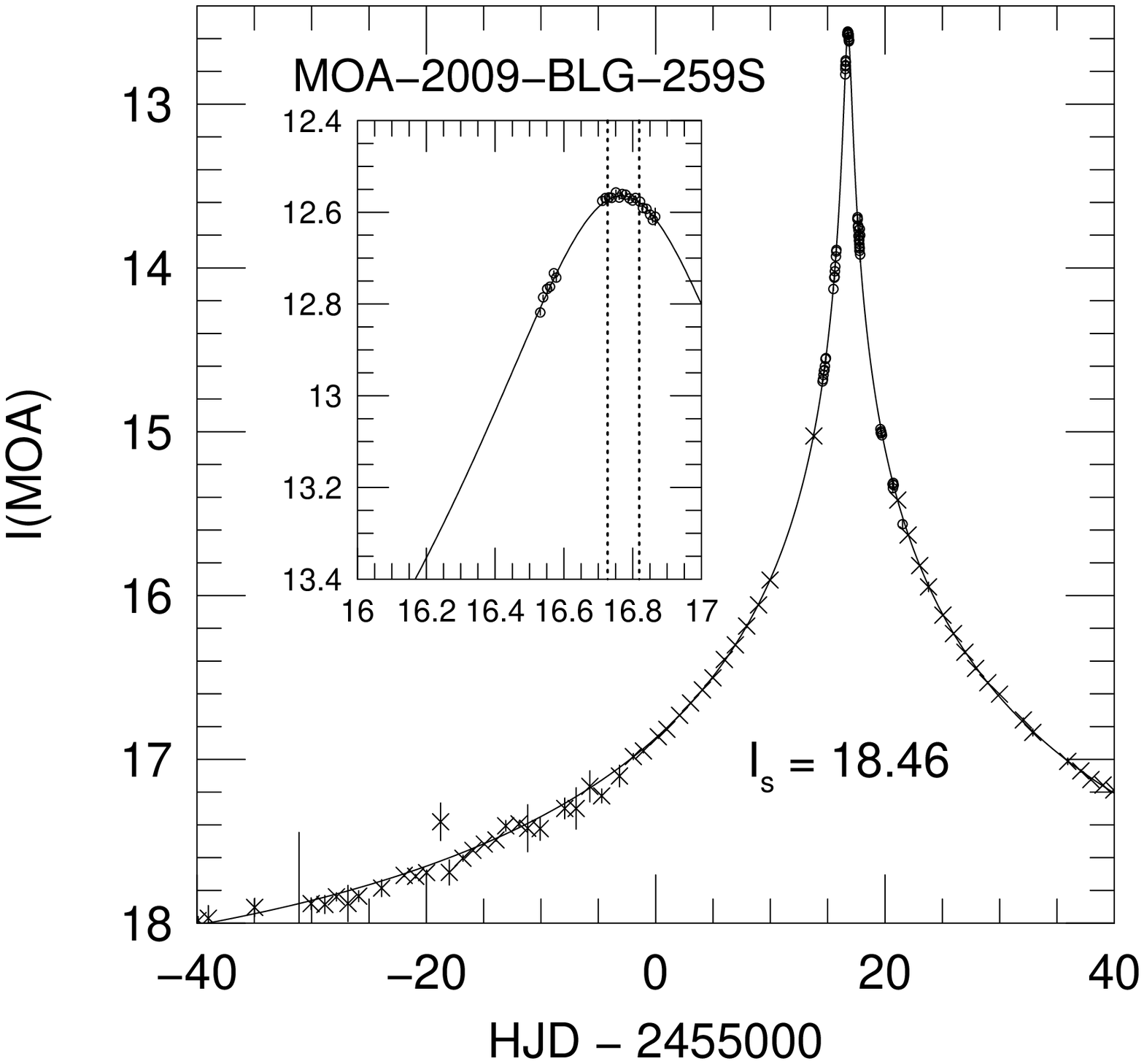}
\includegraphics[bb=60 40 530 550,clip]{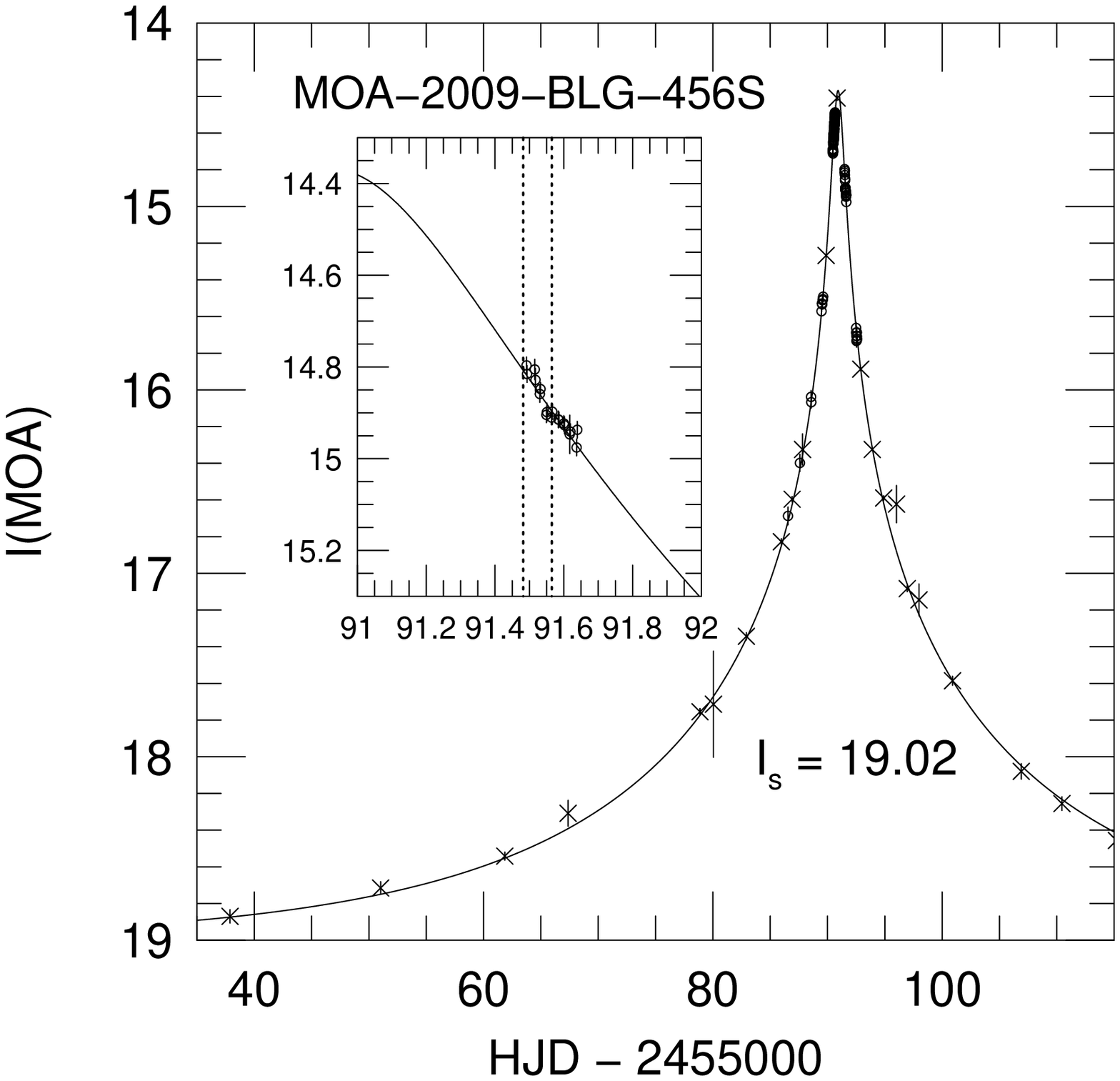}
}
\resizebox{\hsize}{!}{
\includegraphics[bb=10 30 520 540,clip]{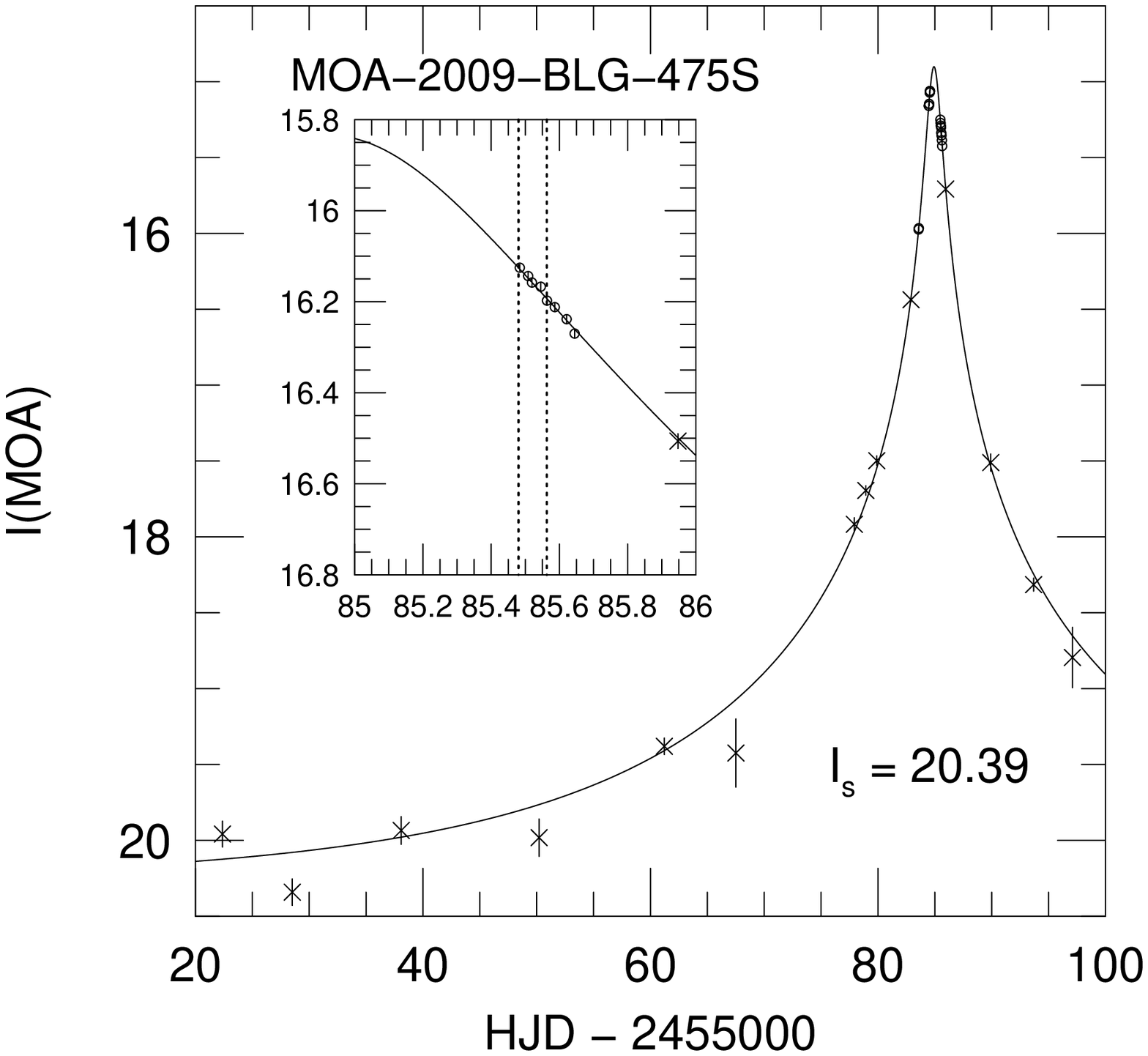}
\includegraphics[bb=60 30 520 540,clip]{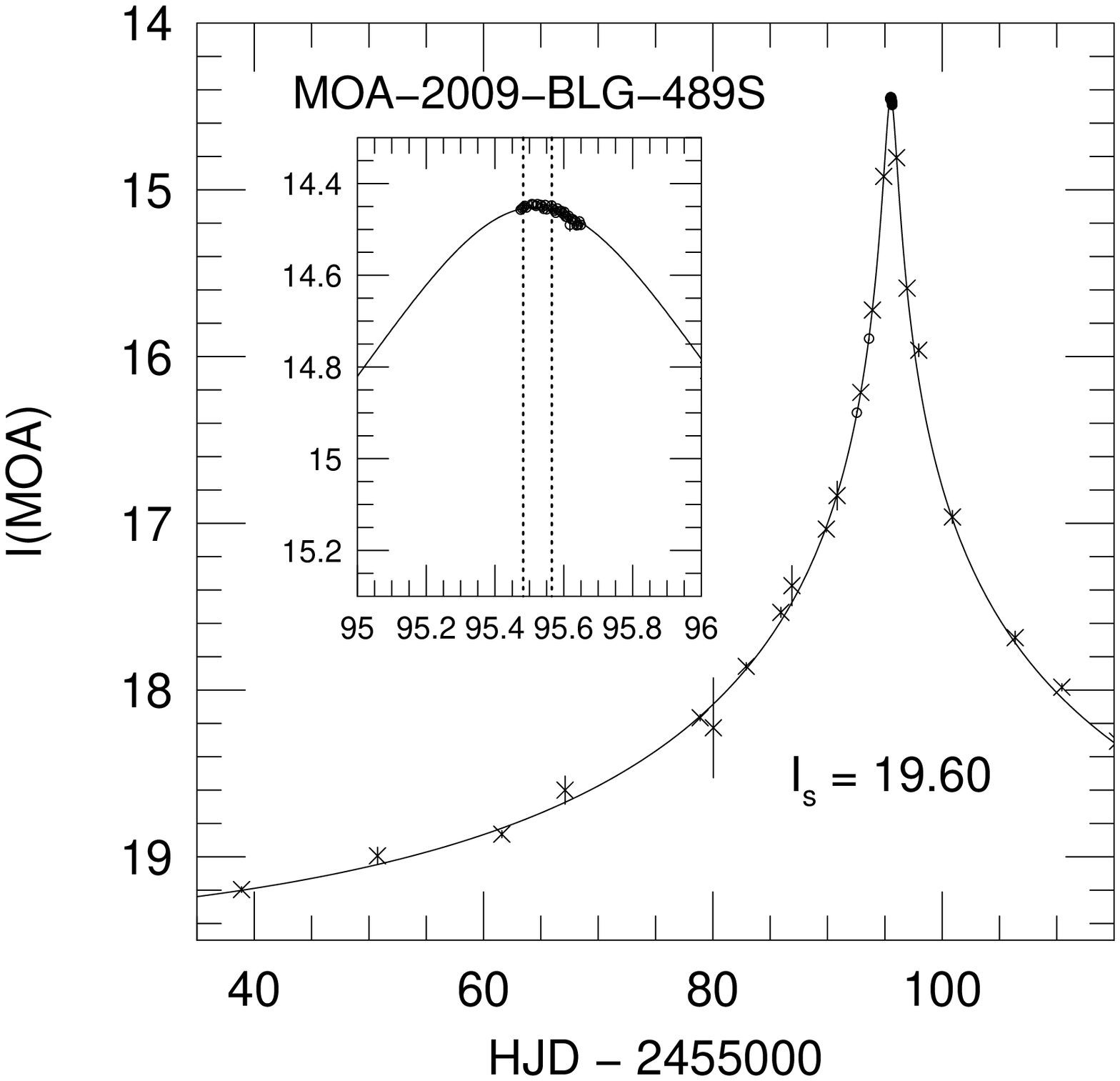}
\includegraphics[bb=60 30 520 540,clip]{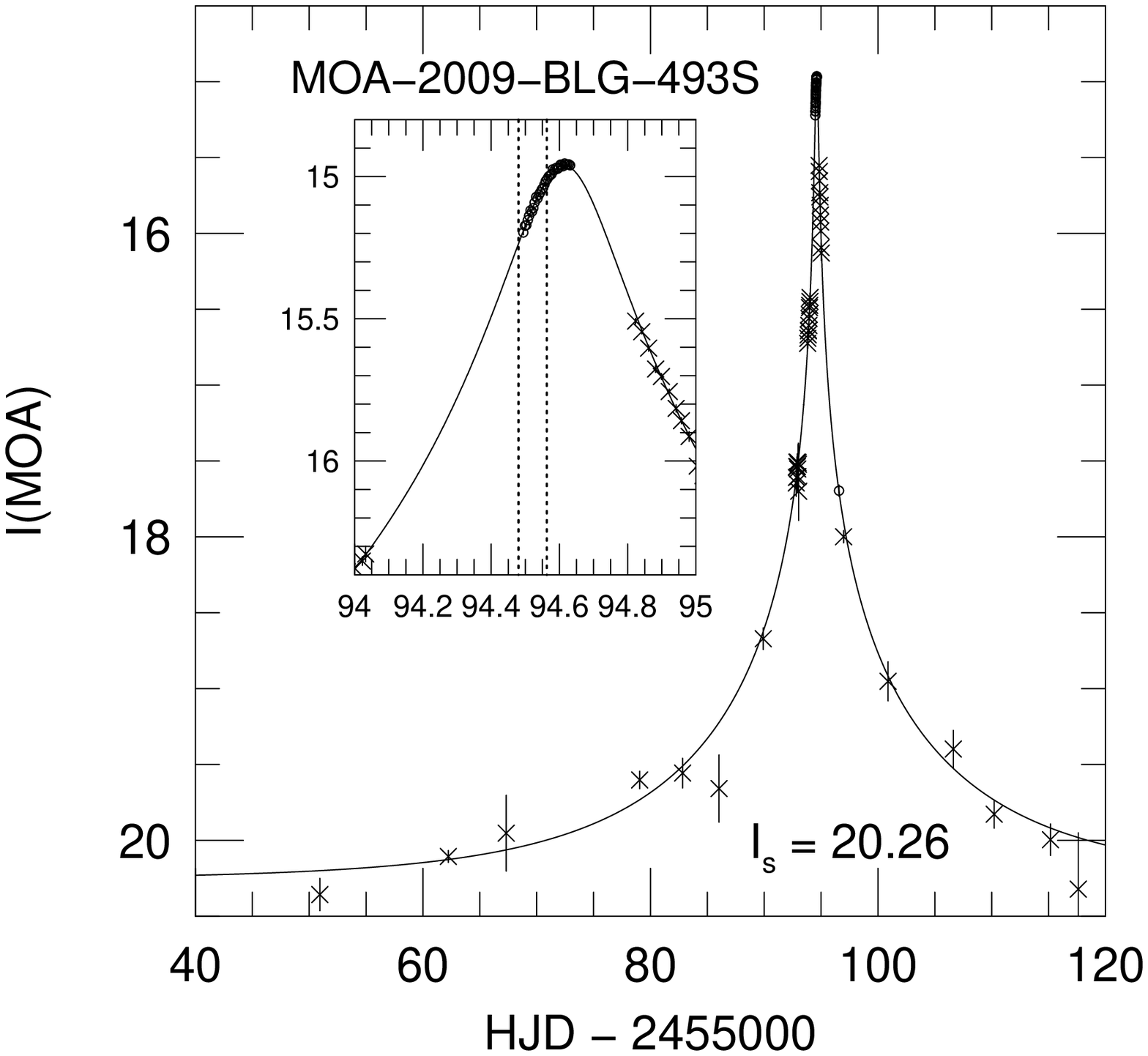}
\includegraphics[bb=60 30 530 540,clip]{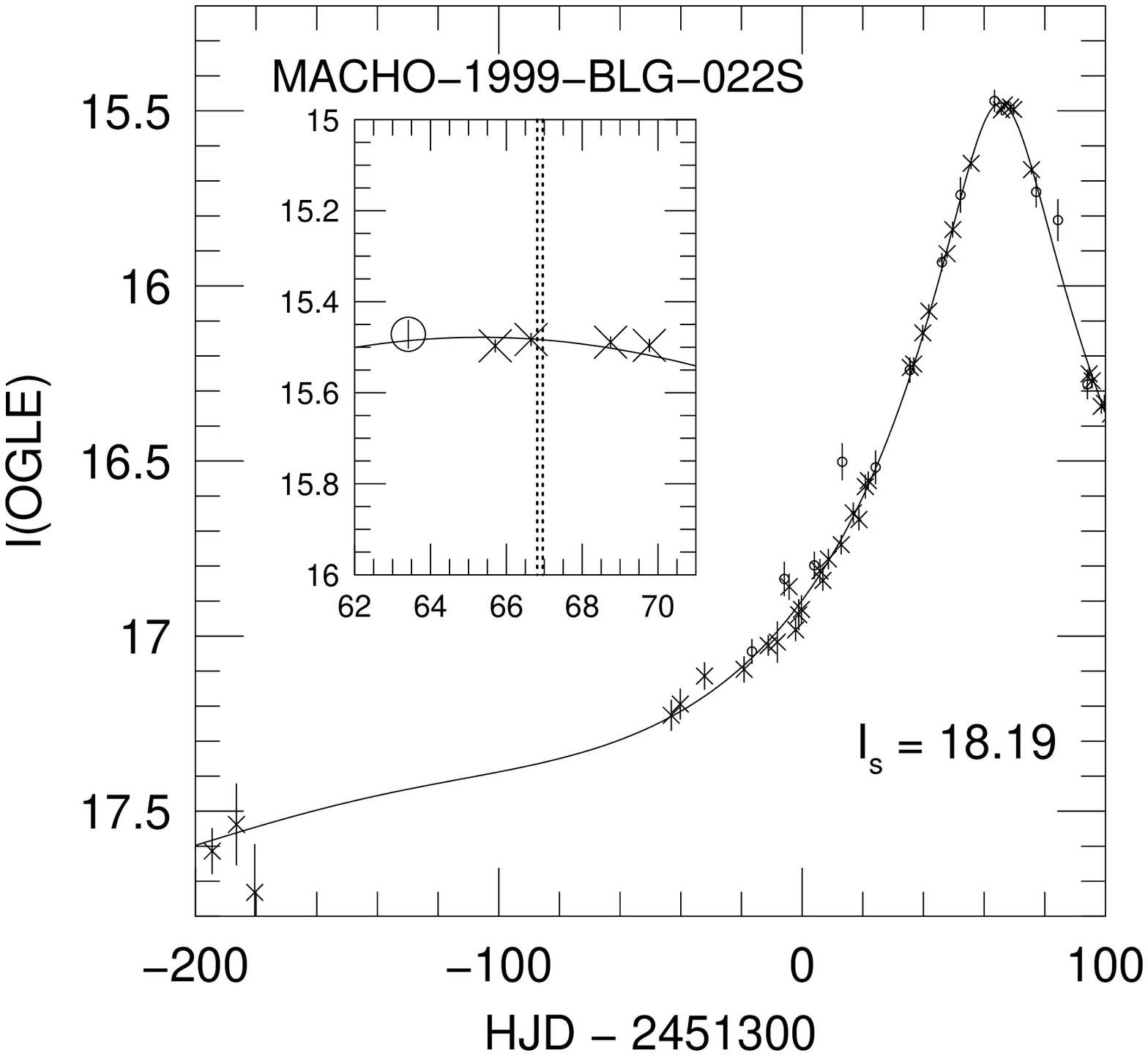}
}
\caption{Light curves for the eight new microlensing events. 
The photometry comes from the surveys indicated by their names 
(MOA or OGLE), except for  MACHO-1999-BLG-022S that has data from 
from both OGLE (circles) and binned MACHO data (crosses). Each plot 
has a zoom window, showing the time intervals when the source stars 
were observed with high-resolution spectrographs. In each plot the 
un-lensed magnitude of the source star is also given ($I_{\rm S}$).
\label{fig:events}
}
\end{figure*}

\subsection{The 2009 events observed with UVES}

On March 21, 2009, the OGLE early warning system identified 
OGLE-2009-BLG-076 to be a possible high-magnification microlensing 
event toward the Galactic bulge. As the magnitude of the source star 
before the microlensing event indicated that it was either a dwarf or  
a subgiant star, we triggered observations with the UVES spectrograph 
\citep{dekker2000} located on UT2 at the ESO Very Large Telescope on 
Paranal. Due to the limited visibility of the Bulge in March, the 
target had to be observed towards the end of the night. Hence, 
OGLE-2009-BLG-076S was observed on March 26, a few hours after 
reaching peak brightness (see Fig.~\ref{fig:events}). A few weeks 
later, on April 11, we observed the MOA-2009-BLG-133, this time first 
alerted by the MOA collaboration, also with UVES. The third event was
observed in the beginning of July, MOA-2009-BLG-259S. This object 
was observed during the UVES red arm upgrade, so we could only 
obtain a spectrum with the blue CCD that has a limited wavelength 
coverage of 3700-5000\,{\AA}. In September, the end of ESO observing 
period P83, we saw an explosion of microlensing events and another 
four source stars were observed with UVES: MOA-2009-BLG-475S on Sep 10, 
MOA-2009-BLG-456S on Sep 16, MOA-2009-BLG-493S on Sep 19, and 
MOA-2009-BLG-489S on Sep 20. The UVES red arm was now back with two 
new red CCDs with increased sensitivities. 

For the 2009 observations with UVES listed above, each target was 
observed for a total of two hours, split into either four 30 min or 
three 40 min exposures. Using UVES with dichroic number 2, each 
observation resulted in spectra with wavelength coverage between 
3760-4980\,{\AA} (blue CCD), 5680-7500\,{\AA} (lower red CCD), and 
7660-9460\,{\AA} (upper red CCD). In all cases a slit width of 1" 
was used, giving a resolving power of $R\approx45\,000$. 

On all occasions, right before or right after the observations 
of the microlensed targets in the Bulge, a rapidly rotating B star, 
either HR\,6141 or HR\,8431, was observed at an airmass similar to 
that of the Bulge stars. The featureless spectra from these B 
stars were used to divide out telluric lines in the spectra of 
the Bulge stars. Also, at the beginning of the night of 
April 11 we obtained a solar spectrum by observing the asteroid 
Pallas. 

Data taken before the upgrade of the UVES red CCD in mid-July 
were reduced with the UVES pipeline (CPL version 3.9.0), while the 
data taken after the upgrade were reduced with version 4.4.5. 
Typical signal-to-noise ratios per pixel at 6400\,{\AA} are given 
in Table~\ref{tab:events}. 

The light curves for the seven microlensing events 
\citep[including OGLE-2009-BLG-076S from][]{bensby2009letter} 
observed with UVES in 2009 are shown
in Fig.~\ref{fig:events}, in which we have also indicated the time 
interval during which they were observed with high-resolution 
spectrographs. Positions, amplifications, times of observation, and 
exposure times are given in Table~\ref{tab:events}.

\subsection{Los MACHOs}

\cite{cavallo2003} presented the first detailed elemental abundance 
study of microlensed dwarf stars in the Bulge.
Their analysis was of a ``preliminary" nature, 
so we decided to re-analyse the stars 
that they classified as either dwarf or subgiant stars. There are four 
such stars: MACHO-1997-BLG-045S, MACHO-1998-BLG-006S, 
MACHO-1999-BLG-001S, and MACHO-1999-BLG-022S.  

The observations of these stars were carried out from 1997 to 1999 
with the HIRES spectrograph on the Keck I telescope on Hawaii. By 
using a 1.148" wide slit and a 2x2 binning, spectra with a 
resolution of $R\approx 29\,000$ were obtained. 
These data were obtained when the HIRES detector had only a
single CCD chip. The data are now 
publicly available and we gathered science and associated calibration 
files from the Keck Observatory Archive\footnote{Available at 
\url{http://koa.ipac.caltech.edu}}. 

The data reduction was carried out using the LONG and ECHELLE contexts 
of the MIDAS\footnote{ESO-MIDAS is the acronym for the European Southern
Observatory  Munich Image Data Analysis System which is developed and
maintained by the  European Southern Observatory.} software.
Because the bias level is changing on a time scale of a few minutes, 
we used the over-scan region to compensate for the observed variations 
and to bring all the raw data to an effectively homogenised bias level. 
Master calibration frames were created by averaging the relevant 
frames obtained close in time to the science observations. 
The data were then bias, dark and background-illumination subtracted 
using standard procedures of the ECHELLE context. The orders were 
traced directly on the science images and a 5-pixel window was used 
to extract the object spectra. Sky spectra were extracted from two 
smaller windows on both sides of the object window. Flat-field and 
wavelength calibration spectra were extracted using the exact 
same windows as the one used for object and sky extraction. The science 
and sky spectra were then flat-fielded and sky subtracted. Finally, 
because of little or no overlap between the orders, the 
wavelength calibration was performed individually for each order. 
In total, 27 orders were observed  yielding  an effective wavelength 
coverage from 4670 to 7180\,\AA, although with some gaps between 
the orders.

Only two of these four stars could be analysed. The 
reduced spectrum for MACHO-1997-BLG-045S was not of sufficient quality 
to allow for any measurements of equivalent widths or line synthesis 
necessary for a proper abundance analysis, and 
MACHO-1999-BLG-001S appears to be a spectroscopic binary.
Also, as MACHO-1999-BLG-006S turned out to be a low-luminosity giant 
after our re-analysis ($\log g\approx 2-3$), the results for this star 
will be presented together with the other similar low-luminosity giant 
stars observed at ESO in a subsequent study.
The light curve for MACHO-1997-BLG-022S is shown in Fig.~\ref{fig:events}
and event data given in Table~\ref{tab:events}.

\begin{figure*}
\sidecaption
\includegraphics[width=12cm, bb=20 195 570 520,clip]{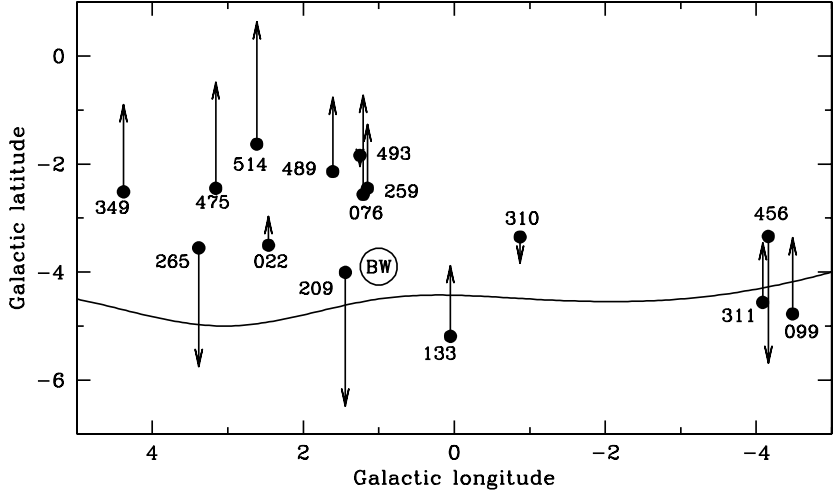}
\caption{
	Positions and radial velocities for
	the 15 microlensed stars.  
	The arrows represent measured radial velocities and one degree
	corresponds to 70\,$\kms$.
	Upward pointing arrows indicate positive velocities. 
	The curved line shows 
	the outline of the southern Bulge based on observations with 
	the COBE satellite \citep{weiland1994}. The location of Baade's 
	window (BW) is marked by the larger open circle. Stars
	have been labelled with the last number in their IDs.
        \label{fig:allevents}}
\end{figure*}

\section{Bulge membership}

\subsection{Positions on the sky and radial velocities}

The locations on the sky of the events are shown in 
Fig.~\ref{fig:allevents}. All the stars have negative Galactic 
latitudes because OGLE and MOA currently only monitor fields with 
$b < 0$. The angular distances to the Galactic plane are similar for 
all events, between 2-5$^{\circ}$. The measured radial velocities 
for the stars have been indicated in Fig.~\ref{fig:allevents}.
Arrows upward means positive radial velocities, arrows downward
negative velocities, and the scale in the figure is 70\,$\kms$ 
per degree. The high variation in $v_{\rm r}$ for the 
microlensed stars is consistent with the high velocity dispersion
seen for Bulge giant stars (compare, e.g., the recent BRAVA 
radial velocity survey of red giants in the Bulge by \citealt{howard2008}).

\subsection{Microlensing toward the Galactic bulge}

Because we need to observe the microlensing events wherever they 
occur in the central regions of the Galaxy, we cannot choose stars 
along, e.g., the minor axis to maximise the contribution of the 
Bulge, leading to possible confusion about whether these stars are 
Bulge, disc or halo stars.

Our current approach to this is to regard the division into Bulge and 
disc for stars within 1\,kpc of the Galactic centre as a semantic 
division. There is no evidence for a cold rotating extended disc that 
close to the Galactic centre for the fields that have been studied 
so far \citep{howard2009}, and the creation of the Bulge from the 
Galactic disc is one of the scenarios we wish to test. So the question
then becomes whether the microlensed dwarf and subgiant stars are
located in the Bulge region, or in the disc on either this
side or the far side of the Bulge.  \cite{nair1999} estimate that about 
15\,\% of the events toward the Bulge could have source stars belonging 
to the far side of disc, more than 3\,kpc away from the Galactic centre. 
On the other hand, more recent theoretical calculations of the 
distance to microlensed sources, assuming a constant disc density and
an exponential bulge, show that the distance to the sources
is strongly peaked in the Bulge, with the probability of having
$D < 7$ kpc very small \citep{kane2006}.

Another argument that these are Bulge stars, rather than disc stars,
are the large radial velocities for stars close to the Galactic centre 
\citep[e.g.][]{epstein2009} as well as the fact that there are stars 
with radial velocities in opposite directions on the same side of 
the Galactic centre \citep[e.g.,][]{cohen2008}.

In addition, we have prior knowledge of the colours and magnitudes
of the source stars when unmagnified. The stars we observed were all identified as dwarf or subgiant stars at the distance of the Bulge 
based on OGLE or MOA instrumental colours and magnitudes, estimated 
from the offsets from the red clump stars in the same field.
The parameters for the stars determined in this manner have
been repeatedly tested against the parameters derived spectroscopically
and the overall consistency between these results 
\citep[e.g.][]{johnson2007,johnson2008} again shows that these stars 
are likely located at the distance of the
Bulge and not the near or far disc. The fact that the
latitudes of the stars are closer to the Galactic plane may make
disc contamination more likely for unmagnified sources; however,
the combination of kinematics,
colour-magnitude diagrams and microlensing statistics indicate that
we are studying a stellar population belonging in the Bulge.

\section{Analysis}

\begin{figure*}
\resizebox{\hsize}{!}{
\includegraphics[angle=-90,bb=160 28 530 810,clip]{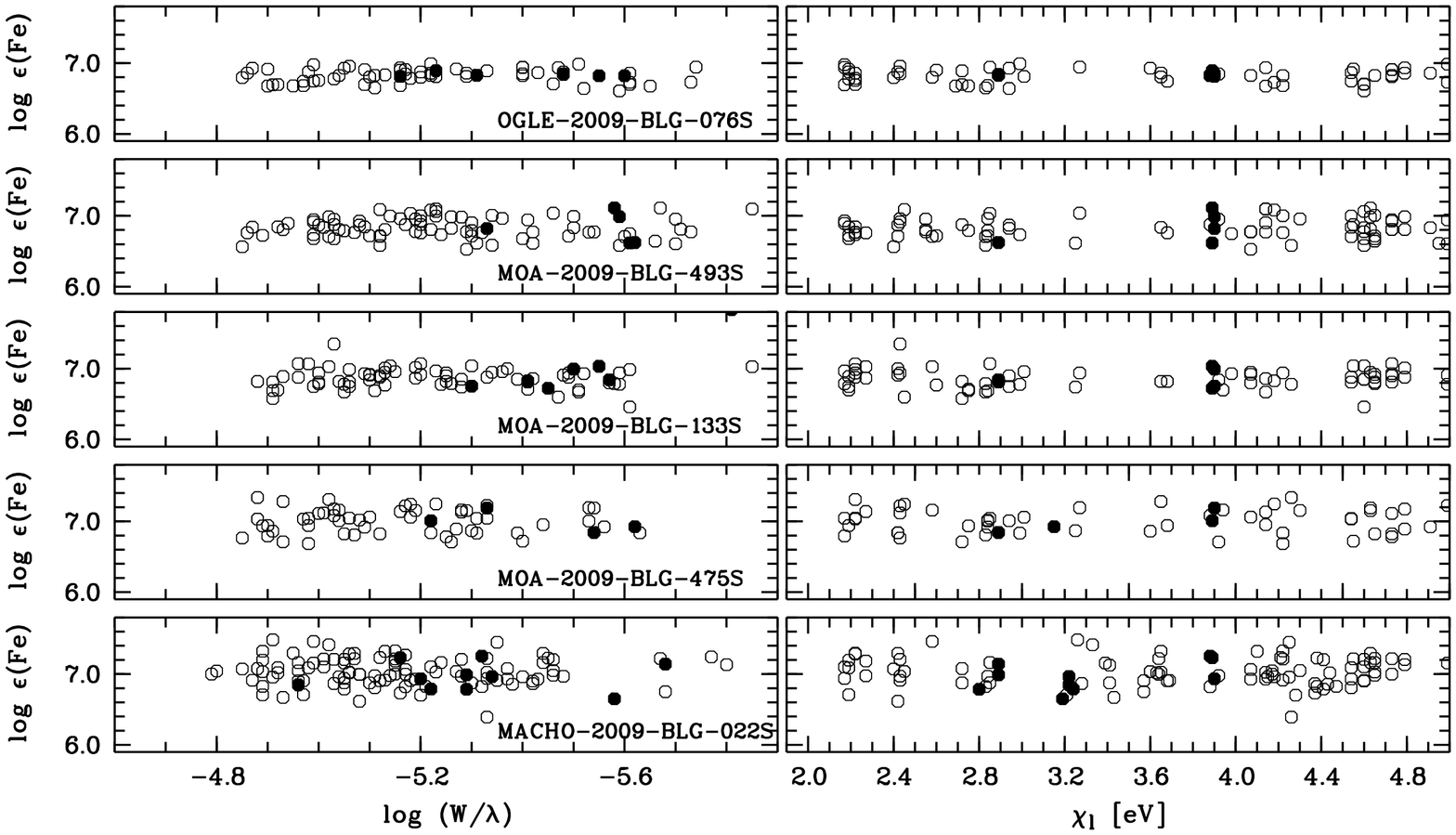}}
\resizebox{\hsize}{!}{
\includegraphics[angle=-90,bb=102 28 560 810,clip]{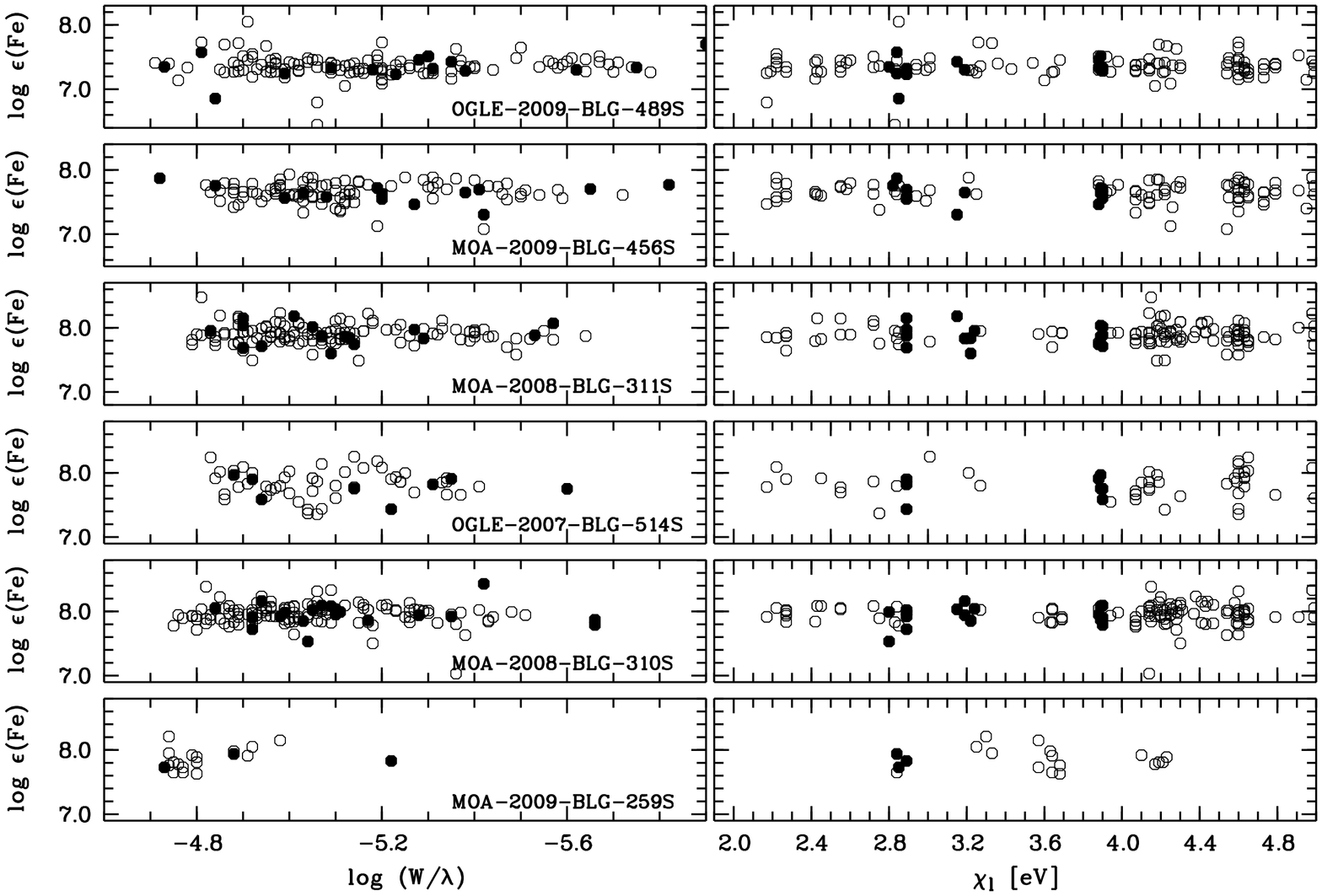}}
\caption{
\label{fig:fetrends}
Diagnostic plots showing absolute Fe abundances 
versus reduced line strength and lower excitation potential.
Open circles indicate abundances from \ion{Fe}{i} lines
and filled circles from \ion{Fe}{ii} lines.
Similar plots for MOA-2006-BLG-099S, OGLE-2006-BLG-265S,
OGLE-2007-BLG-349S, and OGLE-2008-BLG-209S can be found in
\cite{bensby2009}. Note the limited number of lines
for MOA-2009-BLG-259S due to that this star was observed when
only the UVES blue CCD was available. 
}
\end{figure*}

\begin{table*}
\centering
\caption{
Stellar parameters, ages, and radial velocities for the sample of microlensed dwarf stars.$^{\dagger}$
\label{tab:parameters}
}
\setlength{\tabcolsep}{2.5mm}
\begin{tabular}{rlllccrr}
\hline\hline
\noalign{\smallskip}
  \multicolumn{1}{c}{Object}                     &
  \multicolumn{1}{c}{$T_{\rm eff}$}              & 
  \multicolumn{1}{c}{$\log g$}                   &  
  \multicolumn{1}{c}{$\xi_{\rm t}$}              &  
  \multicolumn{1}{c}{[Fe/H]}                     &
  $N_{\ion{Fe}{i}}$,\,\, $N_{\ion{Fe}{ii}}$                              &
  \multicolumn{1}{c}{Age}                        &
  \multicolumn{1}{c}{$v_{\rm r}$}                \\
                     & 
  \multicolumn{1}{c}{$\rm [K]$}                  &  
  \multicolumn{1}{c}{$\rm [cgs]$}                 &   
  \multicolumn{1}{c}{$\rm [\kms]$}               &    
                     &
                     &
  \multicolumn{1}{c}{[Gyr]}                      &
  \multicolumn{1}{c}{[$\kms$]}                   \\  
\noalign{\smallskip}
\hline
\noalign{\smallskip}
 OGLE-2009-BLG-076S & $5877\pm96$  & $4.30\pm0.18$ & $1.61\pm0.16$ &  $-0.72\pm0.07$ &  \phantom{3}66,\,\,\,   \phantom{3}7 &  $11.7 \pm 1.9$    & $+128.7$ \\
  MOA-2009-BLG-493S & $5457\pm98$  & $4.50\pm0.22$ & $0.83\pm0.30$ &  $-0.70\pm0.14$ &  \phantom{3}80,\,\,\,   \phantom{3}5 &  $ 9.1 \pm 4.0$    & $ -14.5$ \\
  MOA-2009-BLG-133S & $5597\pm92$  & $4.40\pm0.27$ & $1.15\pm0.27$ &  $-0.64\pm0.17$ &  \phantom{3}68,\,\,\,   \phantom{3}7 &  $ 9.4 \pm 4.0$    & $ +91.6$ \\
  MOA-2009-BLG-475S & $5843\pm173$ & $4.40\pm0.30$ & $1.31\pm0.31$ &  $-0.54\pm0.17$ &  \phantom{3}53,\,\,\,   \phantom{3}4 &  $ 9.1 \pm 3.7$    & $+137.8$ \\
MACHO-1999-BLG-022S & $5650\pm106$ & $4.05\pm0.20$ & $0.30\pm0.45$ &  $-0.49\pm0.14$ &  \phantom{3}97,\,\,\,             10 &  $11.7 \pm 2.3$    & $ +37.6$ \\
 OGLE-2008-BLG-209S & $5243\pm65$  & $3.82\pm0.16$ & $1.01\pm0.11$ &  $-0.30\pm0.06$ &            146,\,\,\,             19 &  $ 9.5 \pm 3.1$    & $-173.6$ \\
  MOA-2009-BLG-489S & $5634\pm89$  & $4.30\pm0.18$ & $0.68\pm0.17$ &  $-0.18\pm0.11$ &            114,\,\,\,             15 &  $11.0 \pm 2.3$    & $ +96.5$ \\
  MOA-2009-BLG-456S & $5700\pm93$  & $4.24\pm0.18$ & $1.00\pm0.20$ &  $+0.12\pm0.09$ &  \phantom{3}91,\,\,\,             14 &  $ 7.9 \pm 1.6$    & $-164.6$ \\
 OGLE-2007-BLG-514S & $5644\pm130$ & $4.10\pm0.28$ & $1.55\pm0.29$ &  $+0.27\pm0.09$ &  \phantom{3}49,\,\,\,   \phantom{3}8 &  $ 6.9 \pm 1.6$    & $+158.8$ \\
  MOA-2009-BLG-259S & $5000\pm400$ & $4.50\pm0.50$ & $0.50\pm0.75$ &  $+0.33\pm0.40$ &  \phantom{3}16,\,\,\,   \phantom{3}3 &  $ 9.4 \pm 4.0$    & $ +83.2$ \\
  MOA-2008-BLG-311S & $5944\pm68$  & $4.40\pm0.13$ & $1.17\pm0.12$ &  $+0.36\pm0.07$ &            118,\,\,\,             16 &  $ 2.3 \pm 1.0$    & $ -34.1$ \\
  MOA-2008-BLG-310S & $5704\pm65$  & $4.30\pm0.12$ & $1.05\pm0.11$ &  $+0.42\pm0.08$ &            122,\,\,\,             20 &  $ 4.5 \pm 1.0$    & $ +77.5$ \\
 OGLE-2007-BLG-349S & $5229\pm63$  & $4.18\pm0.13$ & $0.78\pm0.13$ &  $+0.42\pm0.08$ &            103,\,\,\,             18 &  $13.6 \pm 1.0$    & $+113.0$ \\
  MOA-2006-BLG-099S & $5741\pm87$  & $4.47\pm0.15$ & $0.84\pm0.14$ &  $+0.44\pm0.10$ &            119,\,\,\,             21 &  $ 2.8 \pm 1.3$    & $ +99.0$ \\
 OGLE-2006-BLG-265S & $5486\pm70$  & $4.24\pm0.15$ & $1.17\pm0.12$ &  $+0.47\pm0.06$ &  \phantom{3}92,\,\,\,             14 &  $ 7.9 \pm 1.3$    & $-154.0$ \\
\noalign{\smallskip}
\hline
\end{tabular}
\flushleft
{\tiny
$^{\dagger}$ 
For the four stars analysed in \cite{bensby2009}: OGLE-2008-BLG-209S,  MOA-2006-BLG-099S, OGLE-2006-BLG-265S, and OGLE-2007-BLG-349S,
we have here updated the estimations of the uncertainties using the method outlined in \cite{epstein2009}.
}
\end{table*}


\subsection{Stellar parameters and elemental abundances}

The determination of stellar parameters and calculation of elemental
abundances were carried out as described in method~1 of
\cite{bensby2009}. Briefly, this method is based on equivalent width 
measurements ($W_{\lambda}$) and one-dimensional LTE model stellar 
atmospheres calculated with the Uppsala MARCS code 
\citep{gustafsson1975,edvardsson1993,asplund1997}. The spectral line 
list is an expanded version of the list used by 
\cite{bensby2003,bensby2005} and is in full given in 
Bensby et al.~(in prep.). Equivalent widths were measured using the 
IRAF\footnote{IRAF is distributed by the National Optical Astronomy 
Observatory, which is operated by the Association of Universities for 
Research in Astronomy, Inc., under co-operative agreement with the 
National Science Foundation.} task {\sc splot}. Gaussian line profiles 
were fitted to the observed lines, but in special cases of strong 
Mg, Ca, Si and Ba lines, Voigt profiles were used to better account for 
the extended wing profiles of these lines.

The effective temperature ($\teff$) is found by requiring excitation 
balance of abundances from \ion{Fe}{i} lines, the microturbulence 
parameter ($\xi_{\rm t}$) by requiring zero slope in the graph where 
abundances from \ion{Fe}{i} lines are plotted versus the reduced 
strength ($\log (W_{\lambda}/\lambda)$) of the spectral lines, and the 
surface gravity ($\log g$) from ionisation balance, i.e., requiring that 
the average abundances from \ion{Fe}{i} and \ion{Fe}{ii} lines are 
equal. Only \ion{Fe}{i} and \ion{Fe}{ii} lines with measured 
equivalent widths smaller than 90\,m{\AA} are used in the 
determination of the stellar parameters. Figure~\ref{fig:fetrends} 
shows the diagnostic plots, $\rm\log \epsilon(Fe)$ versus 
$\log (W_{\lambda}/\lambda)$ and lower excitation potential 
($\chi_{\rm e}$), for the stars.

To relate the elemental abundances to those in the Sun we determine 
our own solar abundances. 
The equivalent widths we measure in the solar spectrum that 
was obtained by observing the asteroid Pallas with UVES on April 11, 
2009, show very good agreement with the equivalent widths of 
several solar spectra (average of Ganymede, Ceres, Vesta, Moon, 
and sky spectra) in Bensby et al.~(in prep.).
On average the measurements of the Pallas solar spectrum is only 
0.3\,\% larger, which is truly negligible. Hence, to ensure
that the normalised abundances for the microlensed dwarf stars are
on the same baseline as the sample of $\sim$\,700 thin and thick disc
dwarf stars in Bensby et al.~(in prep.) we use the average equivalent
widths based on measurements in all solar spectra 
(see Bensby et al.~in prep.).
 
Final abundances are normalised on a line-by-line 
basis and then we take the median value for each element.
In a few cases when the equivalent width of an Fe line in the Sun 
was larger than 90\,m{\AA}, or when a Ti or Cr line were larger
than 110\,m{\AA}, and these lines were measured in the Bulge dwarf 
star, we normalised the abundance for that line with the average 
abundance from all other lines that were measured in the solar 
spectrum for that element. These cases are marked by ``av" in col.~7 
in Table~\ref{tab:abundances}.

Final stellar parameters for our targets are given in
Table~\ref{tab:parameters}.
All measured equivalent widths and elemental abundances for individual 
spectral lines are given in Table~\ref{tab:abundances}, while
Table~\ref{tab:abundances2} gives the normalised abundance ratios.

\begin{figure*}
\resizebox{\hsize}{!}{
\includegraphics[angle=-90,bb=150 30 520 450,clip]{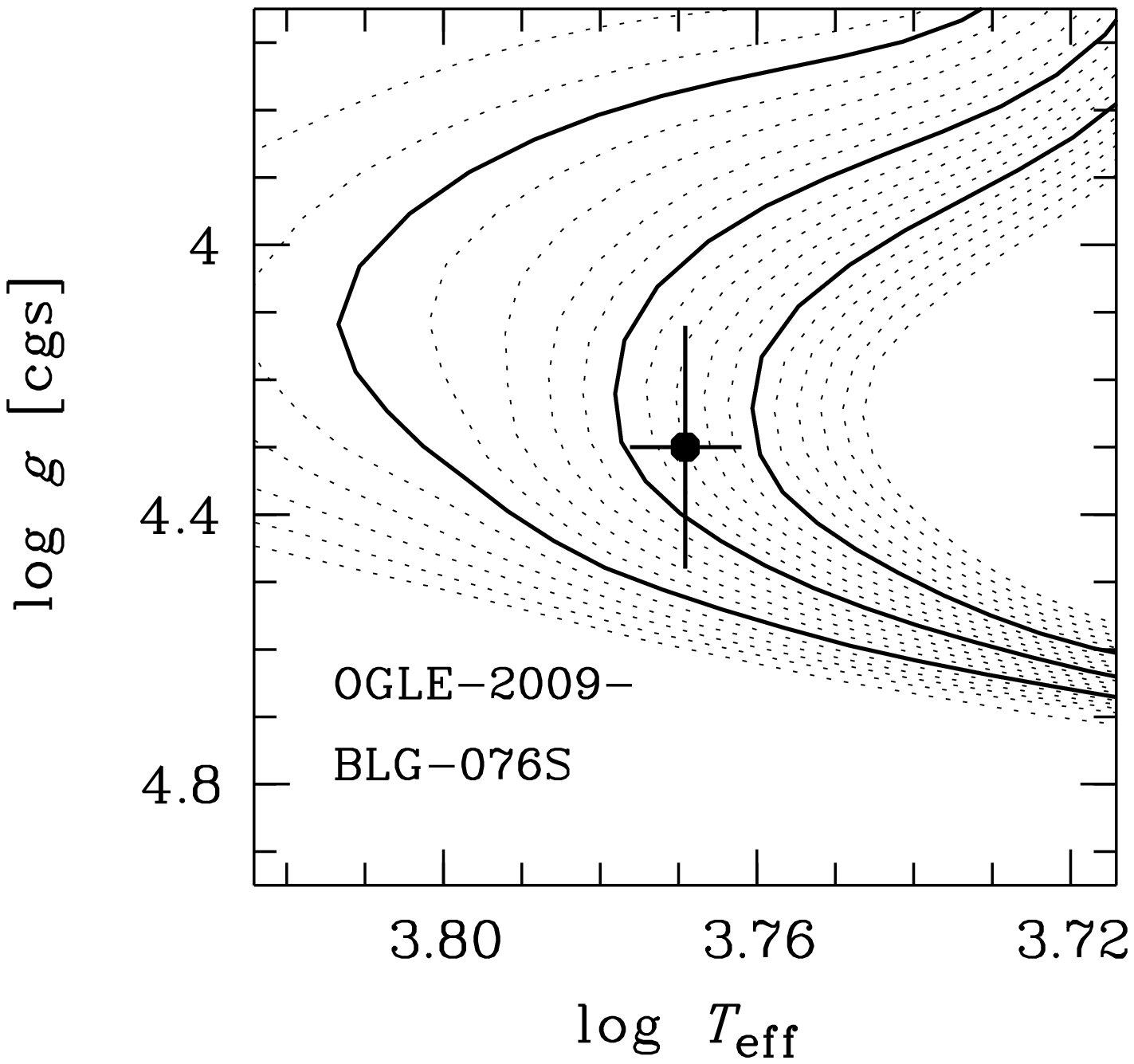}
\includegraphics[angle=-90,bb=150 85 520 450,clip]{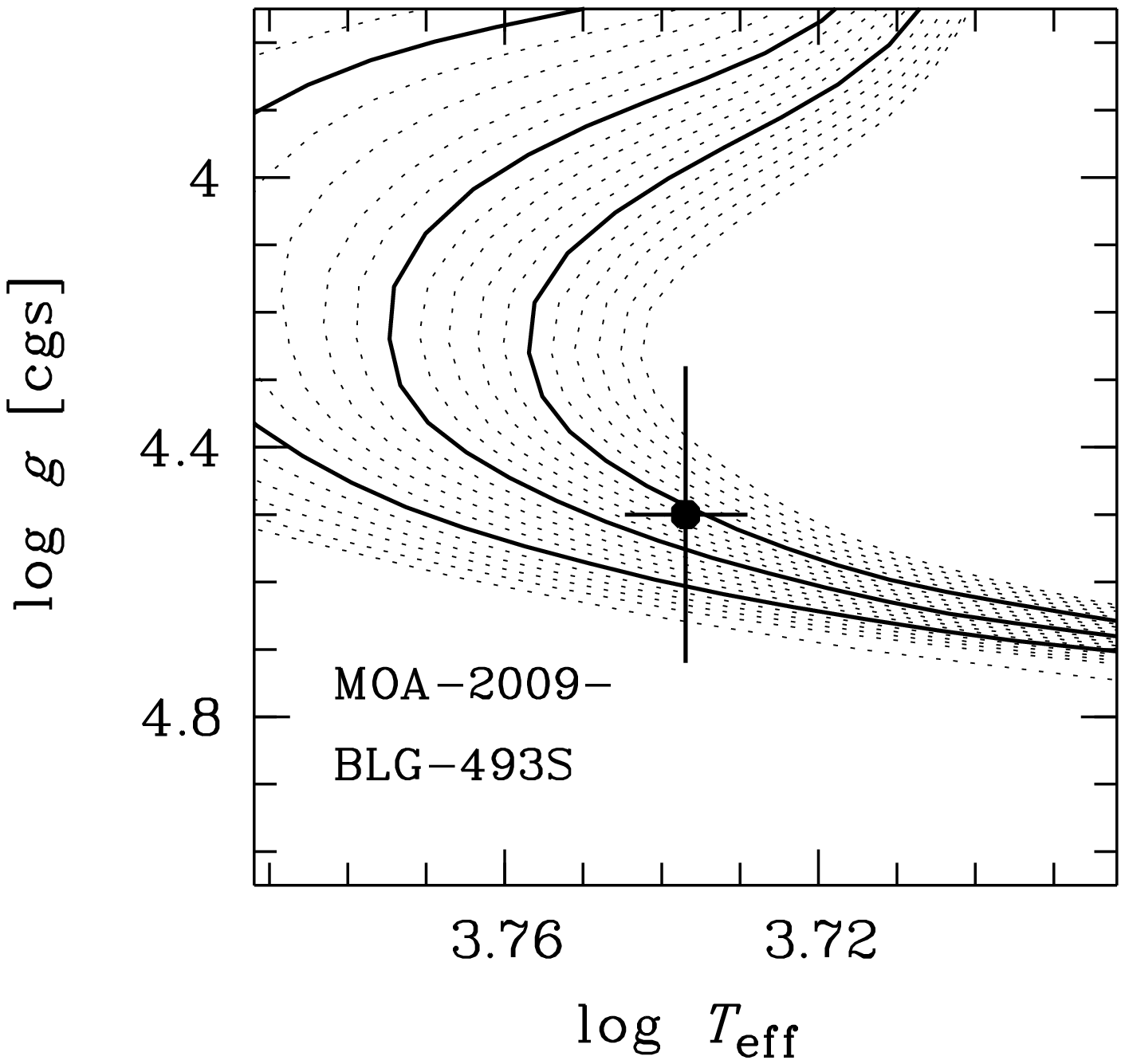}
\includegraphics[angle=-90,bb=150 85 520 450,clip]{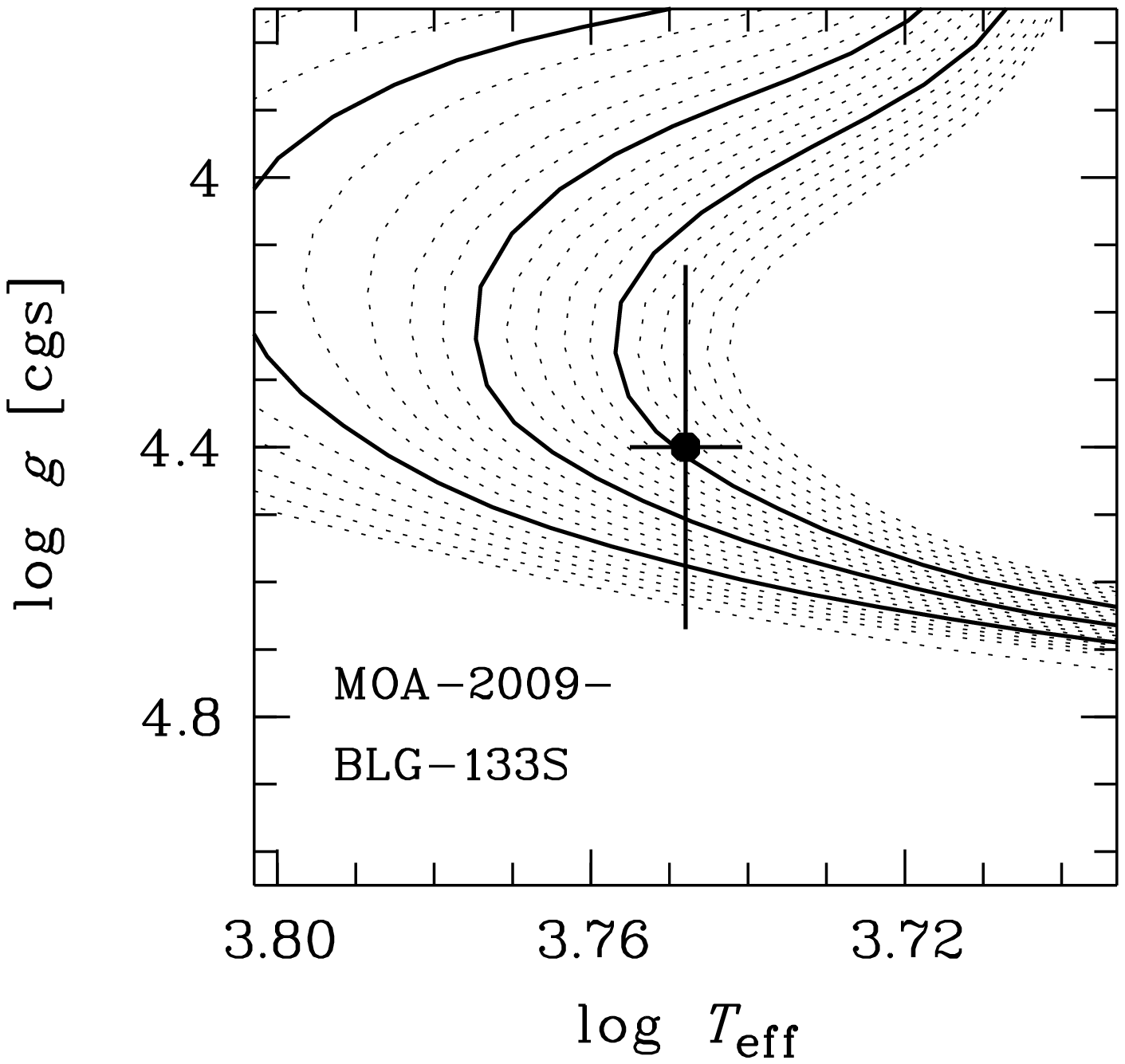}
\includegraphics[angle=-90,bb=150 85 520 485,clip]{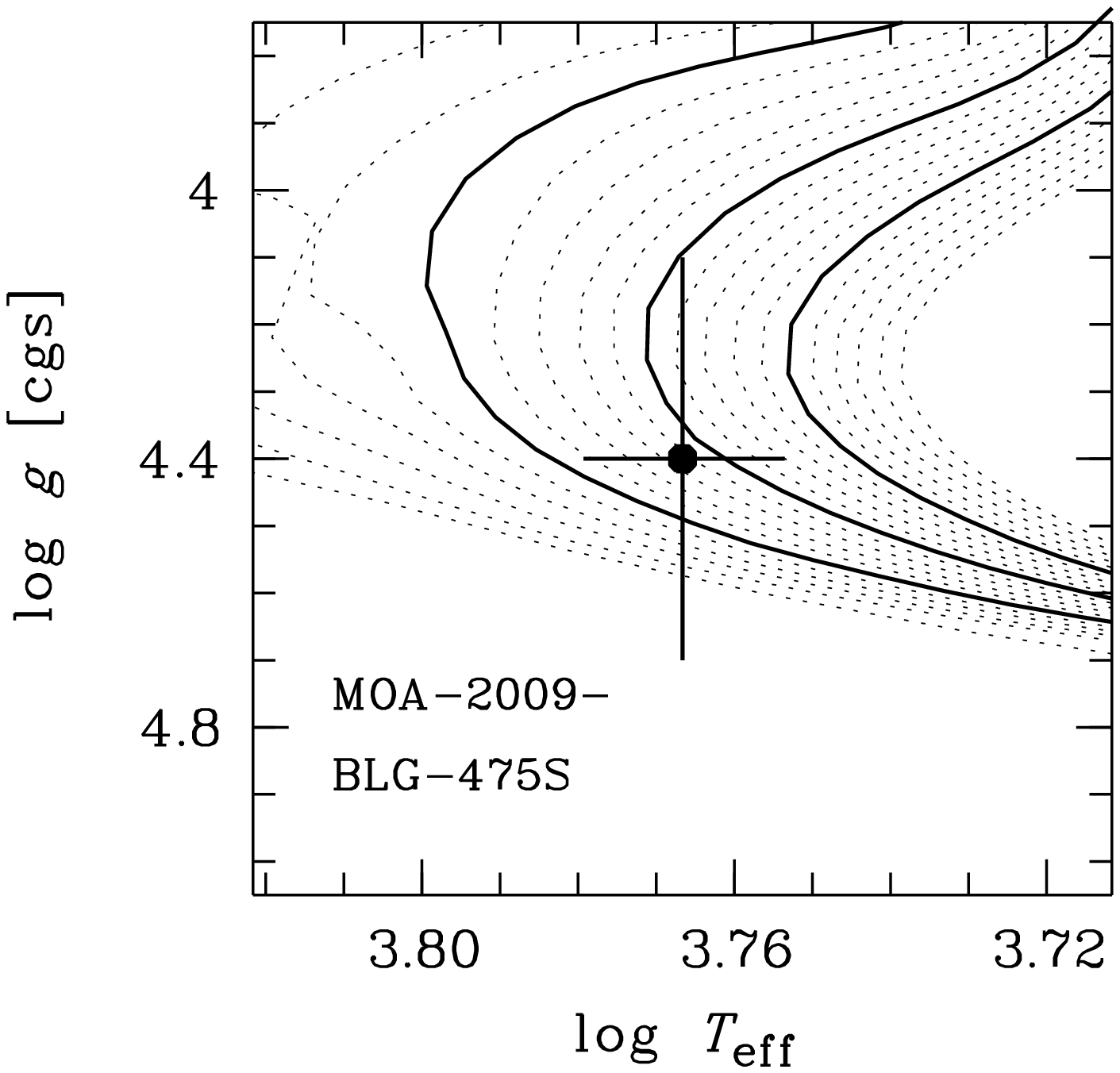}
}
\resizebox{\hsize}{!}{
\includegraphics[angle=-90,bb=150 30 520 450,clip]{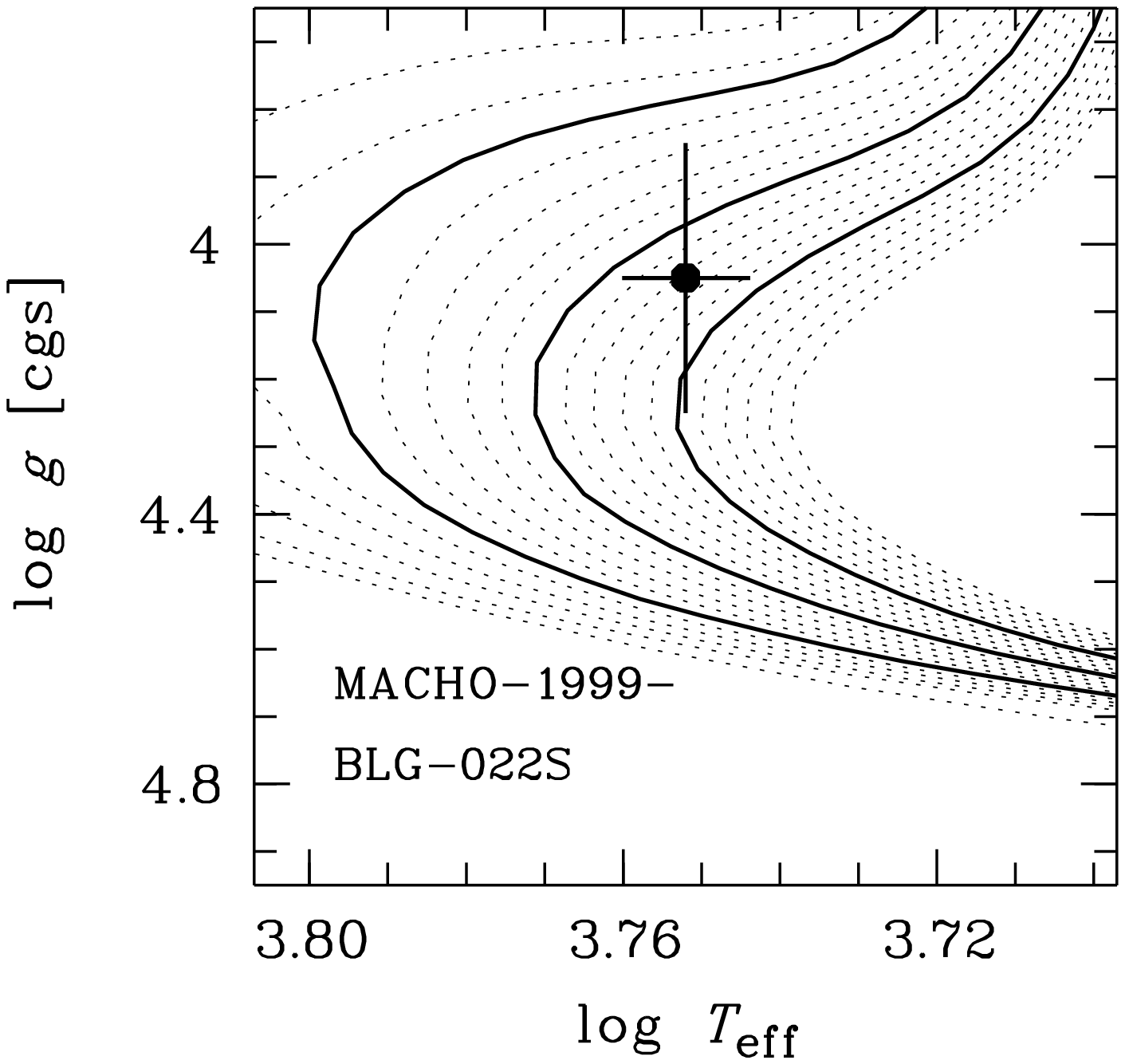}
\includegraphics[angle=-90,bb=150 85 520 450,clip]{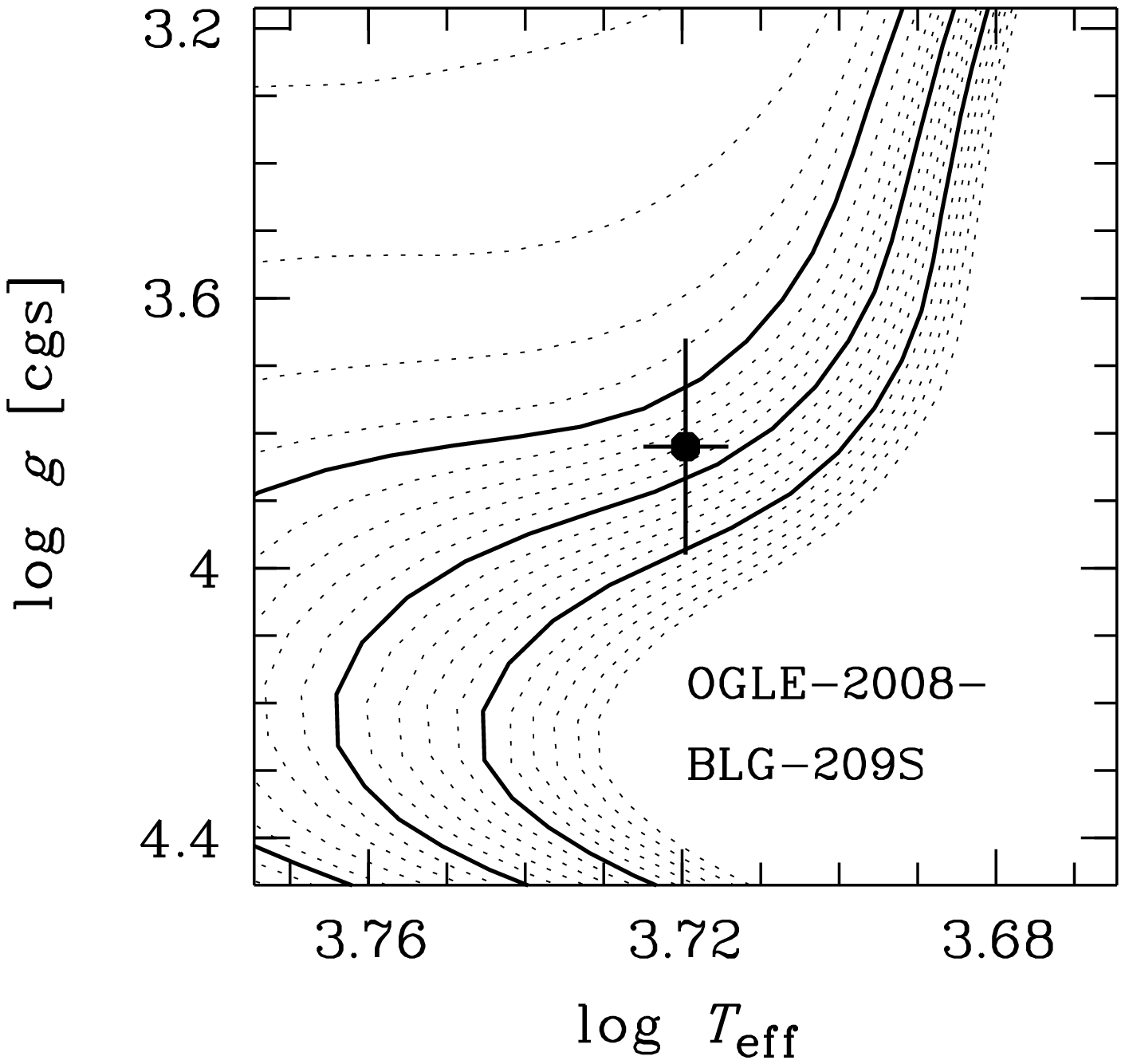}
\includegraphics[angle=-90,bb=150 85 520 450,clip]{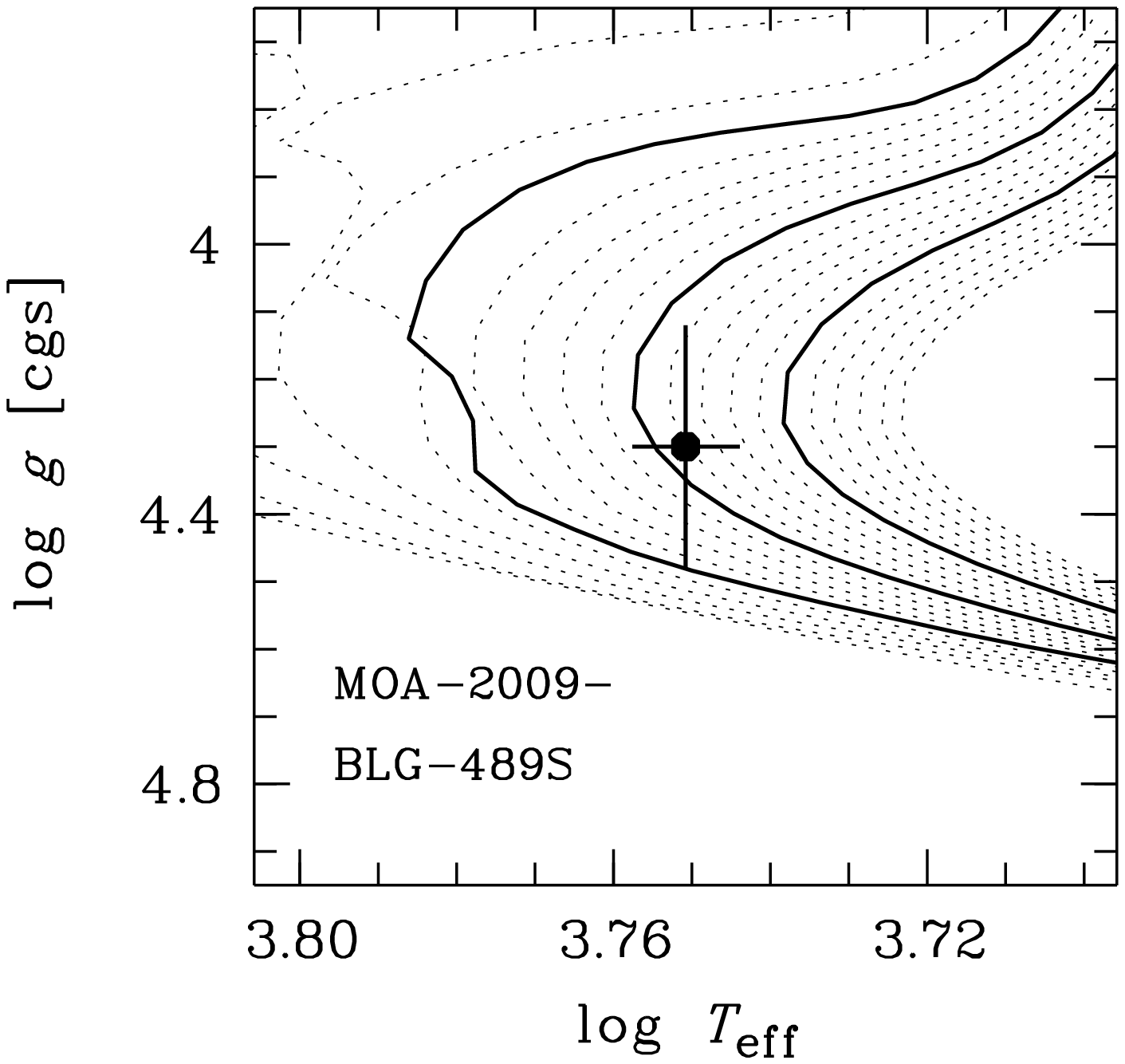}
\includegraphics[angle=-90,bb=150 85 520 485,clip]{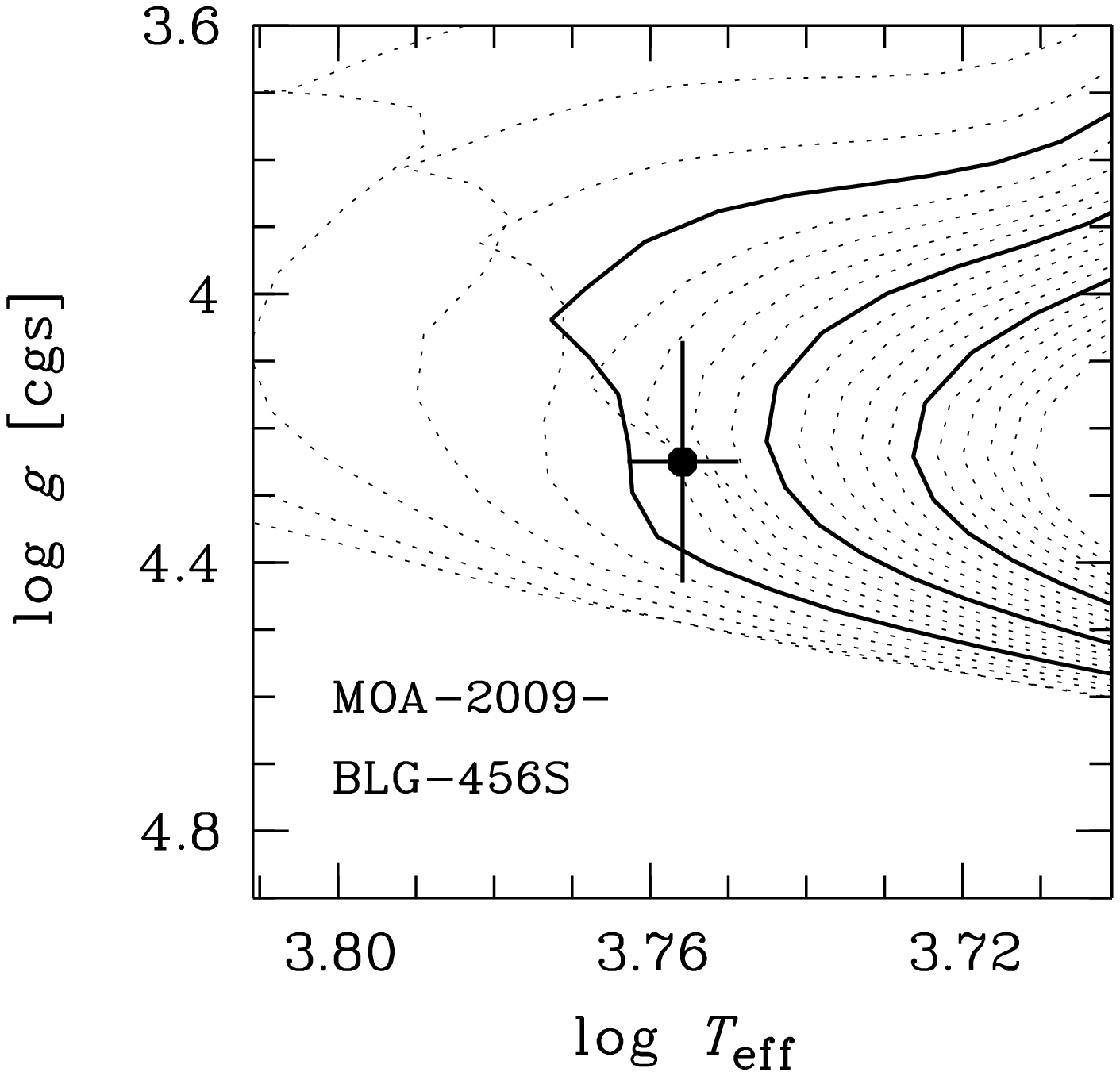}
}
\resizebox{\hsize}{!}{
\includegraphics[angle=-90,bb=150 30 520 450,clip]{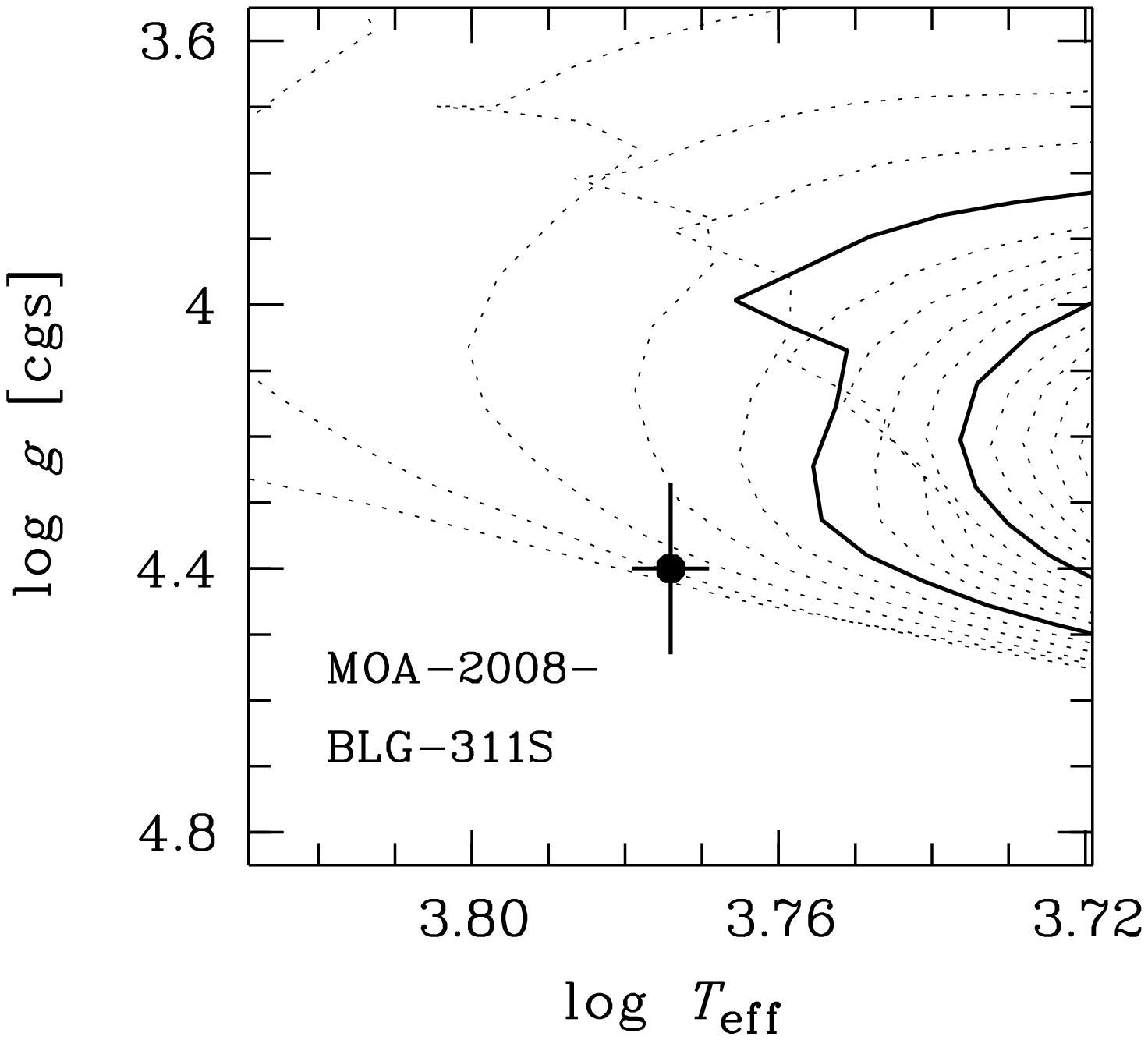}
\includegraphics[angle=-90,bb=150 85 520 450,clip]{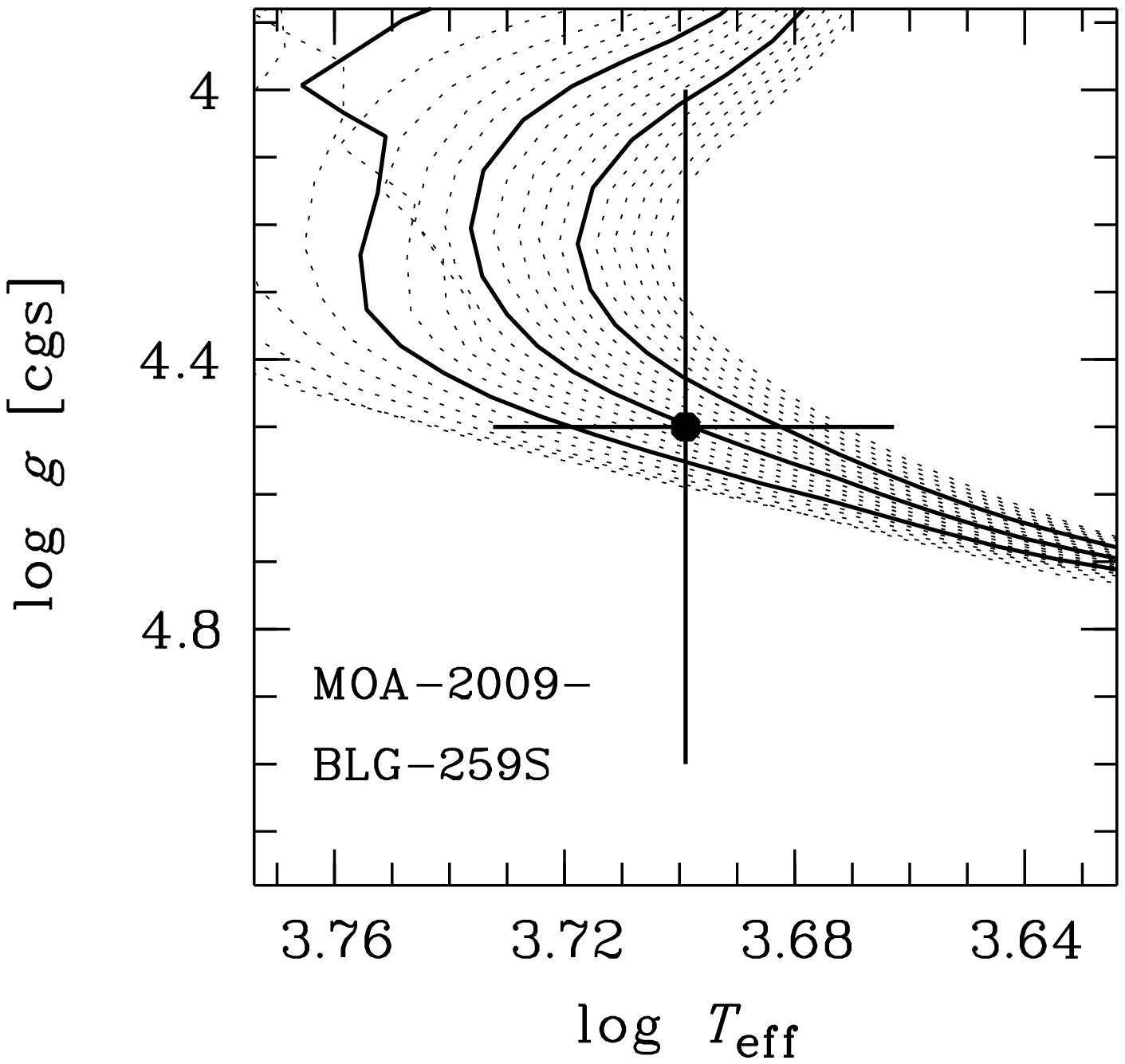}
\includegraphics[angle=-90,bb=150 85 520 450,clip]{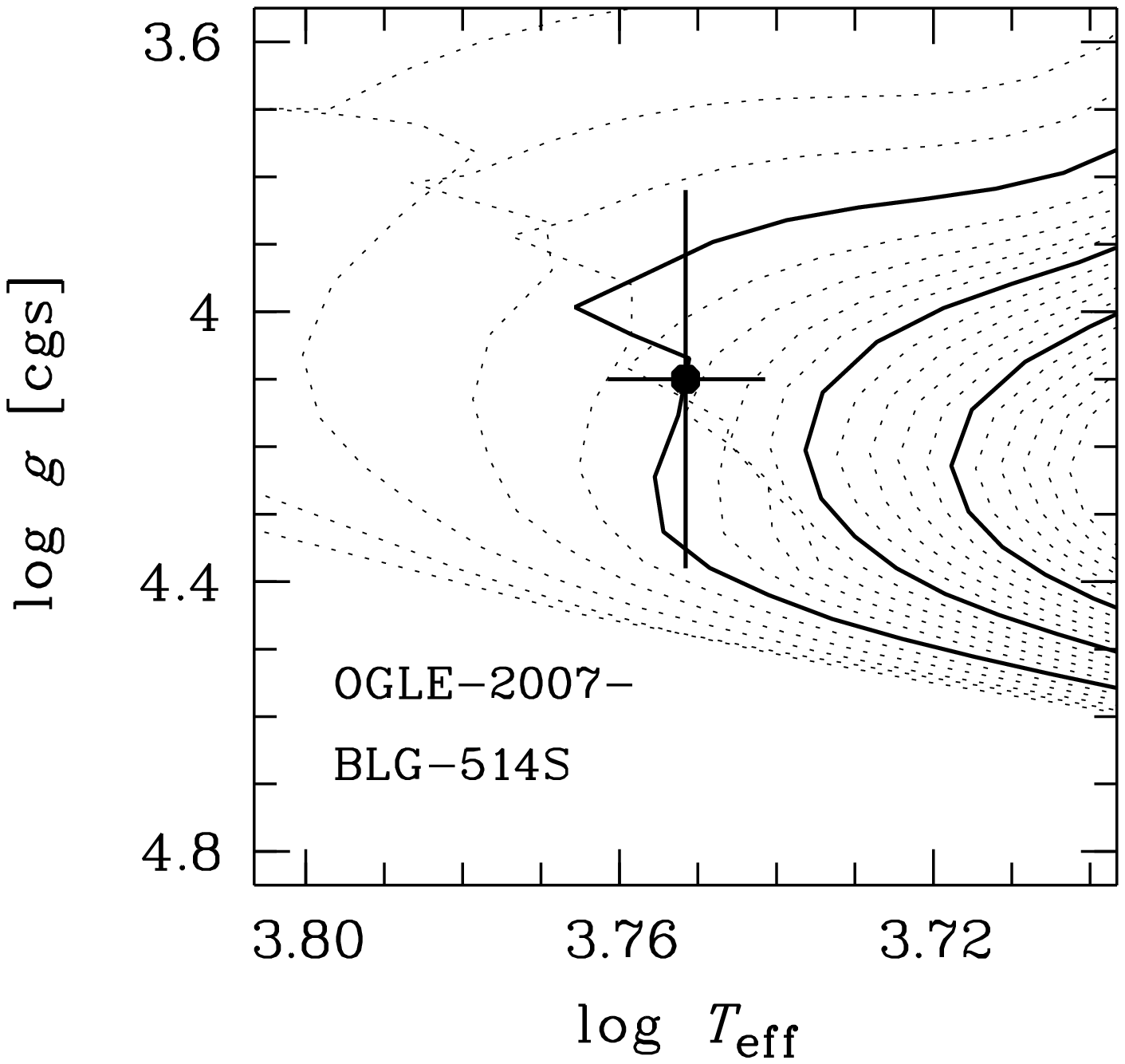}
\includegraphics[angle=-90,bb=150 85 520 485,clip]{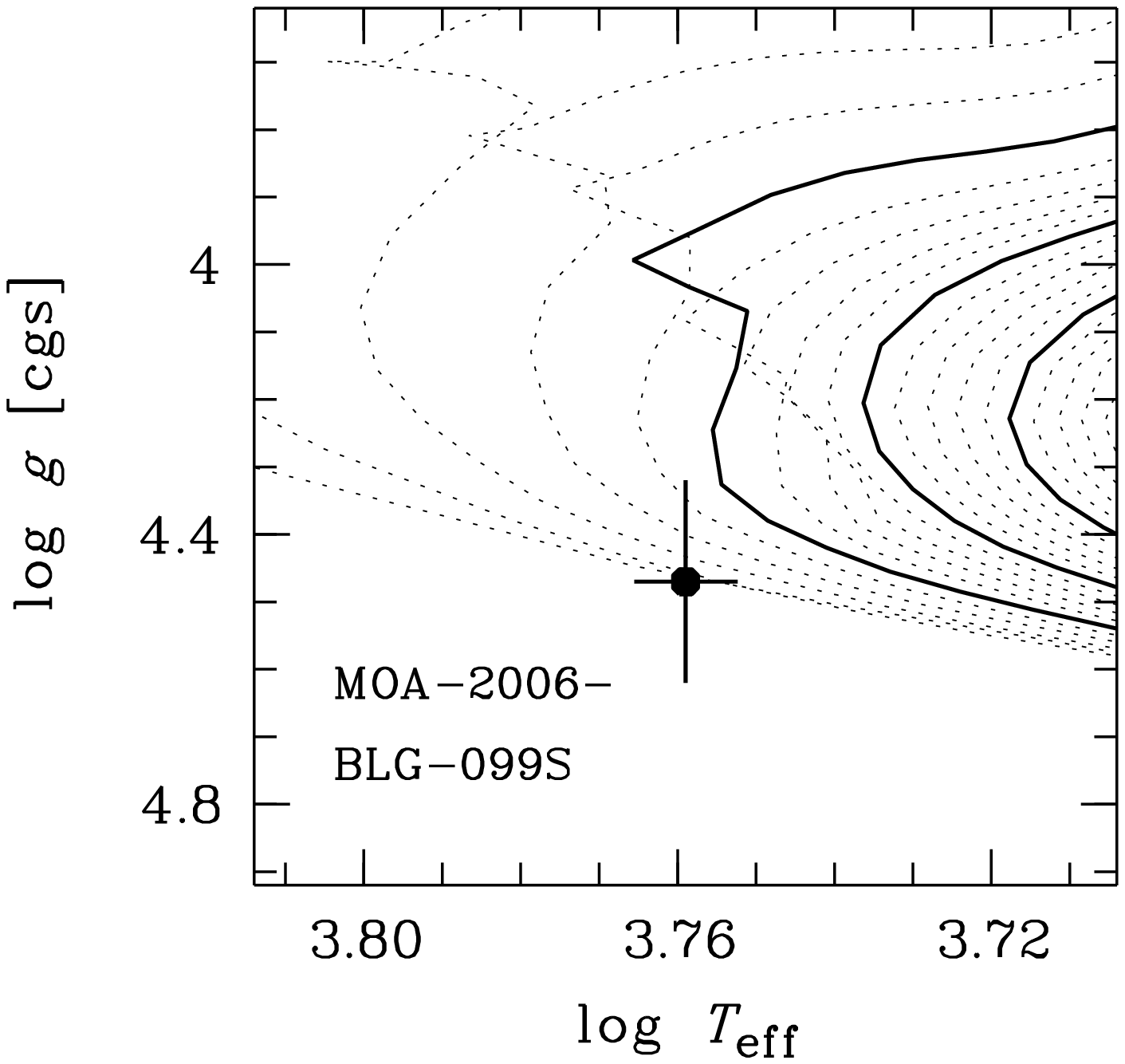}
}
\resizebox{\hsize}{!}{
\includegraphics[angle=-90,bb=150 30 550 450,clip]{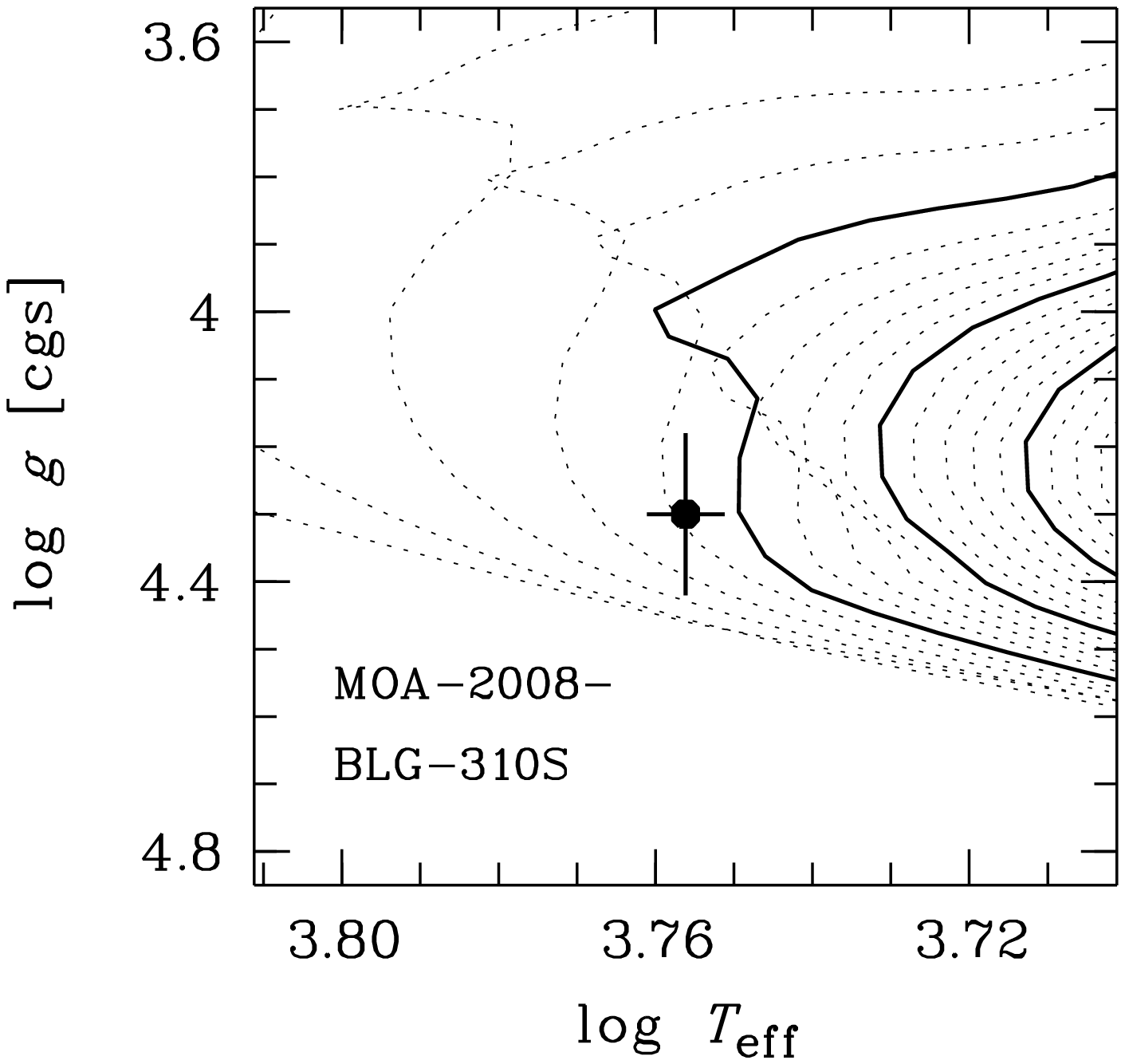}
\includegraphics[angle=-90,bb=150 85 550 450,clip]{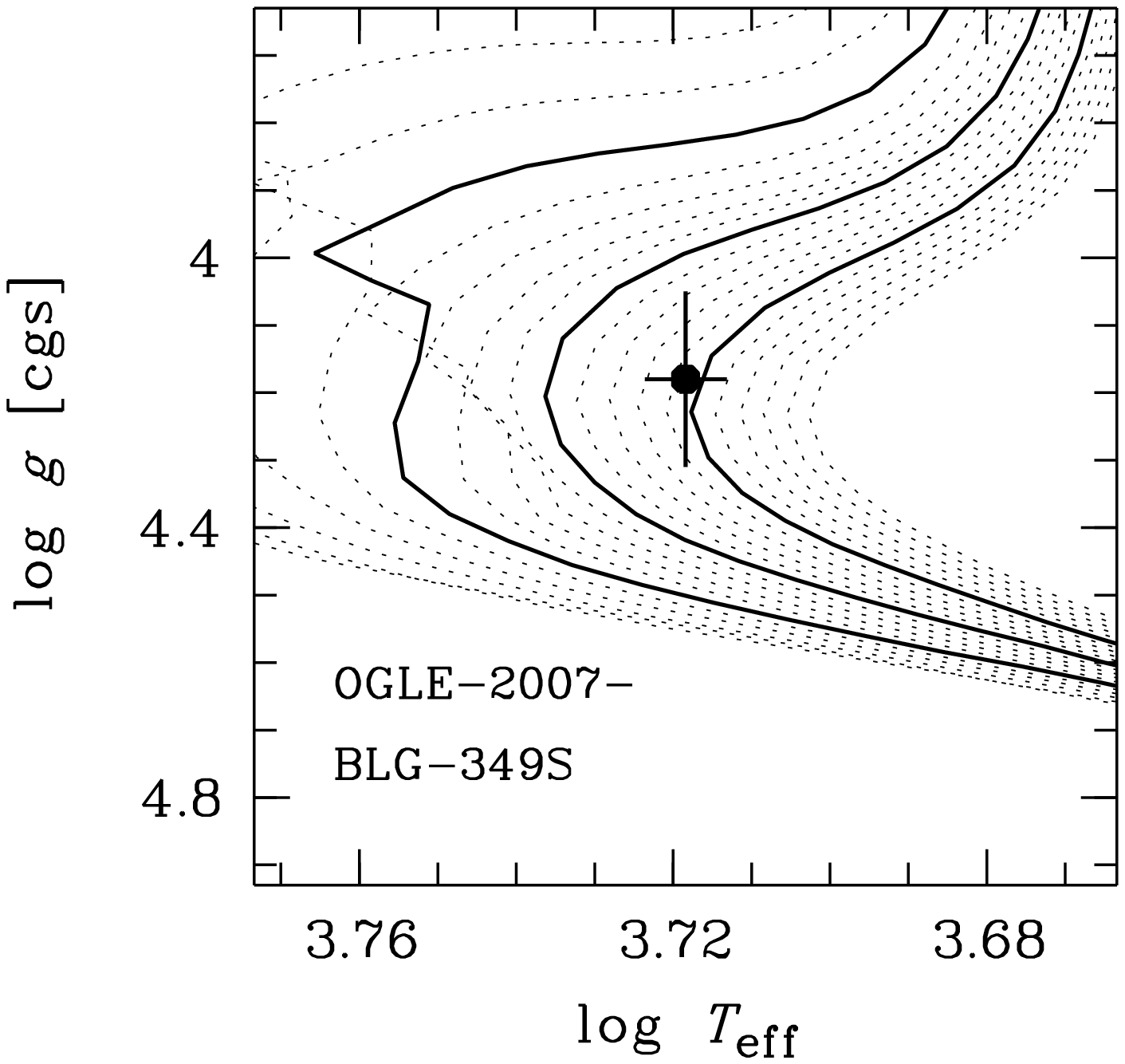}
\includegraphics[angle=-90,bb=150 85 550 450,clip]{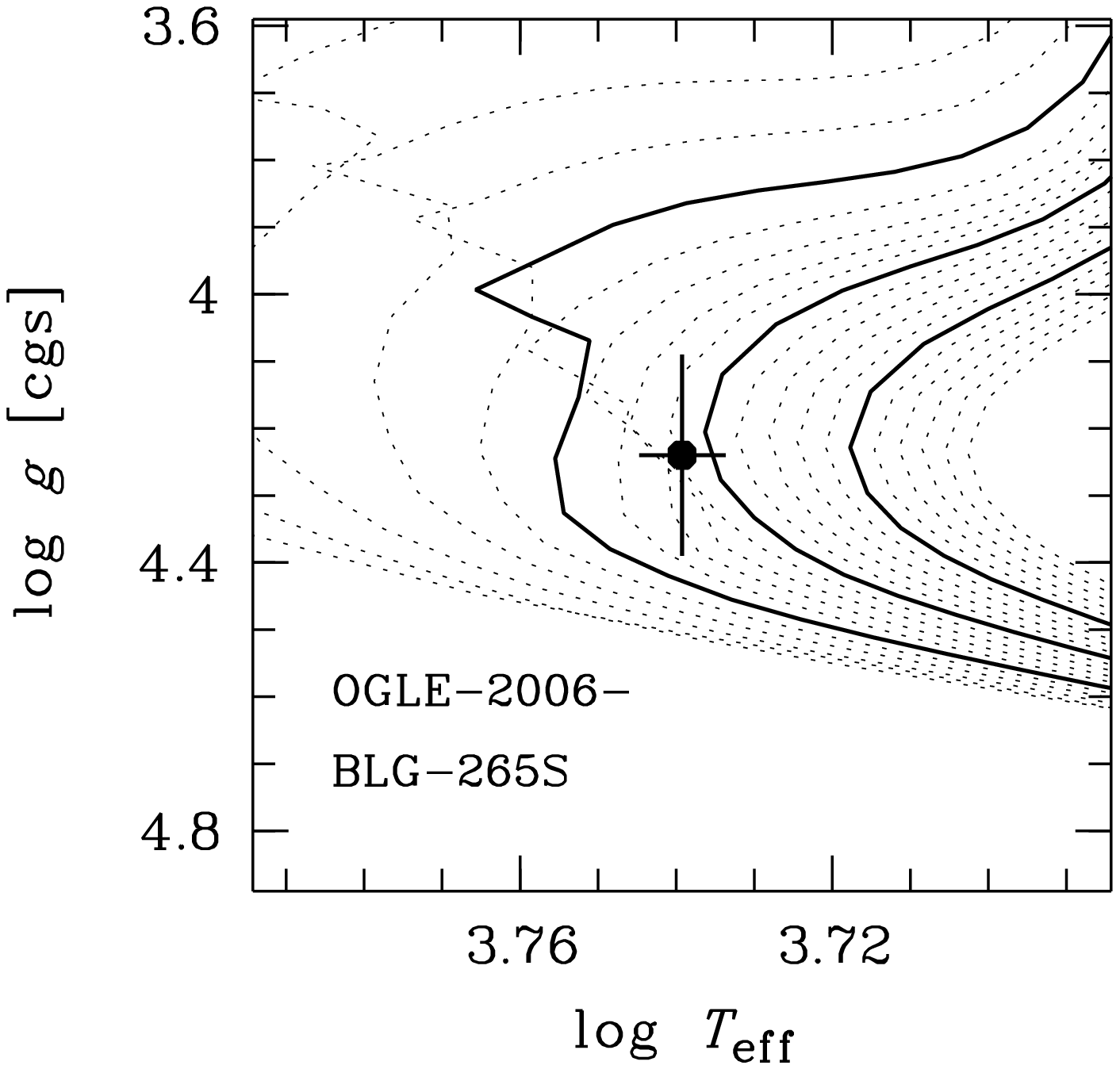}
\includegraphics[angle=-90,bb=150 85 151 485,clip]{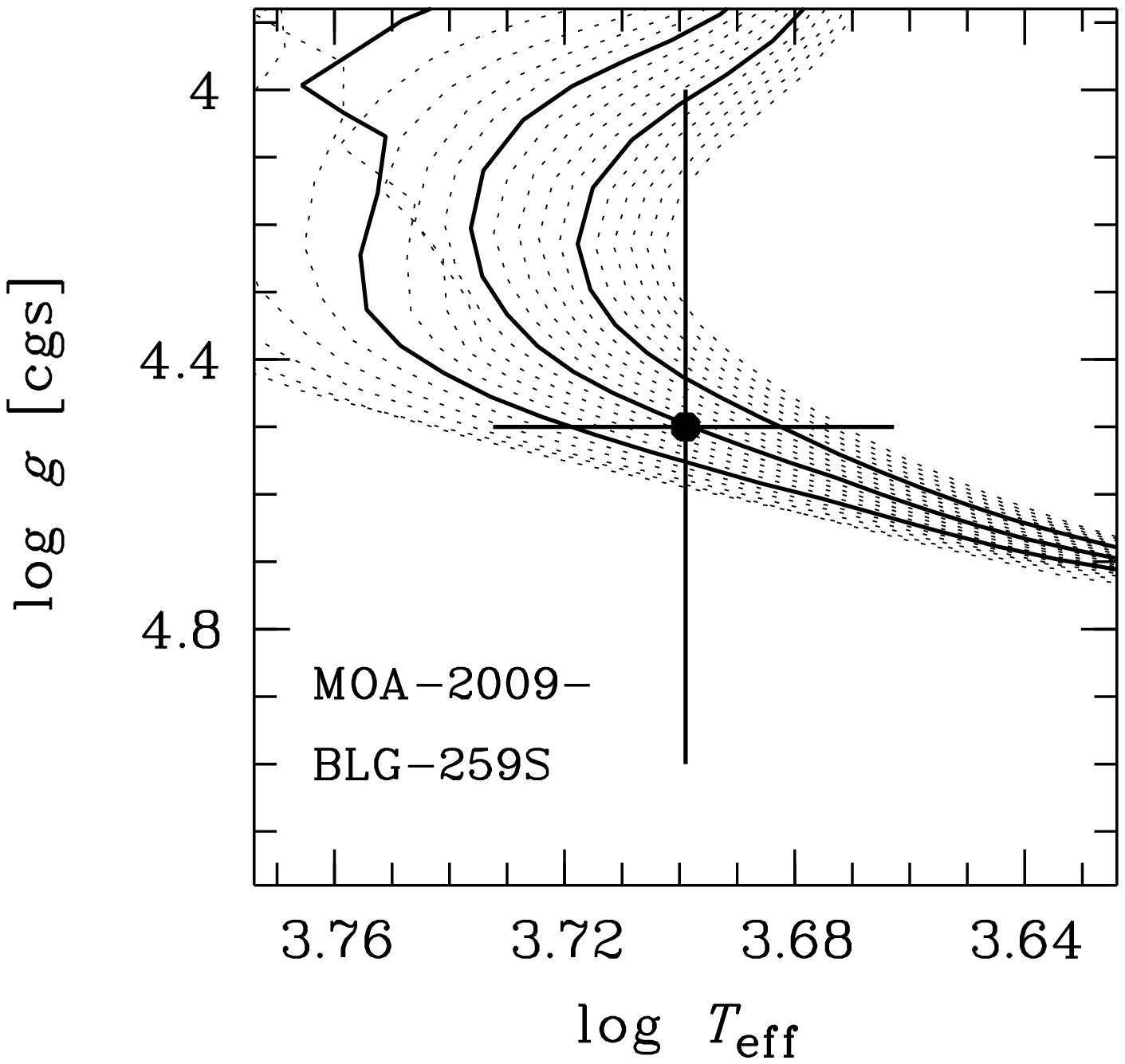}
}
\caption{Illustration of the estimation of stellar ages using the 
$\alpha$-enhanced 
isochrones from \cite{demarque2004}. Each set of isochrones have been
calculated with the same metallicity and $\alpha$-enhancement as 
derived for the stars. In each plot the solid lines represent
isochrones with ages of 5, 10, and 15\,Gyr (from left to right).
Dotted lines are isochrones in steps of 1\,Gyr, ranging from 0.1\,Gyr
to 20\,Gyr. Error bars represent the uncertainties
in $\teff$ and $\log g$ as given in Table~\ref{tab:parameters}.
\label{fig:ages}
}
\end{figure*}

\subsection{Additional dwarf stars}

In order to increase the sample size of microlensed dwarf stars, we
include the two stars MOA-2009-BLG-310S, and --311S, recently
published by \cite{cohen2009}, and OGLE-2007-BLG-514S by
\cite{epstein2009}. The metallicities that were found for these three
stars by \cite{cohen2009} and \cite{epstein2009} are 
$\rm [Fe/H]=+0.41$, $\rm [Fe/H]=+0.26$, and 
$\rm [Fe/H]=+0.33$, respectively.

The spectra for these stars were kindly provided by the authors
and we have re-analysed them using our methods in order to have
all 15 microlensed dwarf stars on the same baseline. 
The values we find for these stars are listed in 
Table~\ref{tab:parameters}, and they are generally in good agreement 
with what were found in \cite{cohen2009} and \cite{epstein2009}. 
The main differences are that we derive a 240\,K higher $\teff$ and 
0.2\,dex higher $\log g$ 
for MOA-2009-BLG-311S, and a 0.2\,dex lower $\log g$ for 
OGLE-2007-BLG-514S. The other differences are within the 
estimated uncertainties.

\subsection{Error analysis}
\label{sec:errors}

A rigorous error analysis as outlined in \cite{epstein2009} 
has been performed for the microlensed dwarf stars.
This method takes into account the uncertainties in the four 
observables that were used to find the stellar parameters, i.e. the
uncertainty of the slope in the graph of \ion{Fe}{i} abundances 
versus lower excitation potential; the uncertainty of the slope in 
the graph of
\ion{Fe}{i} abundances versus line strength; the uncertainty
in the difference between \ion{Fe}{i} and \ion{Fe}{ii} abundances; 
and the uncertainty in the difference
between input and output metallicities. The method also accounts for
abundance spreads (line-to-line scatter) as well as how the
abundances for each element reacts to changes in the stellar parameters.

The resulting errors in the stellar parameters are given together with 
the best fit values of the stellar parameters in 
Table~\ref{tab:parameters}.
The errors in the abundance ratios are given in 
Table~\ref{tab:abundances2}.

\subsection{Stellar ages}
\label{sec:ages}

Stellar ages were determined as described in \cite{melendez2010}. 
We interpolated a fine grid of $\alpha$-enhanced Yonsei-Yale  (Y$^2$) 
isochrones by \cite{demarque2004}, adopting $\rm [\alpha/Fe] = 0$ for 
$\rm [Fe/H] > 0$, $\rm [\alpha/Fe] = -0.3 \times[Fe/H]$ for 
$\rm -1\leq [Fe/H] \leq0$, and $\rm [\alpha/Fe] = +0.3$ for 
$\rm [Fe/H] < -1$. At a given metallicity, we searched for all 
solutions allowed by the error bars in $\teff$, $\log g$ and [Fe/H], 
adopting as final result the median age and as error the standard 
deviation. Column~6 in Table~\ref{tab:parameters} give the median age, 
and column~7 the error.  Figure~\ref{fig:ages} shows the 15 microlensed 
dwarf stars together with the Y$^2$ isochrones in the 
$\log\teff - \log g$ plane.

\begin{table}
\centering
\caption{
Comparison of colours and effective temperatures as determined
from spectroscopy and microlensing techniques$^\dag$. 
\label{tab:photometry}
}
\setlength{\tabcolsep}{1.4mm}
\begin{tabular}{rccc|cc}
\hline\hline
\noalign{\smallskip}
  \multicolumn{1}{c}{Object}                 &
  \multicolumn{1}{c}{$M_{I}$}                &  
  \multicolumn{1}{c}{$(V$--$I)_{0}$}                &  
  \multicolumn{1}{c|}{$\teff^{phot}$}             &
  \multicolumn{1}{c}{$\teff$}                &
  \multicolumn{1}{c}{$(V$--$I)_{0}^{spec}$}      \\
\noalign{\smallskip}
\hline
\noalign{\smallskip}
 OGLE-2009-BLG-076S   &  4.19 & 0.67 & 5775  & 5877 & 0.65 \\ 
  MOA-2009-BLG-493S   &  3.34 & 0.79 & 5325  & 5457 & 0.75 \\ 
  MOA-2009-BLG-133S   &  4.19 & 0.68 & 5730  & 5597 & 0.71 \\ 
  MOA-2009-BLG-475S   &  4.25 & 0.59 & 6150  & 5843 & 0.65 \\ 
MACHO-1999-BLG-022S   &   --  &  --  & --    & 5650 & 0.71 \\ 
 OGLE-2008-BLG-209S   &  2.52 & 0.71 & 5670  & 5243 & 0.82 \\ 
  MOA-2009-BLG-489S   &  3.39 & 0.85 & 5200  & 5634 & 0.71 \\ 
  MOA-2009-BLG-456S   &  2.76 & 0.66 & 5870  & 5700 & 0.71 \\ 
 OGLE-2007-BLG-514S   &  4.68 & 0.70 & 5760  & 5644 & 0.73 \\ 
  MOA-2009-BLG-259S   &  2.91 & 0.79 & 5450  & 5000 & 0.97 \\ 
  MOA-2008-BLG-311S   &  3.85 & 0.66 & 5880  & 5944 & 0.65 \\ 
  MOA-2008-BLG-310S   &  3.46 & 0.69 & 5780  & 5704 & 0.72 \\ 
 OGLE-2007-BLG-349S   &  4.54 & 0.78 & 5490  & 5229 & 0.87 \\ 
  MOA-2006-BLG-099S   &  3.81 & 0.74 & 5620  & 5741 & 0.70 \\ 
 OGLE-2006-BLG-265S   &  3.59 & 0.68 & 5840  & 5486 & 0.78 \\ 
\noalign{\smallskip}
\hline
\end{tabular}
\flushleft
{\tiny
$^{\dagger}$
Columns 2 and 3 give the absolute dereddened magnitudes and colours
based on microlensing techniques; based on the colours in 
col.~3; col.~4 gives the inferred effective
temperatures using the  colour--[Fe/H]--$\teff$ calibrations
by \cite{ramirez2005}; col.~5 gives our spectroscopic
temperatures (same as in col.~2 in Table~2); and col.~6
gives the colours, based on the  colour--[Fe/H]--$\teff$ calibrations
by \cite{ramirez2005}, that the spectroscopic temperatures 
in col.~5 gives.
}
\end{table}

\begin{table*}
\centering
\caption{
Measured equivalent widths and calculated elemental abundances for each star.$^\dag$                                 
\label{tab:abundances}
}
\setlength{\tabcolsep}{2.2mm}
\tiny
\begin{tabular}{ccc|ccc|cc|cc|cc|cc|cc|cc|cc}
\hline\hline
\noalign{\smallskip}
                                  &  
$\lambda$                         &
$\chi_{\rm l}$                    &
\multicolumn{3}{c|}{Sun}          &
\multicolumn{2}{c|}{ob09076}      &
\multicolumn{2}{c|}{mb09133}      &
\multicolumn{2}{c|}{mb09456}      &
\multicolumn{2}{c|}{mb09475}      &
\multicolumn{2}{c|}{mb09489}      &
\multicolumn{2}{c|}{mb09493}      & 
\multicolumn{2}{c}{mb99022}      \\
\noalign{\smallskip}
El.                               &
[{\AA}]                           &
[eV]                              &
$W_{\rm \lambda,\odot}$           &
$\epsilon (X)_{\odot}$            &
flg                               &
$W_{\rm \lambda}$                 &
$\epsilon (X)$                    &
$W_{\rm \lambda}$                 &
$\epsilon (X)$                    &
$W_{\rm \lambda}$                 &
$\epsilon (X)$                    &
$W_{\rm \lambda}$                 & 
$\epsilon (X)$                    &
$W_{\rm \lambda}$                 & 
$\epsilon (X)$                    &
$W_{\rm \lambda}$                 & 
$\epsilon (X)$                    &
$W_{\rm \lambda}$                 &
$\epsilon (X)$                    \\
\noalign{\smallskip}
\hline
\noalign{\smallskip}
 \ion{Al}{i} & 5557.06  &  3.14        &  10.4 & 6.44 & -- &       &       &       &      &       &       &       &      &       &             &           &      &      &     \\
 \ion{Al}{i} & 6696.02  &  3.14        &  44.9 & 6.62 & -- &       &       &       &      &       &       &       &      &       &             &           &      &      &     \\
\vdots &
\vdots &
\vdots &
\vdots &
\vdots &
\vdots &
\vdots &
\vdots &
\vdots &
\vdots &
\vdots &
\vdots &
\vdots &
\vdots &
\vdots &
\vdots &
\vdots &
\vdots &
\vdots &
\vdots \\
\noalign{\smallskip}
\hline
\end{tabular}
\flushleft
{\tiny
$^{\dagger}$
For each line we give the $\log gf$ value, lower excitation potential ($\chi_{\rm l}$), measured equivalent widths ($W_{\rm \lambda}$),
derived absolute abundance ($\log \epsilon (X)$).
The table is only available in the online version of the paper and in electronic
form at the CDS via anonymous ftp to {\tt cdsarc.u-strasbg.fr (130.79.125.5)}
or via {\tt http://cdsweb.u-strasbg.fr/Abstract.html}.
}
\end{table*}
\begin{table*}
\centering
\caption{
Elemental abundance ratios, errors in the abundance ratios, and number of lines used,  for 13 of the 15 microlensed dwarf stars$^\dag$.
\label{tab:abundances2}
}
\setlength{\tabcolsep}{1.5mm}
\tiny
\begin{tabular}{rrrrrrrrrrrrrr}
\hline\hline
\noalign{\smallskip}
                    &  [Fe/H] &  [O/Fe]$^\ddag$  &  [Na/Fe] &  [Mg/Fe] &  [Al/Fe] &  [Si/Fe] &  [Ca/Fe] &  [Ti/Fe] &  [Cr/Fe] &  [Ni/Fe] &  [Zn/Fe] &  [Y/Fe] &  [Ba/Fe] \\
\noalign{\smallskip}
\hline
\noalign{\smallskip}
 OGLE-2009-BLG-076S & $-$0.72  & $ $0.50  &  $ $0.11  &  0.36  &  $ $0.30  &  $ $0.24  &  $ $0.34  &  $ $0.30  &  $ $  --  &  $ $0.07  &  $ $  --  &  $ $  --  &  $-$0.07  \\
  MOA-2009-BLG-493S & $-$0.70  & $ $0.43  &  $ $0.13  &  0.37  &  $ $0.28  &  $ $0.18  &  $ $0.17  &  $ $0.34  &  $ $0.01  &  $ $0.04  &  $ $0.25  &  $ $  --  &  $-$0.07  \\
  MOA-2009-BLG-133S & $-$0.64  & $ $0.47  &  $ $0.13  &  0.39  &  $ $0.28  &  $ $0.19  &  $ $0.21  &  $ $0.26  &  $-$0.01  &  $ $0.06  &  $ $  --  &  $ $  --  &  $-$0.19  \\
  MOA-2009-BLG-475S & $-$0.52  & $ $0.33  &  $ $  --  &  0.25  &  $ $0.23  &  $ $0.18  &  $ $0.14  &  $ $0.33  &  $ $  --  &  $ $0.07  &  $ $  --  &  $ $  --  &  $-$0.18  \\
MACHO-1999-BLG-022S & $-$0.49  & $ $ --   &  $ $0.01  &  0.35  &  $ $0.31  &  $ $0.25  &  $ $0.18  &  $ $0.25  &  $-$0.03  &  $ $0.00  &  $ $  --  &  $ $0.01  &  $ $0.15  \\
 OGLE-2008-BLG-209S & $-$0.30  & $ $0.34  &  $ $0.08  &  0.34  &  $ $0.24  &  $ $0.20  &  $ $0.18  &  $ $0.29  &  $ $0.05  &  $ $0.05  &  $ $0.16  &  $-$0.08  &  $ $0.05  \\
  MOA-2009-BLG-489S & $-$0.18  & $ $0.18  &  $-$0.01  &  0.24  &  $ $0.12  &  $ $0.07  &  $ $0.09  &  $ $0.16  &  $ $0.04  &  $ $0.01  &  $ $0.17  &  $-$0.05  &  $-$0.07  \\
  MOA-2009-BLG-456S & $ $0.12  & $-$0.12  &  $-$0.03  &  0.09  &  $ $0.08  &  $-$0.01  &  $ $0.00  &  $ $0.08  &  $-$0.06  &  $-$0.03  &  $-$0.03  &  $ $  --  &  $-$0.08  \\
  MOA-2008-BLG-311S & $ $0.36  & $-$0.27  &  $ $0.12  &  0.03  &  $ $0.08  &  $ $0.05  &  $ $0.00  &  $ $0.04  &  $-$0.01  &  $ $0.12  &  $ $0.06  &  $ $0.11  &  $-$0.15  \\
  MOA-2008-BLG-310S & $ $0.42  & $-$0.27  &  $ $0.08  &  0.08  &  $ $0.10  &  $ $0.02  &  $-$0.03  &  $ $0.09  &  $ $0.05  &  $ $0.06  &  $-$0.03  &  $ $0.06  &  $ $0.03  \\
 OGLE-2007-BLG-349S & $ $0.42  & $-$0.14  &  $ $0.15  &  0.05  &  $ $0.09  &  $ $0.12  &  $-$0.06  &  $ $0.02  &  $-$0.03  &  $ $0.11  &  $ $0.11  &  $-$0.12  &  $-$0.13  \\
  MOA-2008-BLG-099S & $ $0.44  & $-$0.10  &  $ $0.05  &  0.04  &  $-$0.02  &  $ $0.00  &  $-$0.05  &  $-$0.01  &  $-$0.16  &  $-$0.02  &  $ $0.05  &  $-$0.04  &  $ $0.00  \\
 OGLE-2006-BLG-265S & $ $0.47  & $-$0.25  &  $ $0.14  &  0.13  &  $ $0.08  &  $ $0.06  &  $-$0.05  &  $-$0.06  &  $-$0.07  &  $ $0.07  &  $-$0.05  &  $ $0.02  &  $-$0.08  \\
\noalign{\smallskip}
\hline
\noalign{\smallskip}
                      & 
$\sigma_{\rm [Fe/H]}$ &  
$\sigma_{\rm [O/Fe]}$  &  
$\sigma_{\rm [Na/Fe]}$ &  
$\sigma_{\rm [Mg/Fe]}$ &  
$\sigma_{\rm [Al/Fe]}$ &  
$\sigma_{\rm [Si/Fe]}$ &  
$\sigma_{\rm [Ca/Fe]}$ &  
$\sigma_{\rm [Ti/Fe]}$ &  
$\sigma_{\rm [Cr/Fe]}$ & 
$\sigma_{\rm [Ni/Fe]}$ &  
$\sigma_{\rm [Zn/Fe]}$ &  
$\sigma_{\rm [Y/Fe]}$ &  
$\sigma_{\rm [Ba/Fe]}$ \\
\noalign{\smallskip}
\hline
\noalign{\smallskip}
\noalign{\smallskip}
 OGLE-2009-BLG-076S &  0.07   &   0.17   &   0.04   &   0.06   &   0.05   &   0.06   &   0.05   &  0.03    &   --     &   0.09   &   --     &  --     &  0.10    \\
  MOA-2009-BLG-493S &  0.14   &   0.51   &   0.07   &   0.10   &   0.08   &   0.19   &   0.30   &  0.08    &   0.10   &   0.17   &   0.38   &  --     &  0.18    \\
  MOA-2009-BLG-133S &  0.17   &   0.62   &   0.13   &   0.21   &   0.07   &   0.12   &   0.36   &  0.06    &   0.16   &   0.15   &   --     &  --     &  0.51    \\
  MOA-2009-BLG-475S &  0.17   &   0.39   &   --     &   0.09   &   0.09   &   0.10   &   0.18   &  0.08    &   --     &   0.15   &   --     &  --     &  0.31    \\
MACHO-1999-BLG-022S &  0.14   &   --     &   0.07   &   0.13   &   0.10   &   0.13   &   0.13   &  0.24    &   0.14   &   0.10   &   0.19   &  0.31   &  0.12    \\
 OGLE-2008-BLG-209S &  0.06   &   0.22   &   0.07   &   0.10   &   0.04   &   0.08   &   0.12   &  0.08    &   0.09   &   0.07   &   0.18   &  0.20   &  0.08    \\
  MOA-2009-BLG-489S &  0.11   &   0.25   &   0.12   &   0.07   &   0.05   &   0.09   &   0.12   &  0.08    &   0.06   &   0.08   &   0.09   &  0.34   &  0.10    \\
  MOA-2009-BLG-456S &  0.09   &   0.21   &   0.05   &   0.08   &   0.05   &   0.07   &   0.10   &  0.05    &   0.09   &   0.08   &   0.17   &  --     &  0.09    \\
  MOA-2008-BLG-311S &  0.07   &   0.18   &   0.12   &   0.09   &   0.05   &   0.04   &   0.11   &  0.10    &   0.07   &   0.05   &   0.13   &  0.19   &  0.11    \\
  MOA-2008-BLG-310S &  0.08   &   0.16   &   0.10   &   0.09   &   0.07   &   0.05   &   0.10   &  0.08    &   0.06   &   0.05   &   0.13   &  0.23   &  0.09    \\
 OGLE-2007-BLG-349S &  0.08   &   0.19   &   0.15   &   0.14   &   0.08   &   0.07   &   0.12   &  0.07    &   0.05   &   0.04   &   0.10   &  0.17   &  0.08    \\
  MOA-2006-BLG-099S &  0.10   &   0.20   &   0.11   &   0.10   &   0.04   &   0.07   &   0.11   &  0.11    &   0.08   &   0.06   &   0.10   &  0.30   &  0.10    \\
 OGLE-2006-BLG-265S &  0.06   &   0.17   &   0.11   &   0.10   &   0.05   &   0.06   &   0.11   &  0.07    &   0.06   &   0.06   &   0.43   &  0.23   &  0.08    \\
\noalign{\smallskip}
\hline
\noalign{\smallskip}
                      &
$\sigma_{\rm [Fe/H]}$ &
$\sigma_{\rm [O/H]}$  &
$\sigma_{\rm [Na/H]}$ &
$\sigma_{\rm [Mg/H]}$ &
$\sigma_{\rm [Al/H]}$ &
$\sigma_{\rm [Si/H]}$ &
$\sigma_{\rm [Ca/H]}$ &
$\sigma_{\rm [Ti/H]}$ &
$\sigma_{\rm [Cr/H]}$ &
$\sigma_{\rm [Ni/H]}$ &
$\sigma_{\rm [Zn/H]}$ &
$\sigma_{\rm [Y/H]}$ &
$\sigma_{\rm [Ba/H]}$ \\
\noalign{\smallskip}
\hline
\noalign{\smallskip}
\noalign{\smallskip}
 OGLE-2009-BLG-076S &  0.07   &  0.10    &  0.05    &  0.06    &   0.05   &   0.05   &  0.10    &   0.08   &   --     &   0.06   &   --     &   --    &  0.11    \\
  MOA-2009-BLG-493S &  0.14   &  0.38    &  0.09    &  0.16    &   0.07   &   0.06   &  0.43    &   0.13   &   0.07   &   0.05   &   0.26   &   --    &  0.18    \\
  MOA-2009-BLG-133S &  0.17   &  0.46    &  0.05    &  0.36    &   0.18   &   0.06   &  0.52    &   0.12   &   0.07   &   0.04   &   --     &   --    &  0.36    \\
  MOA-2009-BLG-475S &  0.17   &  0.23    &  --      &  0.17    &   0.13   &   0.08   &  0.31    &   0.14   &   --     &   0.07   &   --     &   --    &  0.21    \\
MACHO-1999-BLG-022S &  0.14   &  --      &  0.14    &  0.23    &   0.06   &   0.06   &  0.25    &   0.16   &   0.11   &   0.08   &   0.10   &   0.22  &  0.15    \\
 OGLE-2008-BLG-209S &  0.06   &  0.16    &  0.11    &  0.15    &   0.07   &   0.03   &  0.17    &   0.06   &   0.06   &   0.03   &   0.16   &   0.16  &  0.08    \\
  MOA-2009-BLG-489S &  0.11   &  0.15    &  0.13    &  0.14    &   0.07   &   0.03   &  0.21    &   0.09   &   0.07   &   0.04   &   0.06   &   0.29  &  0.09    \\
  MOA-2009-BLG-456S &  0.09   &  0.14    &  0.07    &  0.12    &   0.09   &   0.04   &  0.15    &   0.10   &   0.14   &   0.04   &   0.15   &   --    &  0.10    \\
  MOA-2008-BLG-311S &  0.07   &  0.12    &  0.17    &  0.14    &   0.10   &   0.06   &  0.17    &   0.07   &   0.06   &   0.04   &   0.07   &   0.13  &  0.07    \\
  MOA-2008-BLG-310S &  0.08   &  0.10    &  0.14    &  0.13    &   0.09   &   0.04   &  0.15    &   0.08   &   0.05   &   0.05   &   0.08   &   0.21  &  0.12    \\
 OGLE-2007-BLG-349S &  0.08   &  0.13    &  0.19    &  0.18    &   0.08   &   0.04   &  0.17    &   0.09   &   0.07   &   0.06   &   0.10   &   0.17  &  0.12    \\
  MOA-2006-BLG-099S &  0.10   &  0.11    &  0.18    &  0.18    &   0.08   &   0.05   &  0.20    &   0.07   &   0.06   &   0.07   &   0.09   &   0.25  &  0.09    \\
 OGLE-2006-BLG-265S &  0.06   &  0.12    &  0.14    &  0.14    &   0.07   &   0.04   &  0.14    &   0.08   &   0.07   &   0.04   &   0.43   &   0.21  &  0.08    \\
\noalign{\smallskip}
\hline
\noalign{\smallskip}
\noalign{\smallskip}
                    &  
$N_{\rm \ion{Fe}{i}}$ &  
$N_{\rm O}$    &  
$N_{\rm Na}$   &  
$N_{\rm Mg}$   &  
$N_{\rm Al}$   &  
$N_{\rm Si}$   &  
$N_{\rm Ca}$   &  
$N_{\rm Ti}$   &  
$N_{\rm Cr}$   &  
$N_{\rm Ni}$   &  
$N_{\rm Zn}$   &  
$N_{\rm Y}$    &  
$N_{\rm Ba}$   \\
\noalign{\smallskip}
\hline
\noalign{\smallskip}
 OGLE-2009-BLG-076S &   57  &  3  &  1  &  5  &  4  &  15  &  10  &   2  &  --  &  11  & --  & --  &  2   \\
  MOA-2009-BLG-493S &   80  &  3  &  1  &  6  &  6  &  19  &  10  &   5  &   2  &  21  &  1  & --  &  4   \\
  MOA-2009-BLG-133S &   68  &  2  &  1  &  4  &  6  &  17  &  10  &   4  &   1  &  15  & --  & --  &  3   \\
  MOA-2009-BLG-475S &   53  &  1  & --  &  5  &  3  &  12  &   9  &   2  &  --  &  14  & --  & --  &  3   \\
MACHO-1999-BLG-022S &   97  & --  &  3  &  1  &  2  &   7  &  16  &  14  &   2  &  21  & --  &  2  &  3   \\
 OGLE-2008-BLG-209S &  146  &  3  &  4  &  5  &  7  &  27  &  19  &  36  &  10  &  40  &  3  &  4  &  4   \\
  MOA-2009-BLG-489S &  114  &  3  &  2  &  6  &  6  &  27  &  13  &  16  &   6  &  37  &  3  &  2  &  4   \\
  MOA-2009-BLG-456S &   91  &  3  &  2  &  5  &  6  &  26  &  11  &   7  &   3  &  33  &  3  & --  &  2   \\
  MOA-2008-BLG-311S &  118  &  3  &  4  &  5  &  5  &  26  &  17  &  14  &   9  &  36  &  1  &  2  &  3   \\
  MOA-2008-BLG-310S &  122  &  3  &  4  &  4  &  7  &  27  &  17  &  22  &  13  &  42  &  1  &  4  &  3   \\
 OGLE-2007-BLG-349S &  103  &  3  &  4  &  3  &  3  &  23  &  18  &  24  &   9  &  39  &  3  &  5  &  4   \\
  MOA-2006-BLG-099S &  119  &  3  &  4  &  4  &  4  &  25  &  18  &  30  &  12  &  37  &  3  &  5  &  4   \\
 OGLE-2006-BLG-265S &   92  &  3  &  4  &  2  &  3  &  23  &  16  &  13  &   8  &  30  &  2  &  2  &  3   \\
\noalign{\smallskip}
\hline
\end{tabular}
\flushleft
{\tiny
$^\dagger$
No abundances are given for MOA-2009-BLG-259S as the errors in the stellar parameters were too large, and for OGLE-2007-BLG-514S we
only redeteremined stellar parameters and [Fe/H]. Abundance ratios for OGLE-2007-BLG-514S can be found in \cite{epstein2009}.\\
$^\ddag$ Note that the abundance ratios for oxygen have been corrected for NLTE effects according to the empirical
formula given in \cite{bensby2004}.
}
\end{table*}


\subsection{Check 1: temperatures from microlensing techniques}

De-reddened colours and magnitudes of the sources can be estimated 
using standard microlensing techniques \citep[e.g.][]{yoo2004}. 
The method for determining the colour does not make any assumption 
about the absolute reddening, nor about the ratio of selective to 
total extinction.  It only assumes that the reddening toward the 
microlensed source is the same as the reddening toward the red clump, and 
that the red clump in the Bulge has $(V-I)_{0} = 1.05$ and $I_{0}=14.32$
\citep[e.g.,][]{johnson2008,epstein2009}.
The absolute de-reddened magnitude and colour are 
then derived from the offsets between the microlensing source and the 
red clump in the instrumental colour-magnitude diagram (CMD).
The absolute de-reddened magnitudes and colours for 14 of the
15 microlensed stars are given in Table~\ref{tab:photometry}.
Photometry for MACHO-1999-BLG-022S could not be recovered at
this time.

\begin{figure}
\resizebox{\hsize}{!}{
\includegraphics[bb=20 160 600 540,clip]{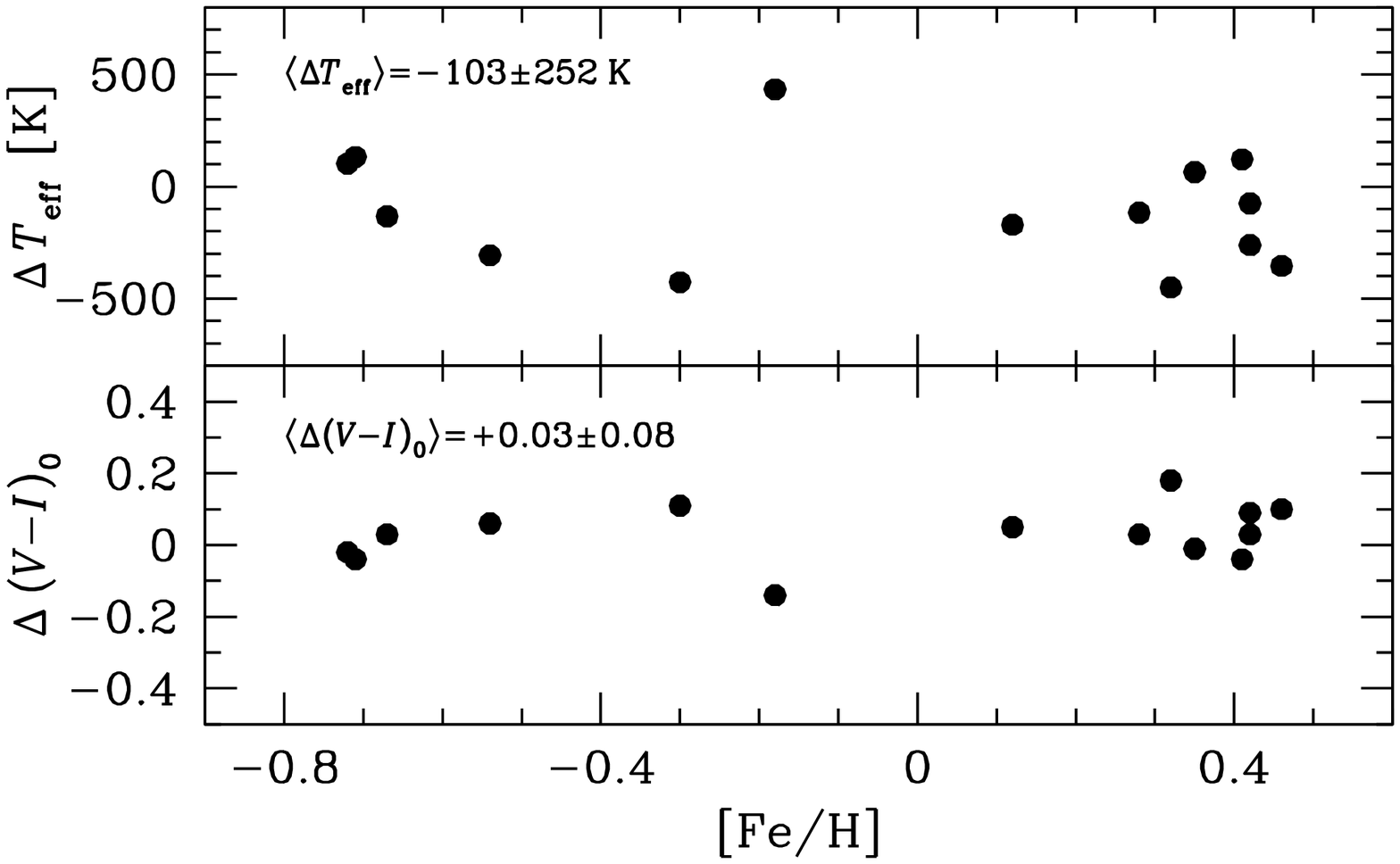}}
\caption{
{\sl Top panel} shows a comparison of our spectroscopic $\teff$:s to 
those implied from the  colour--[Fe/H]--$\teff$ calibrations by 
\cite{ramirez2005} versus [Fe/H]. 
{\sl Bottom panel} shows a comparison between the colours that 
our spectroscopic $\teff$:s imply (using the colour--[Fe/H]--$\teff$ 
calibrations by \citealt{ramirez2005}) and the colours based
on microlensing techniques. Differences are in both cases
given as $spectroscopic - photometric$.
\label{fig:specphot}}
\end{figure}

From the colour--[Fe/H]--$\teff$ calibrations by \cite{ramirez2005} we 
check what temperature we should expect given the de-reddened colour 
and the metallicity we determined. On average we find that the 
spectroscopic temperatures are 103\,K lower than the ones based on
the colour--[Fe/H]--$\teff$ relationships. The top panel of 
Fig.~\ref{fig:specphot} shows a comparison between the two as a 
function of [Fe/H]. No obvious trends can be seen.

It is also possible to use the the same calibrations 
by \cite{ramirez2005} to see what $(V-I)$ colours the spectroscopic 
effective temperatures and metallicities would give. These are listed 
in the last column of Table~\ref{tab:photometry}, and the comparison 
between photometric and ``spectroscopic" $(V-I)$ colours are shown
in the bottom panel of Fig.~\ref{fig:specphot}. On average the
spectroscopic colours are 0.03\,mag higher, with no discernible
trend with metallicity.

The offset that we see between spectroscopic and photometric values
could be a result of the assumed magnitudes and colours of the
red clump in the Bulge. Previously, it was assumed that the red clump 
stars in the Bulge had the same colour as the red clump stars
in the Solar neighbourhood ($(V-I)_{0} = 1.00$). Based on the first 
microlensing events of dwarf stars in the Bulge 
\citep{johnson2008,cohen2008}, and additional observational evidence 
\citep{epstein2009}, this value was revised to $(V-I)_{0} = 1.05$. 
Assuming that the spectroscopic temperatures are the correct ones, 
our results indicate that the $(V-I)_{0}$ colour of the red clump 
stars in the Bulge should be revised upwards by an additional 
few hundredths of a dex to $(V-I)_{0} = 1.08$.

\begin{figure}
\resizebox{\hsize}{!}{
\includegraphics[bb=20 320 600 730,clip]{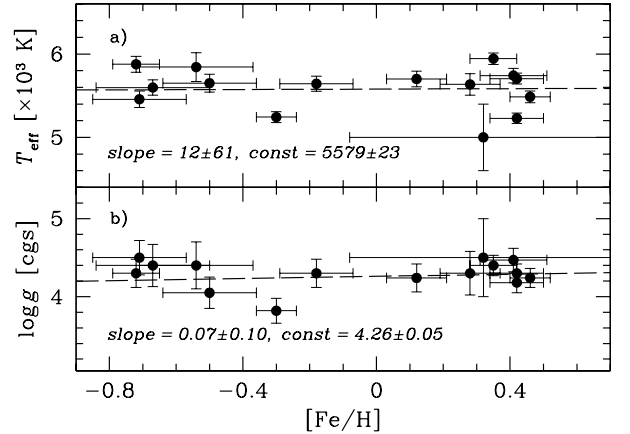}}
\caption{
Effective temperature and surface gravity versus [Fe/H].
In each figure the regression parameters (slope and constant)
are given, as well as their uncertainties. The error bars
in the stellar parameters represent the total uncertainty
(see Table~\ref{tab:parameters} and Sect.~\ref{sec:errors}).       
\label{fig:consistency}}
\end{figure}

\subsection{Check 2: trends with metallicity}

Figure~\ref{fig:consistency} shows how  $\teff$ and $\log g$ 
vary with metallicity. Regression lines, taking the errors in 
both $x$ and $y$ directions into account, are shown, as well 
as the regression parameters and their uncertainties.
No significant trends with metallicity can be seen.

\subsection{Check 3: signal-to-noise  and continuum bias}

The two papers by \cite{bensby2009letter,bensby2009} found
the two first microlensed dwarf stars with sub-solar iron
abundances. In this study we find an additional 5 stars with
sub-solar [Fe/H]. It has been suggested that the reason for
this could be that these spectra have, on average, lower $S/N$
than the other events studied, especially compared to
\cite{cohen2009} which all have very high [Fe/H].
The lower $S/N$ should then result in that the continuum
was set too low and thus the $W_{\lambda}$s underestimated,
resulting in too low [Fe/H].

We looked in greater detail at MOA-2009-BLG-475S, the dwarf star with 
the spectrum that has the lowest $S/N$, and tested if we could make 
the star more metal-rich. This experiment was done by assuming that 
the star is actually metal-rich and
that there are many weak lines that makes it difficult to identify
the level of the continuum. The best way to set the continuum is then
to assume that the high points in the spectrum are indeed the continuum.
We re-measured the star under this assumption and determined
new stellar parameters. On average the equivalent widths became 
10\,m{\AA} larger. This resulted in a change of the metallicity
of $+0.1$\,dex, but the other stellar parameters
($\log g$, $\teff$, and $\xi_{\rm t}$) were intact and did not
change. Hence, we find it unlikely that our low-metallicity stars
could be high-metallicity stars resulting from
an erroneous analysis of their relatively low $S/N$ spectra.

\subsection{In summary}
\label{sec:errorsum}

Our analysis shows that the eight microlensed stars that we observed
were successfully selected to be dwarf stars, 
varying in metallicity from $\rm [Fe/H]=-0.72$ to $+0.54$.
The first results for OGLE-2009-BLG-076S were presented in 
\cite{bensby2009letter} where we found a metallicity of 
$\rm [Fe/H]=-0.76$. The refined analysis in this study gives a
metallicity of $\rm [Fe/H]=-0.72$, and it still holds (barely)
the place as the most metal-poor dwarf star in the Bulge, with 
MOA-2009-BLG-133S just 0.01\,dex higher.

The uncertainties in the stellar parameters are below or
around 100\,K in $\teff$, around 0.2\,dex in $\log g$, and 
around 0.1--0.2\,dex in [Fe/H] (see Table~\ref{tab:parameters}). 
The clear exception is 
MOA-2009-BLG-259S where errors are exceptionally 
large. This is due to difficulties arising
from the very limited wavelength coverage of the UVES spectrum 
that was obtained when only the blue CCD was available. The fact that 
it also turned out to be a metal-rich star at $\rm [Fe/H]=+0.32$, 
further increases the errors due to line blending and uncertainties 
in the placement of the continuum.  Also, the lack of weak lines,
due to the high [Fe/H],  
made it especially difficult to determine the microturbulence parameter
(see Fig.~\ref{fig:fetrends}). A spectrum covering the whole
optical region of MOA-2009-BLG-259S was obtained by another group using 
the HIRES spectrograph, and they find a 0.2\,dex higher metallicity
than what we do \citep{cohen2010puzzle}.
Although the metallicity is in reasonable agreement with
what others have found we think that the errors
in $\teff$ and $\log g$ are so large that this star is not conveying
any information in the [Fe/H], age, or abundance plots.
Therefore we do not include this star in the 
following discussions.

\section{Metallicity distributions of dwarfs and giants}
\label{sec:mdf}

The most recent spectroscopic study of  a large homogeneous sample 
of giant stars in the Bulge is by \cite{zoccali2008}.
Using FLAMES, the multi-fibre spectrograph at the Very Large Telescope,
they studied a sample of 521 giant stars at three 
latitudes in the Bulge: 204 stars in Baade's window at 
$l\approx-4^\circ$; 213 stars at $l\approx-6^\circ$; and 104 stars 
at $l\approx-12^\circ$. 
The 15 microlensed dwarf stars observed so far are 
all located at similar angular distances from the Galactic centre
as Baade's window (see  Fig.~\ref{fig:allevents}).
Therefore, only the 204 stars  in the $l=-4^\circ$ field from 
\cite{zoccali2008} will be used for comparison.

\begin{figure}
\resizebox{\hsize}{!}{
\includegraphics[bb=20 160 590 718,clip]{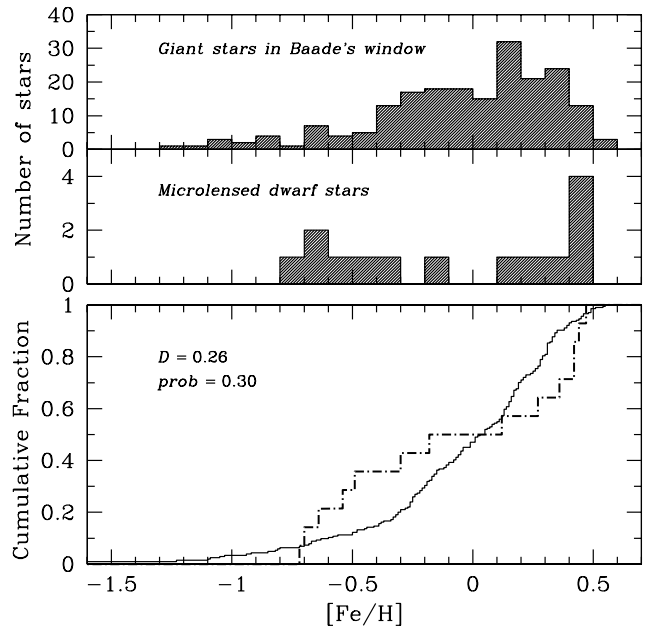}}
\caption{
The top panel shows the MDF for the 204 giant stars 
in Baade's window from \cite{zoccali2008}, and the middle panel
the MDF for the fourteen microlensed dwarf stars (MOA-2009-BLG-259S
excluded).
Their cumulative metallicity distributions (CMDF) are shown
in the bottom panel (giant CMDF marked by solid line, and the
dwarf star CMDF by dash-dotted line).
The two-sample Kolmogorov-Smirnov {\it D} statistic (maximum vertical
distance between two distributions) and the corresponding significance
level, {\it prob}, of {\it D} are indicated.
         \label{fig:kstest}}
\end{figure}

The average metallicity of the 14 microlensed dwarf and subgiant 
stars in the Bulge (MOA-2009-BLG-259S excluded, see 
Sect.~\ref{sec:errorsum}) is $\rm \langle[Fe/H]\rangle=-0.08\pm0.47$.
This is in agreement with the average metallicity of the 204 RGB 
stars in Baade's window that have 
$\rm \langle[Fe/H]\rangle=-0.04\pm0.40$.  However, when comparing 
the two distributions, a two-sided Kolmogorov-Smirnow (KS) test 
gives a significance level of the null hypothesis, that they are 
drawn from the same distribution, of 30\,\% (see Fig.~\ref{fig:kstest}). 
Hence, we can not reject the null hypothesis that the MDF for the 14 
microlensed dwarf stars and the MDF for the 204 Bulge RGB stars from 
\cite{zoccali2008} are identical.

As discussed by, e.g., \cite{santos2009} it is possible that
a systematic shift in [Fe/H] between analyses of dwarf stars and giant
stars, perhaps by as much as 0.2 dex, exists. However, the difference
in the average metallicity between dwarf and giant stars is
essentially zero. If there was a real difference of 0.2 dex
between the dwarf stars and the giant stars, how many dwarf stars
would we need to observe in order to statistically measure that
difference?

\begin{figure}
\resizebox{\hsize}{!}{
\includegraphics[bb=18 420 592 725,clip]{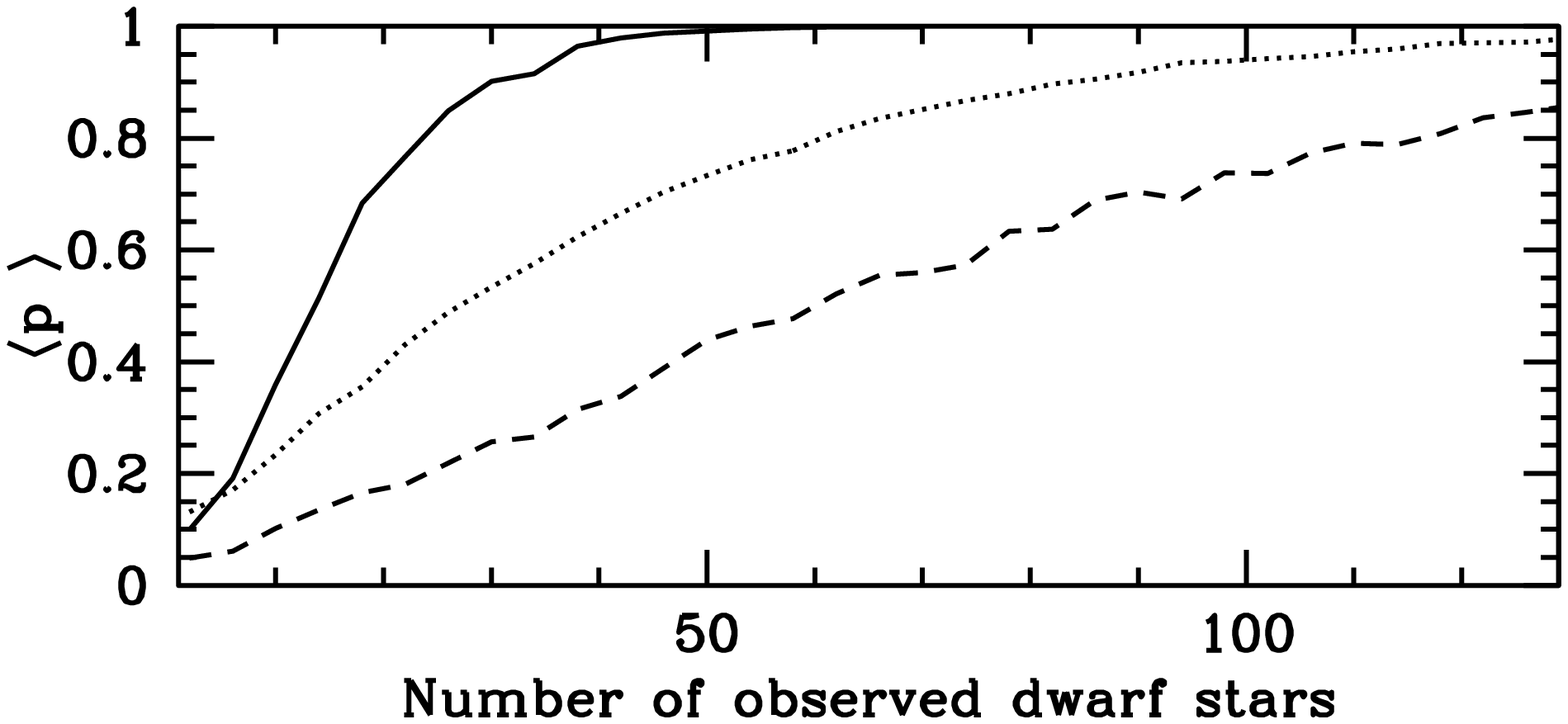}}
\resizebox{\hsize}{!}{
\includegraphics[bb=18 420 592 690,clip]{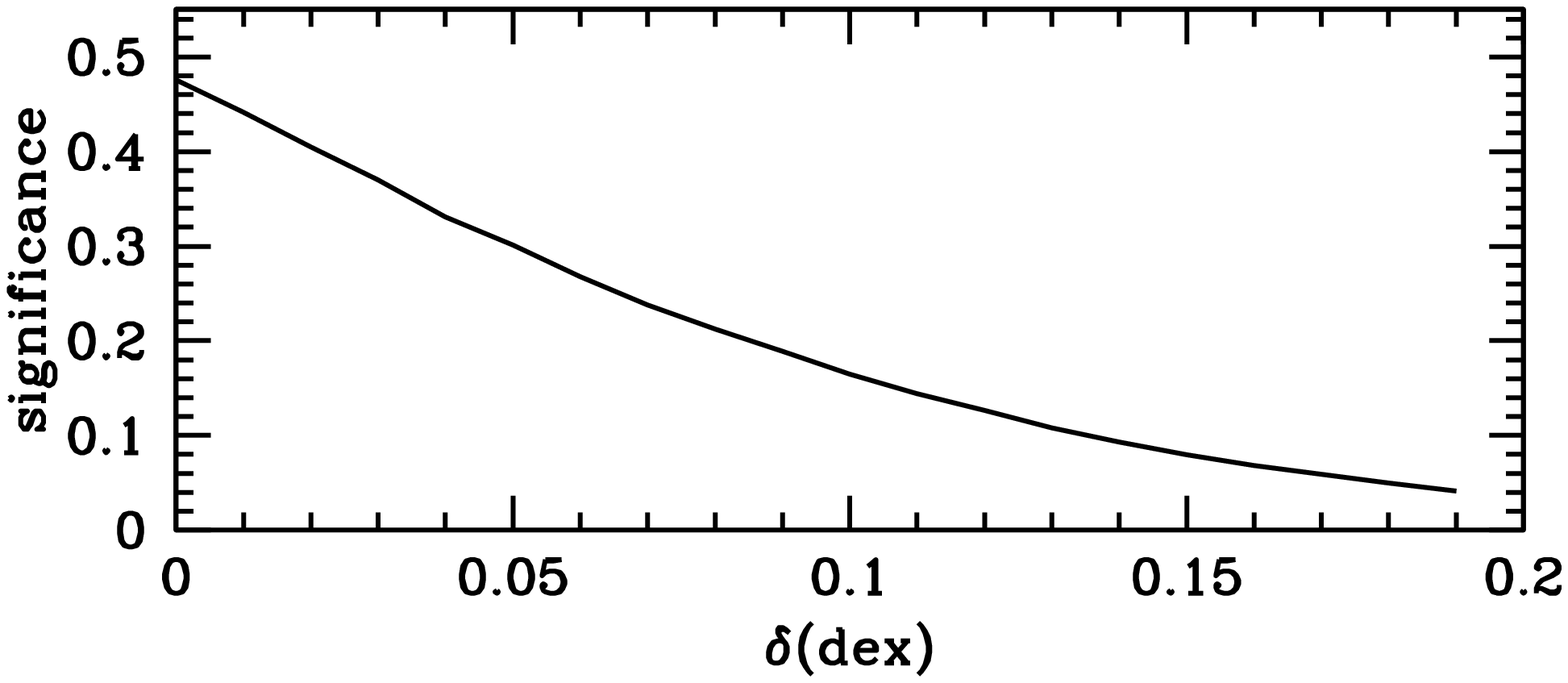}}
\caption{
{\sl Upper panel:} Probability of detecting a difference between the 
Bulge RGB MDF and
the microlensed dwarf MDF as a function of the number of stars in the
dwarf sample. Solid and dashed line indicate the probability if the
difference is 0.2 and 0.1 dex, respectively, in the mean of the
MDFs. Dotted line indicates the probability if the difference is that
the microlensed dwarf MDF has more metal rich, $\rm [Fe/H]>0.0$, 
stars (i.e, a more symmetric MDF) than the RGB MDF.
{\sl Lower panel:} The significance of the agreement between the 
average of the metallicities of the RGB MDF and microlensed dwarf
MDF, as a function of a speculated (still undetected) difference. 
         \label{fig:bootstrap}}
\end{figure}

In order to estimate the number of stars required to reject the null
hypothesis, that the MDF for the sample of 204 RGB stars, 
$S_{\rm RGB}$, is not different from the MDF for the observed dwarf 
stars, we will use a non-parametric bootstrap method.  First, we 
construct a sample, $S_{\rm RGB, 0.2}$, of stars that is shifted 
0.2\,dex from the original RGB sample, $S_{\rm RGB}$.  Secondly, 
we bootstrap $n$ number of stars from sample $S_{\rm RGB,0.2}$, thus 
creating a new sample, $S_{\rm n, 0.2}$, that contains $n$ stars.  
Next, we perform a two-sided KS-test between samples $S_{\rm RGB}$ 
and $S_{\rm n, 0.2}$.  If the KS-test yields that the distributions 
are not the same at the 95\,\% confidence level, we reject the null
hypothesis.  We repeat this process $i=10\,000$ times and take the
average, $\langle p \rangle$, of $p_{\rm i}$, where $p_{\rm i}=1$ if
the null hypothesis was rejected, and 0 otherwise (type II error).
Thus, $\langle p \rangle$ is our probability to identify an intrinsic
difference of 0.2\,dex in the mean in the MDF for RGB and
microlensed dwarf stars. Figure~\ref{fig:bootstrap} shows how this
probability varies with the number of stars in the hypothetical dwarf 
star sample. In order to statistically verify a difference of at least
0.2\,dex, at the 95\,\% level, we need to observe around 40
stars. Also in Fig.~\ref{fig:bootstrap} we show how the probability 
varies if we want to verify a difference of only 0.1\,dex between 
dwarfs and giants. Note that the detection of such a small difference 
would require the observation of many more microlensed dwarf stars.

Above we shifted the entire MDF by 0.1 and 0.2 dex and did
not consider other statistical parameters that describe the MDF such
as variance, skewness and kurtosis.  However, these parameters will
most likely make it easier to reject the null hypothesis, that the
distributions are the same, if they are considered in the
test. Additionally, we only considered 0.1 and 0.2\,dex as a
difference between the samples. A larger difference will also make it
easier to reject the null hypothesis. Our estimate of 
$\langle p \rangle$ is therefore a lower limit.

Additionally, the average metallicity for the 14 microlensed dwarf
stars is in agreement with the average metallicity of the RGB
stars. The question is then, what is the significance of this result,
given the low number statistics of the microlensed dwarf stars,
i.e., can we rule out a shift of 0.2\,dex between the MDF for the RGB
stars and the MDF for the dwarf stars?  We estimate the significance
of the agreement between the average metallicities using $10^5$ Monte
Carlo realisations. In each realisation, we draw 14 stars randomly
from the RGB sample. We determine the significance of the agreement
between the average of the metallicities by computing the fraction of
realisations that fail to produce a dwarf star average metallicity
lower or equal to the average metallicity of the RGB stars.  Next, we
shift the MDF for the RGB stars by a small positive amount, $\delta$,
and repeat the above given exercise. Figure~\ref{fig:bootstrap},
lower panel, shows the significance as a function of $\delta$.
We find that for $\delta\sim0.2$,
the significance has dropped to 0.05 which indicates that we can rule
out the possibility that there is a systematic shift in [Fe/H] of 0.2
dex between the MDF for the RGB stars and the MDF for the dwarf
stars. However, we need to put this in relation to the standard
deviation of the difference between the averages given by
\begin{equation}
\sigma_d=\sqrt{ \frac{\sigma_1^2}{n_1} + \frac{\sigma_2^2}{n_2} }
\end{equation}
where subscript 1 and 2 indicate the RGB star sample and dwarf star
sample, respectively, $\sigma$ is the sample standard deviation, and
$n$ is the number of elements in each sample.  We find a $\sigma_d$ of
0.12, which is not even $2\sigma$ from 0.2 dex.  Thus, given the small
number of dwarf stars in our sample, $n_2=14$, we can not rule out a
possible systematic shift of 0.2 dex between the MDF for the RGB stars
and the MDF for the dwarf stars, even though the distributions at this
point have the same average metallicity. Thus, more microlensed events
are needed for the agreement between the averages to be significant.

The above discussion focused on a systematic offset
between the dwarf and giant abundances, caused either by systematic
effects in the data analysis or differences in the surface
compositions of dwarfs and giants because of their differing
evolutionary states. However, instead of an overall shift, the
dwarf MDF may be different from the giant MDF only at the high
metallicity end.
\cite{kalirai2007} suggested that up to 40 per cent of the
stars at the metallicity of NGC6791 ($\rm [Fe/H]=+0.3$) skip the He burning
phase, resulting in a depletion of the HB and AGB phases. Therefore, an
MDF based on giants may not reflect the MDF for the dwarf stars
(i.e. there are metal rich stars missing in the MDF for the RGB
stars).  The question is then, how many microlensed dwarf stars are
required to reject the null hypothesis that the MDF for these dwarfs,
that has an excess of metal-rich stars compared to $S_{\rm RGB}$,
is no different than the MDF for $S_{\rm RGB}$?  Based on this, we
construct a dwarf star sample with more metal-rich stars
than $S_{\rm RGB}$, $S_{\rm RGB, Kalirai}$, under the assumption
that 100\,\% of the dwarfs with [Fe/H]$<0.0$ evolve to RGBs, but
that the number of dwarfs that evolve to RGBs decrease linearly down
to 60\,\% at $\rm [Fe/H]=+0.3$. We extrapolate this linearly for more
metal-rich stars, i.e. going down to $\sim$34\,\% at $\rm [Fe/H]=+0.5$.
Thus, we add stars, more metal-rich than
$\rm [Fe/H]=0.0$, to $S_{\rm RGB}$ to create a hypothetical sample of
dwarf stars in the Bulge, $S_{\rm RGB, Kalirai}$.  
Additionally, we add stars, drawn from a
Gaussian distribution centred on $\rm [Fe/H]=+0.2$ with
$\sigma=0.4$, to $S_{\rm RGB, Kalirai}$ for the metal rich
region where there are no observed RGB stars. Thus, making the sample
more symmetric.  The dotted line in Fig.~\ref{fig:bootstrap} shows how 
$\langle p \rangle$ varies for this analysis with the number of stars 
in $S_{\rm RGB, Kalirai}$. We note that about 100 stars are required 
in order  to verify that $S_{\rm RGB}$ is different than 
$S_{\rm RGB, Kalirai}$.

Also, there has been some claims that the microlensing event itself 
alters the spectrum of the source star, and that this is the reason for 
the very high metallicities of some of the dwarf stars in the Bulge 
\citep[e.g.][]{zoccali2008}. Recently, \cite{johnson2009} investigated 
the effect that differential limb darkening has on abundance analysis 
of microlensed dwarf stars. They do find changes in the 
measured equivalent widths as a result of the differential limb 
darkening. However, the effect is very small, leading to changes in 
$\teff$ less than 45\,K, $\log g$ less than 0.1\,dex, and [Fe/H] less 
than 0.03\,dex. Hence,   a possible differential 
limb darkening can not be responsible 
for the MDF discrepancy (if any) between dwarf and giant stars.

In summary, it is evident that the extremely super-metal-rich MDF 
proposed by \cite{cohen2009}, exclusively based on dwarf stars with
super-solar [Fe/H], has shifted toward lower metallicities.
The MDF of the 14 microlensed dwarf stars is still poorly
determined, currently being double peaked with excesses of 
low- \textit{and} high-metallicity stars. Whether this is 
an effect of  small number statistics or not is unclear.
More microlensed events will certainly help to clarify the 
dwarf star MDF and to refine the comparison with the giant star MDF.  
Also, an outstanding issue is the puzzle presented in \cite{cohen2010puzzle}
of the correlation between $A_{\rm max}$ and [Fe/H], which we hope 
to diagnose as future events are observed.

\section{Ages and metallicities}

\begin{figure}
\resizebox{\hsize}{!}{
\includegraphics[bb=18 170 592 710,clip]{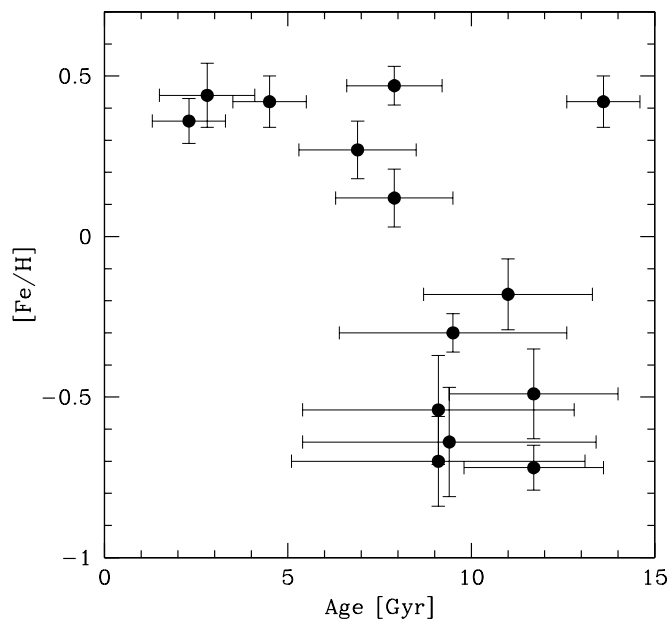}}
\caption{Ages versus [Fe/H] for 14 microlensed dwarf and subgiant 
stars in the Bulge (MOA-2009-BLG-259S has been excluded).
The error bars represent the total uncertainty [Fe/H] and age
(see Table~\ref{tab:parameters} and Sects.~\ref{sec:errors} and
\ref{sec:ages}).
\label{fig:agefe}}
\end{figure}

The 14 microlensed dwarf and subgiant stars (MOA-2009-BLG-259S 
excluded, see Sect~\ref{sec:errorsum})
in the Bulge have an average age of  $8.4\pm3.3$\,Gyr. 
Figure~\ref{fig:agefe} shows the age-metallicity diagram and 
it is evident that stars with sub-solar [Fe/H] all have high ages, 
while at super-solar [Fe/H] there is a large spread in age,  
covering the whole age-range seen in the Galactic disc(s) 
\citep[e.g.,][]{twarog1980b,feltzing2001}. That the large 
age-range seen for super-solar [Fe/H] is real is exemplified by two 
stars: OGLE-2007-BLG-349S and MOA-2008-BLG-310S. The first star has a 
high age while the second has a low age. That the ages are robust can
be seen in Fig.~\ref{fig:ages} and we also note that 
their stellar parameters have small errors and are derived from a 
large number of  \ion{Fe}{i} and \ion{Fe}{ii} lines 
(see Table~\ref{tab:parameters} and Fig.~\ref{fig:fetrends}).

\begin{figure*}
\resizebox{\hsize}{!}{
\includegraphics[bb=18 220 592 470,clip]{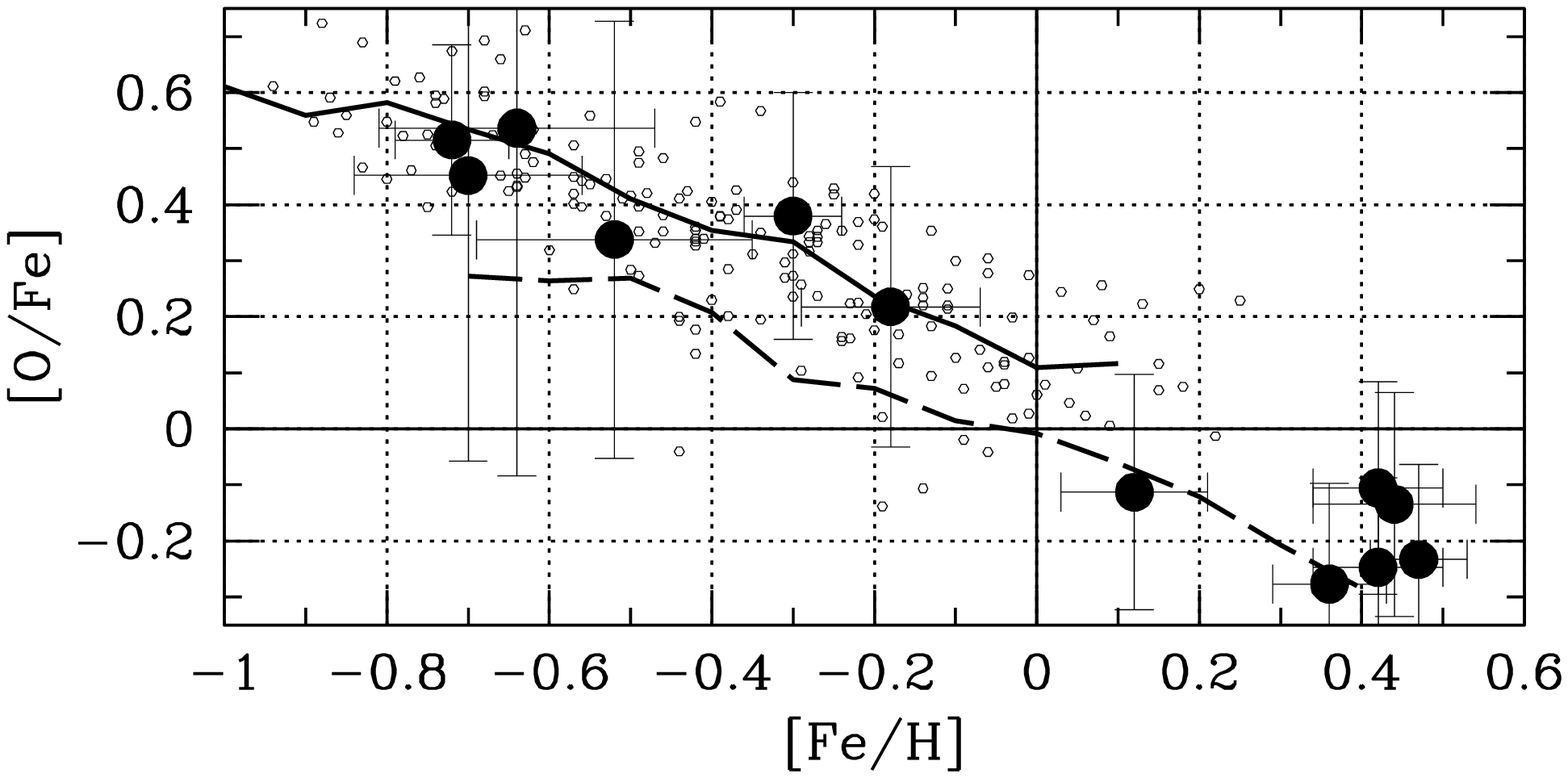}
\includegraphics[bb=18 220 592 470,clip]{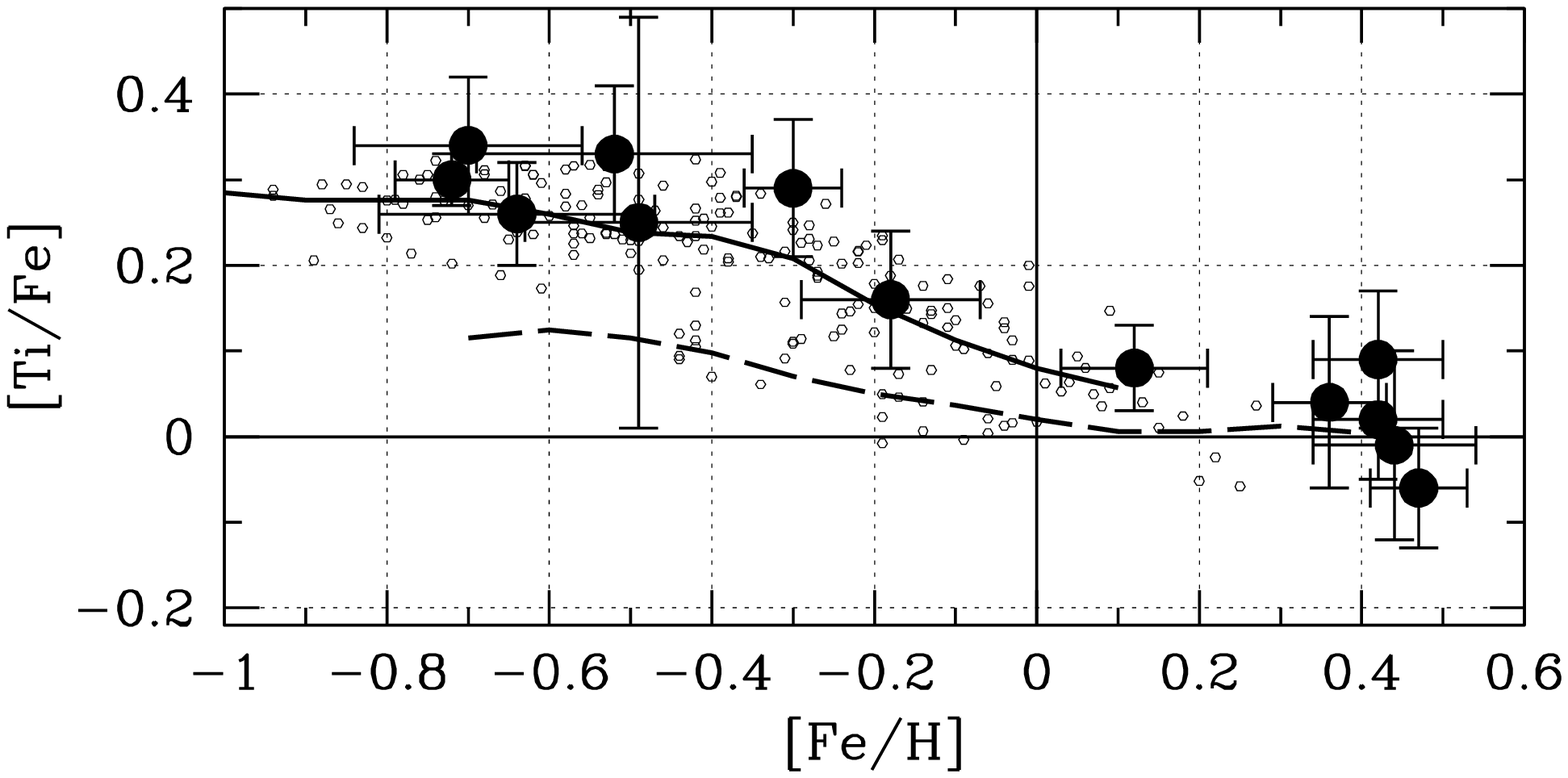}}
\resizebox{\hsize}{!}{
\includegraphics[bb=18 220 592 450,clip]{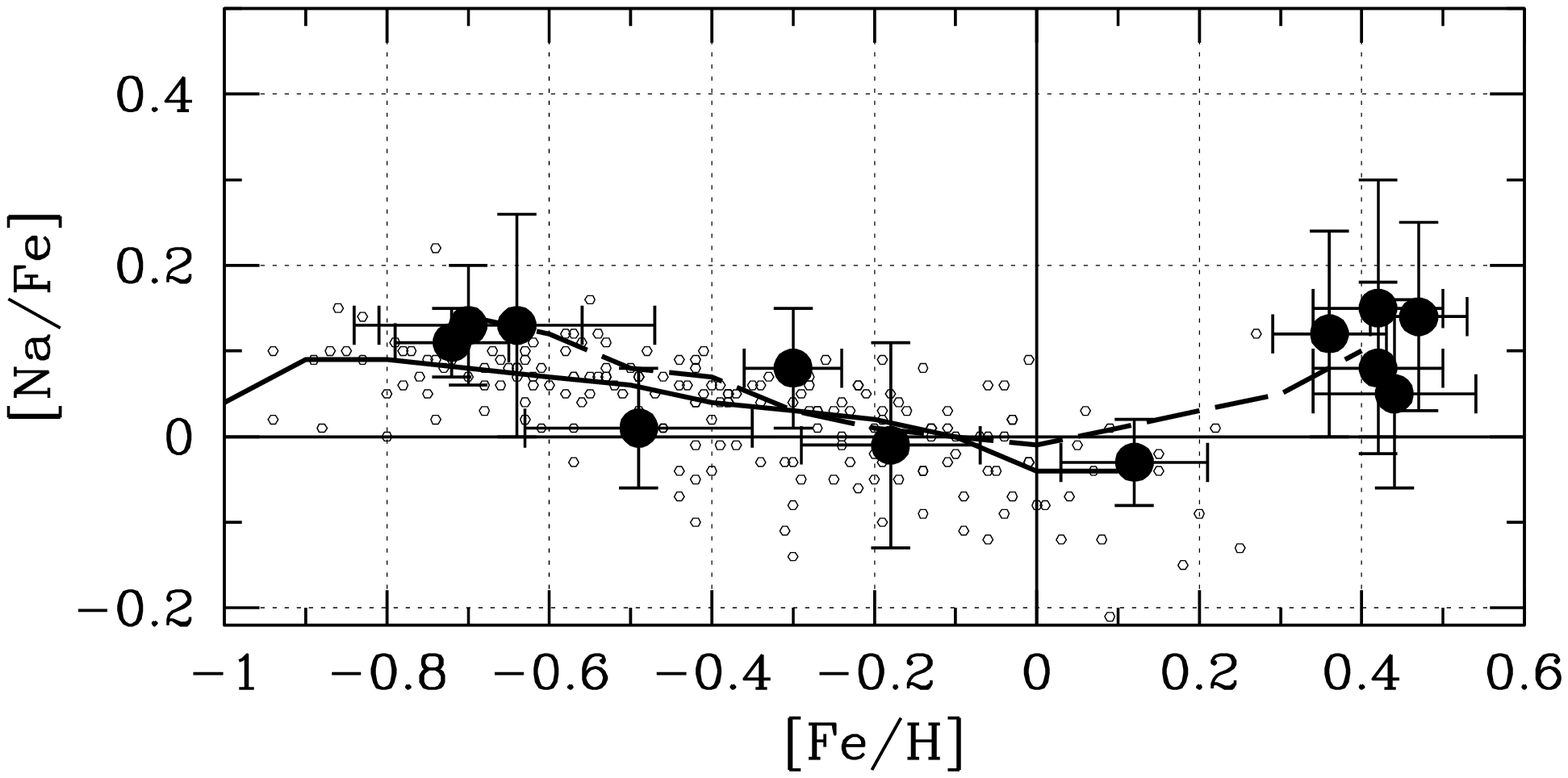}
\includegraphics[bb=18 220 592 450,clip]{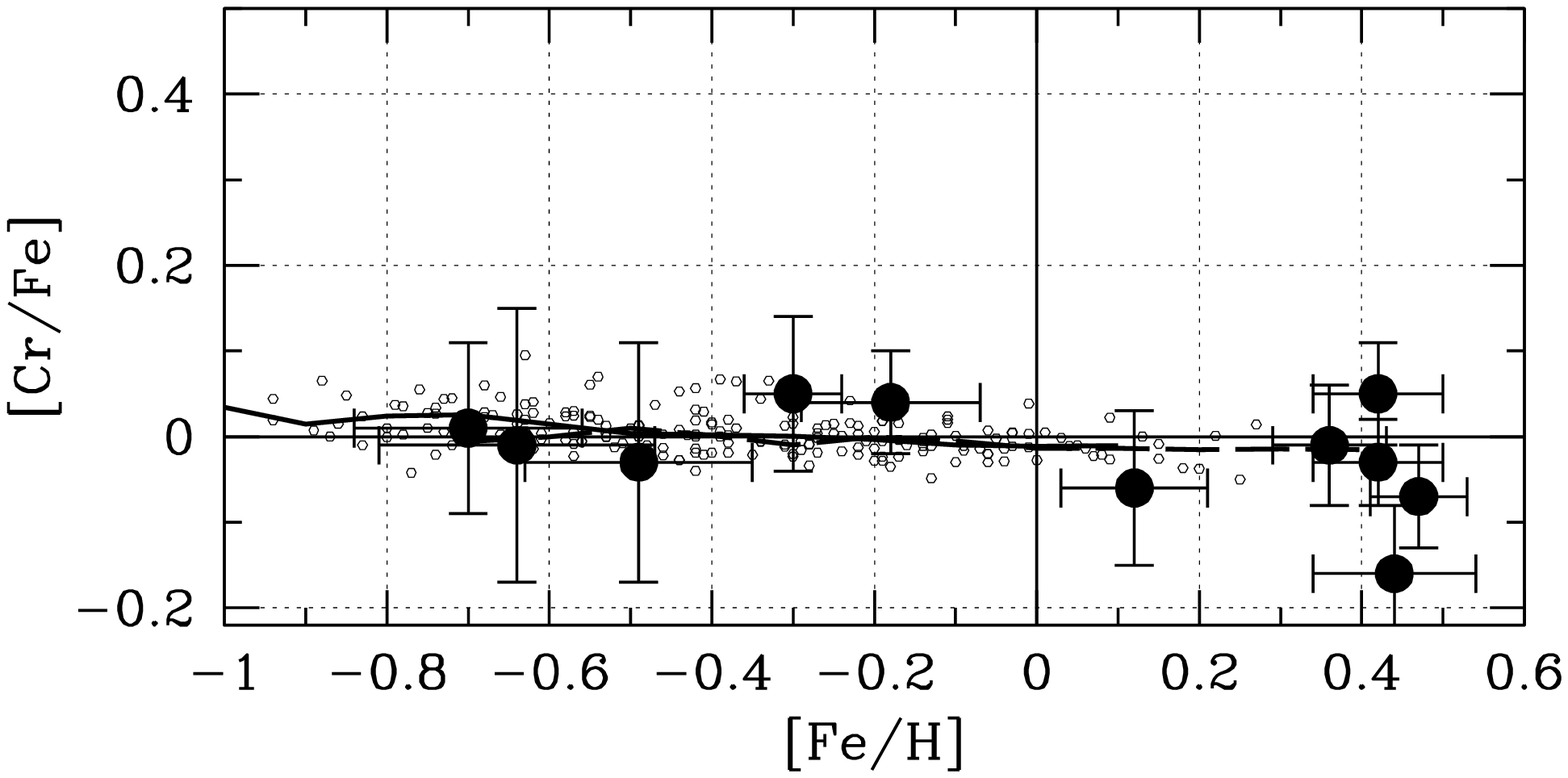}}
\resizebox{\hsize}{!}{
\includegraphics[bb=18 220 592 450,clip]{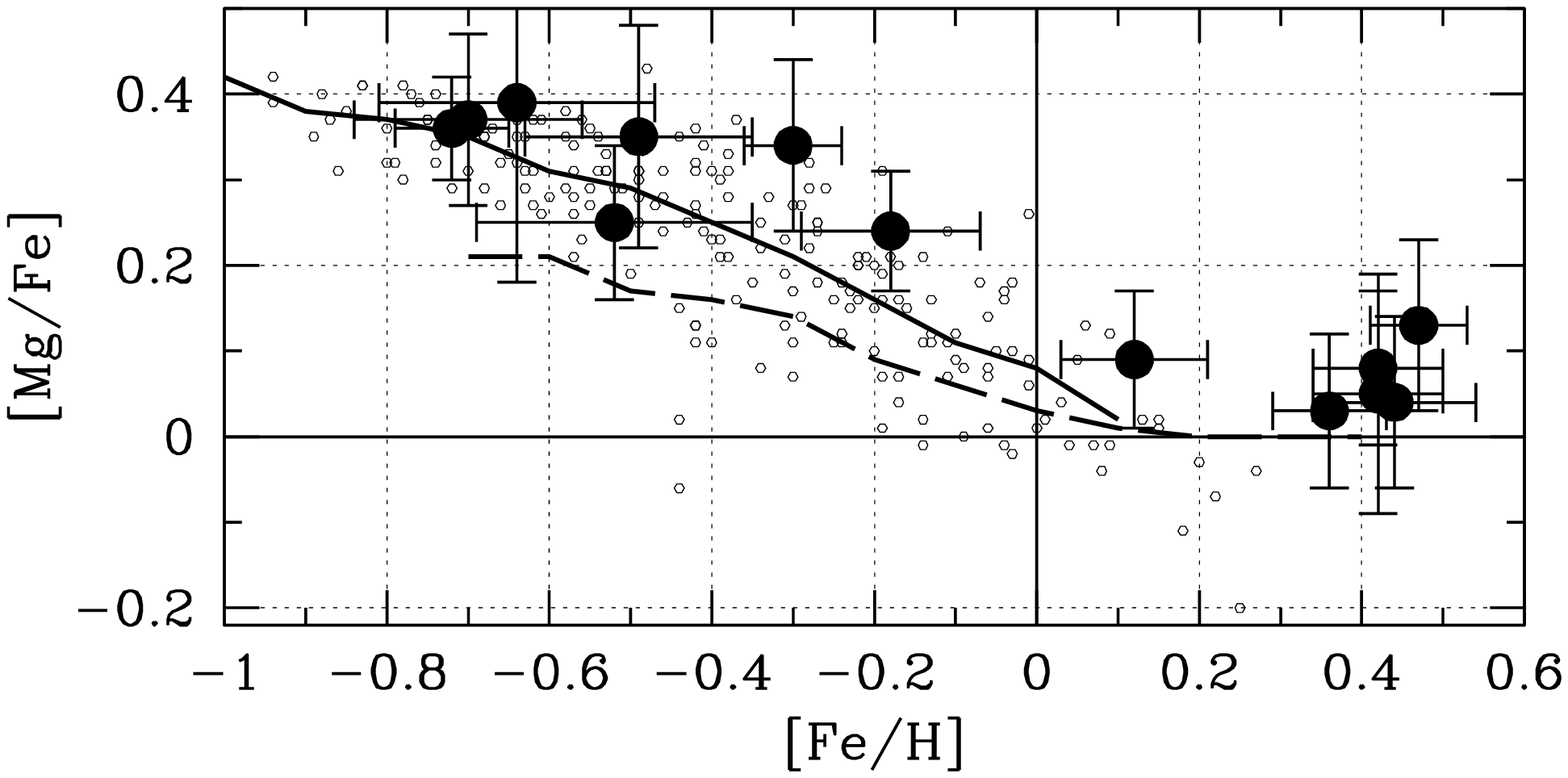}
\includegraphics[bb=18 220 592 450,clip]{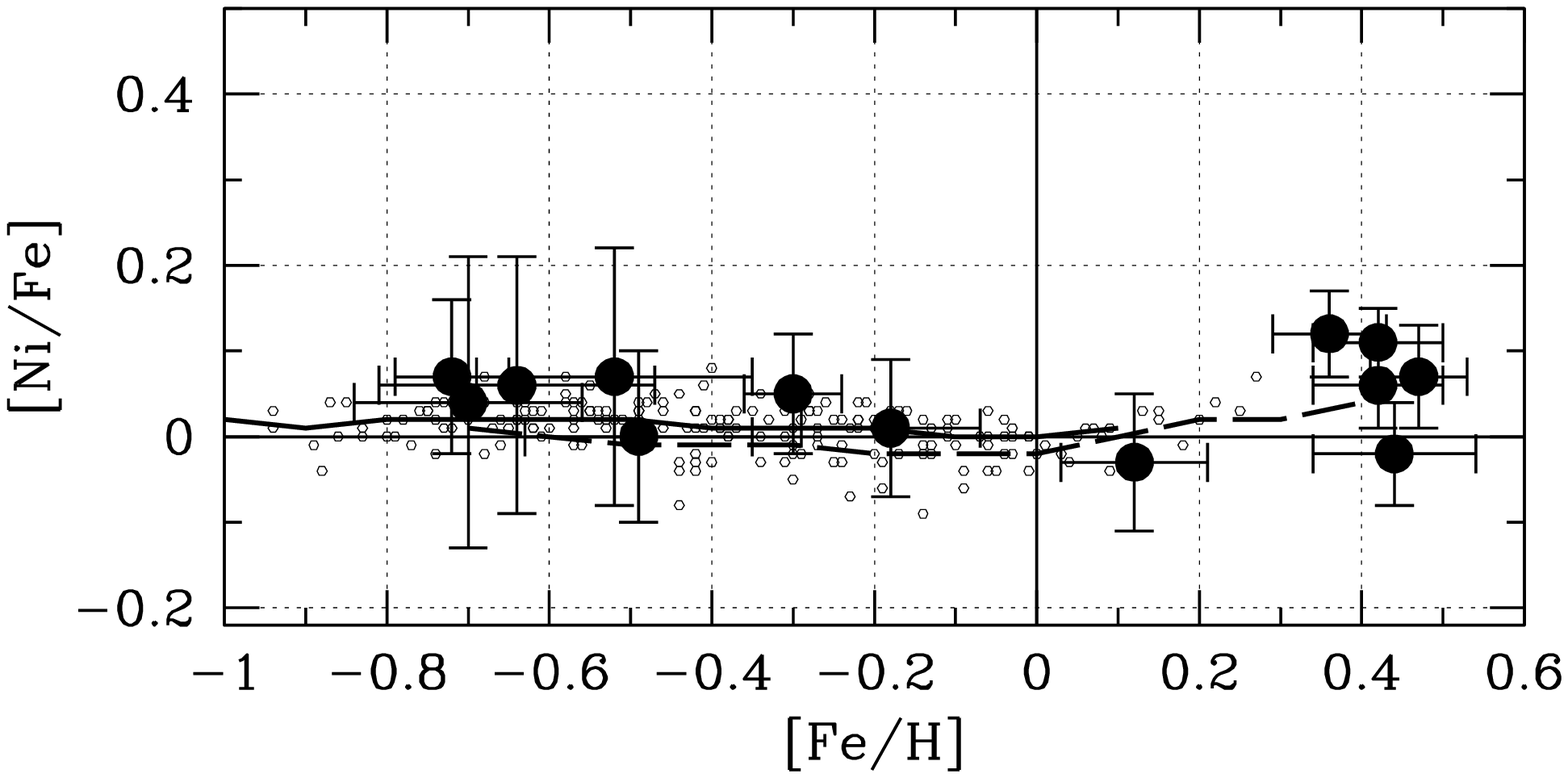}}
\resizebox{\hsize}{!}{
\includegraphics[bb=18 220 592 450,clip]{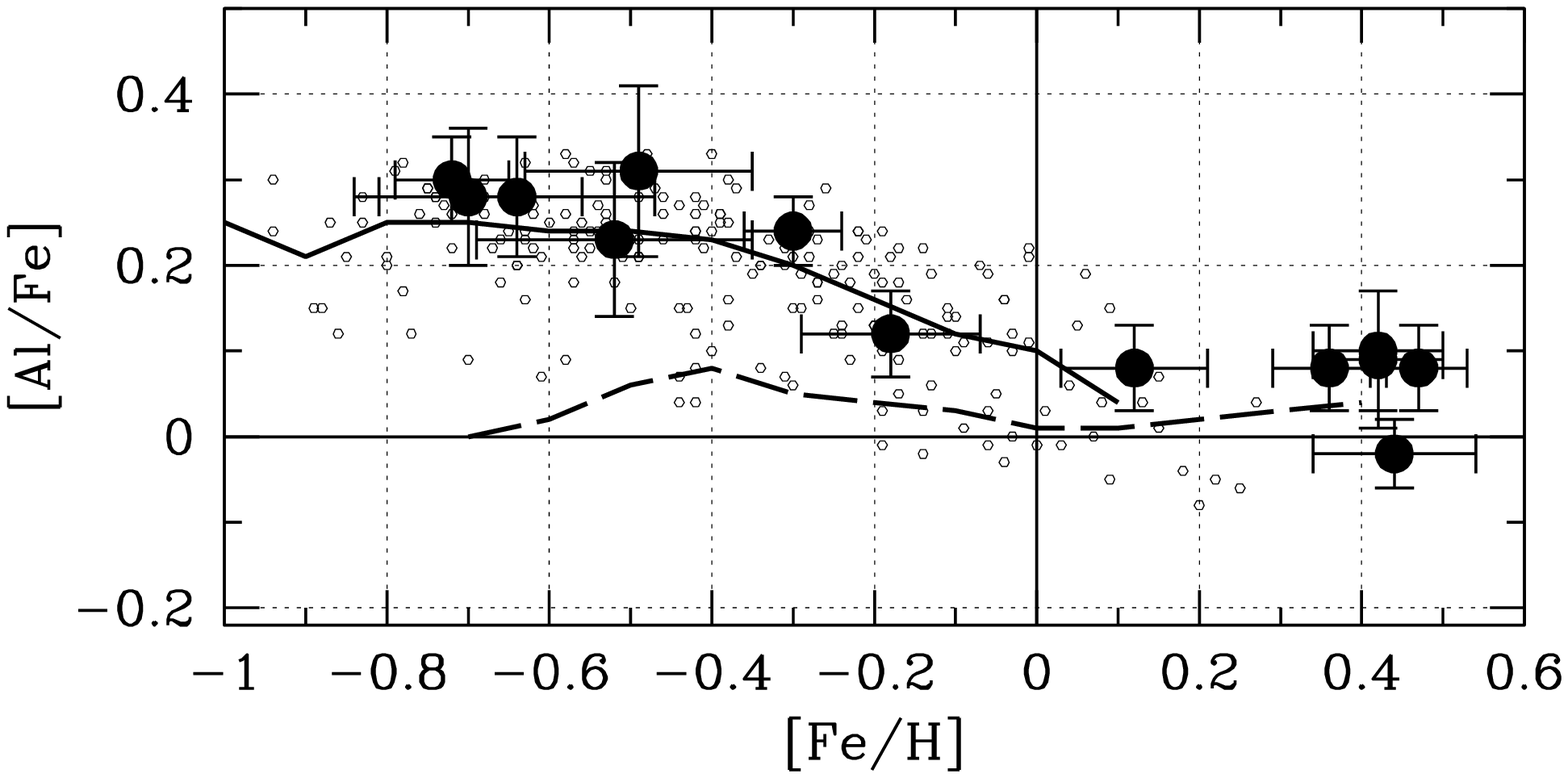}
\includegraphics[bb=18 220 592 450,clip]{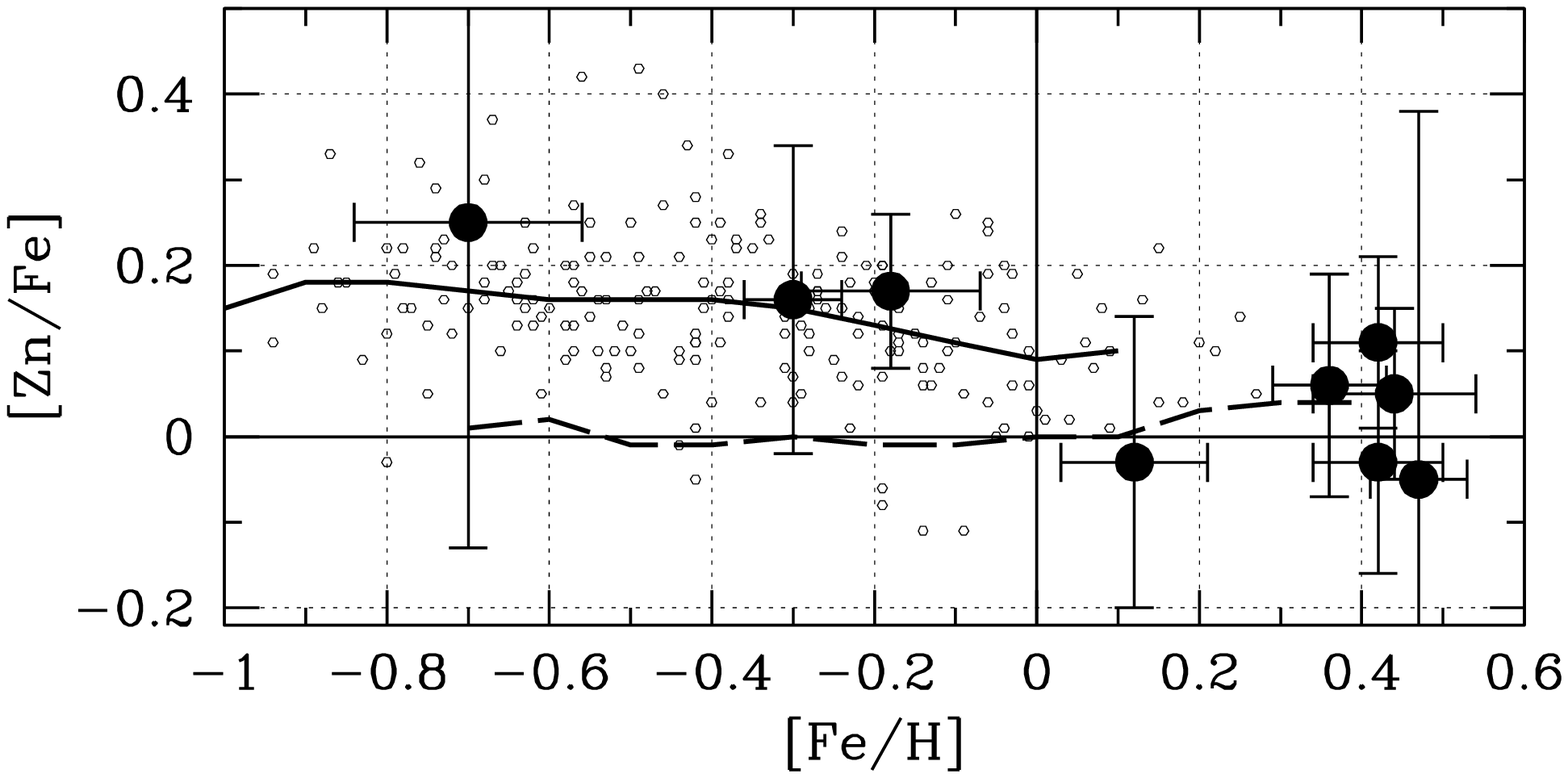}}
\resizebox{\hsize}{!}{
\includegraphics[bb=18 220 592 450,clip]{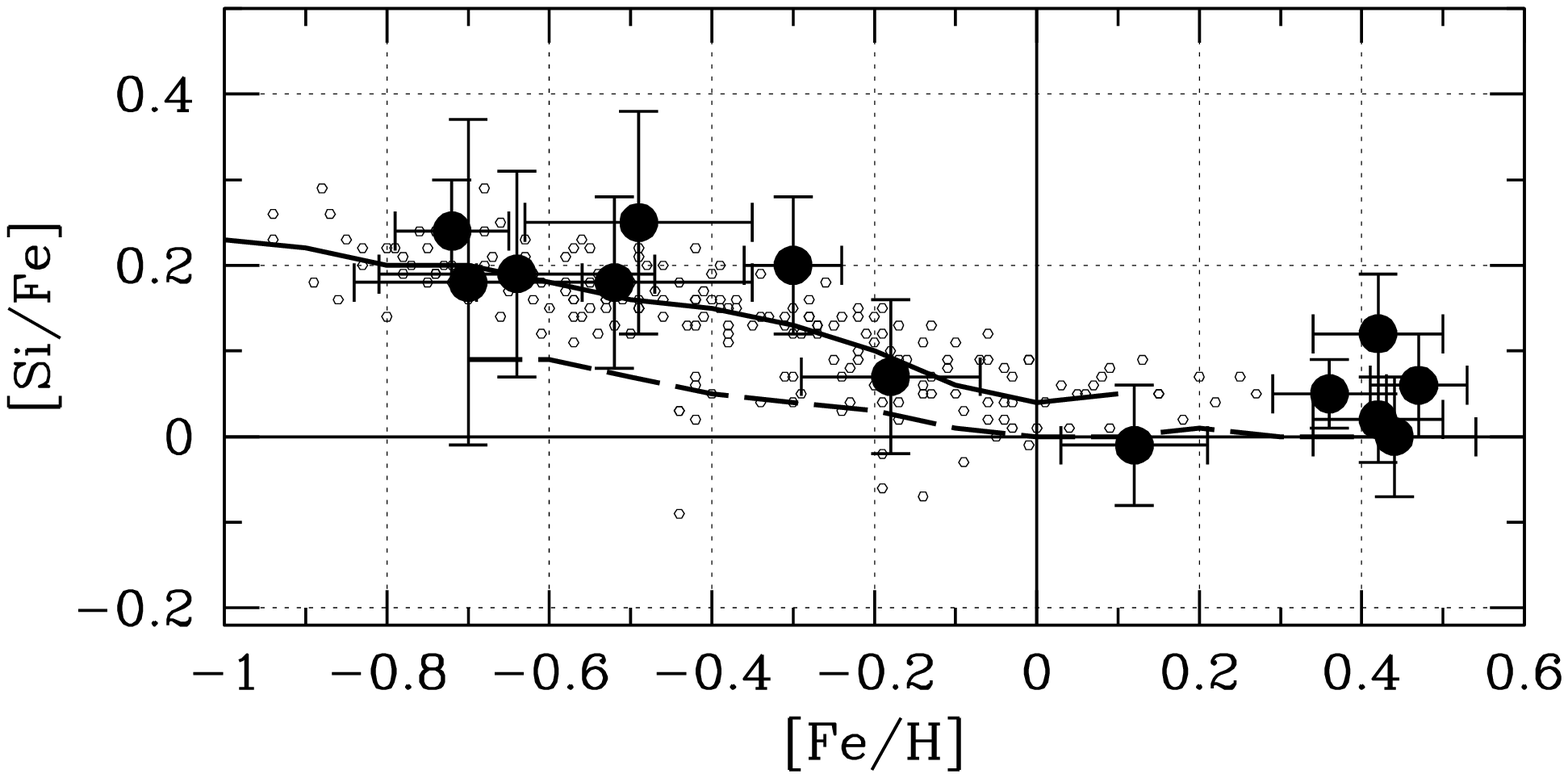}
\includegraphics[bb=18 220 592 450,clip]{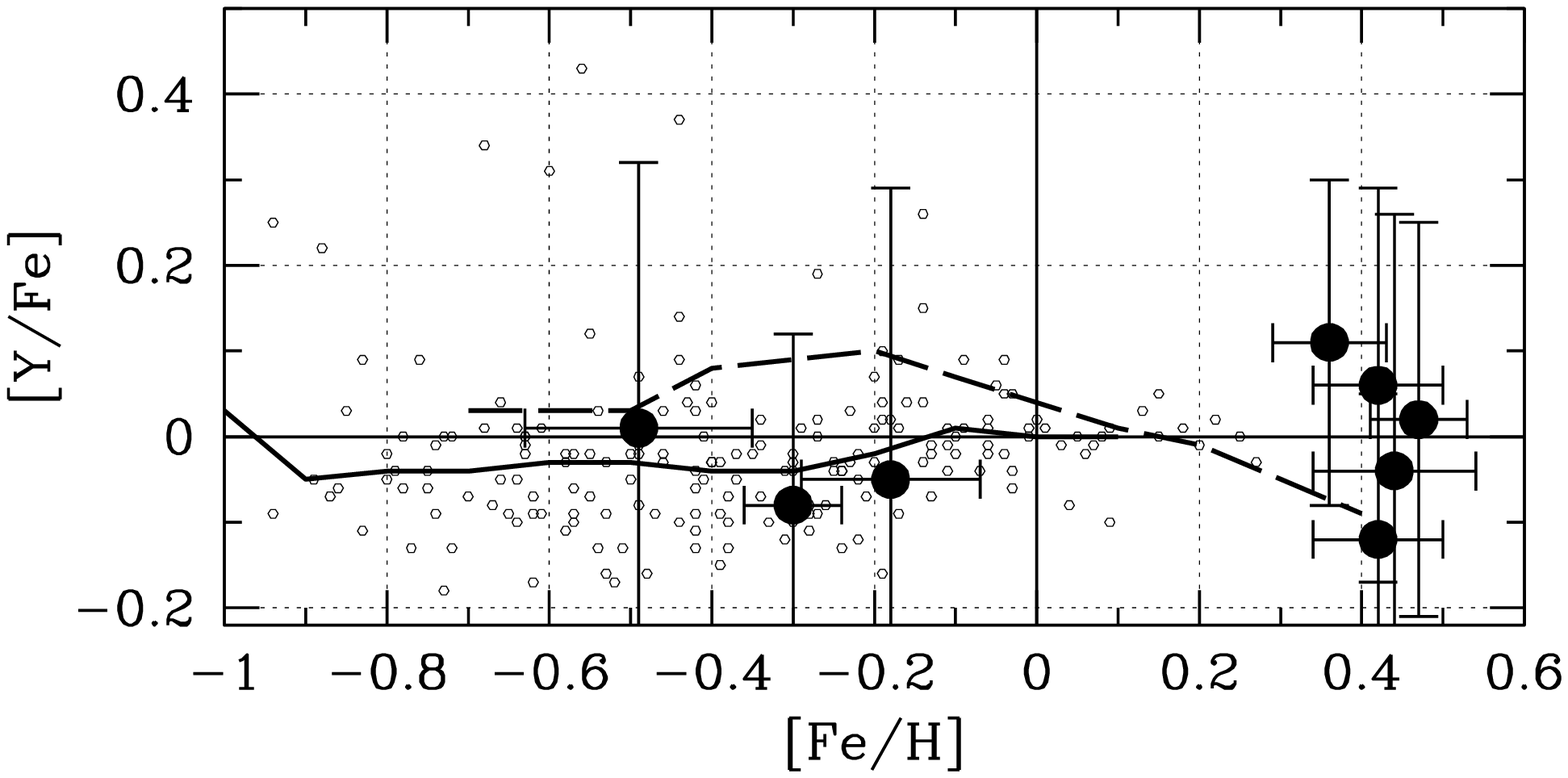}}
\resizebox{\hsize}{!}{
\includegraphics[bb=18 170 592 450,clip]{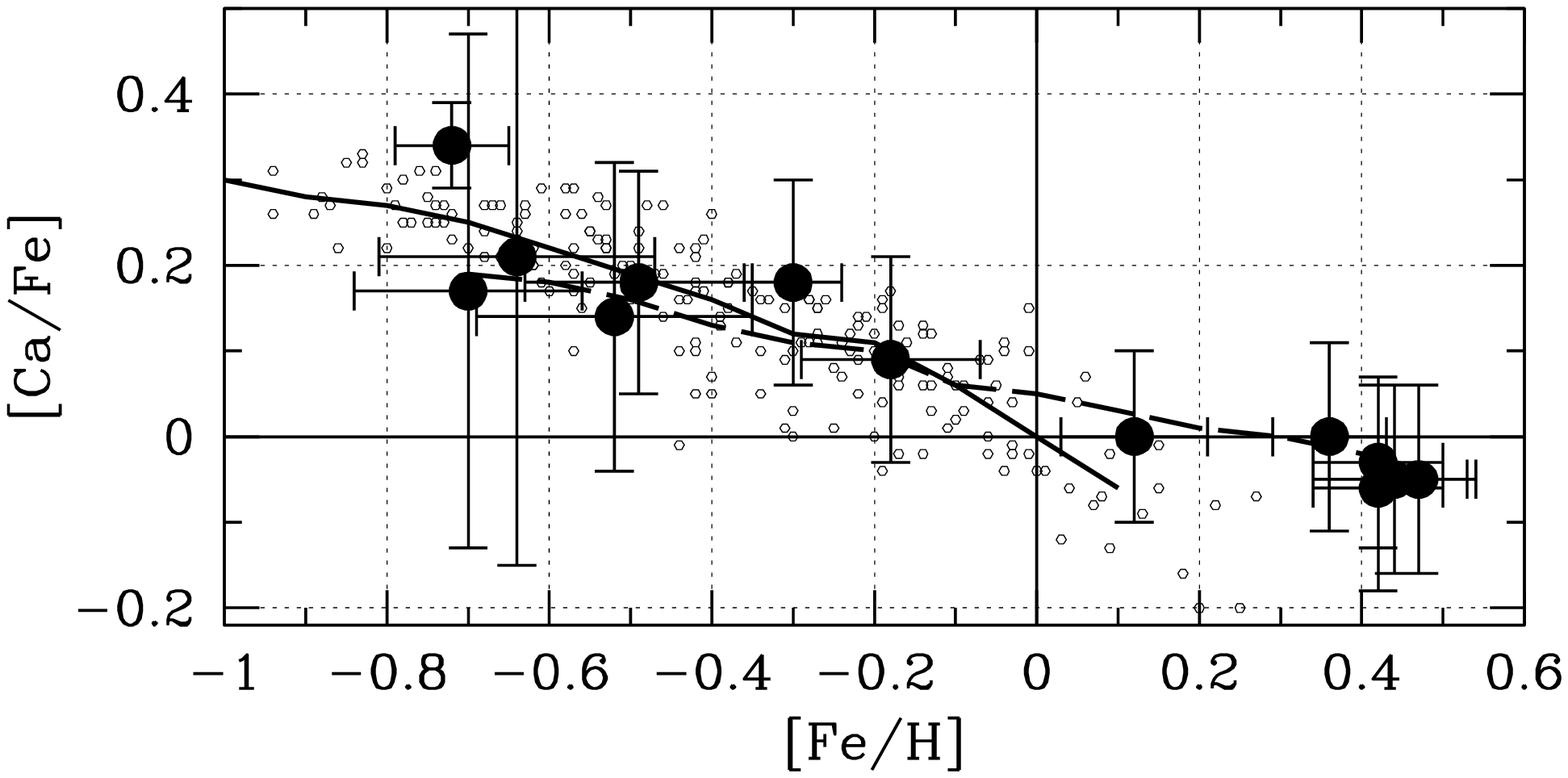}
\includegraphics[bb=18 170 592 450,clip]{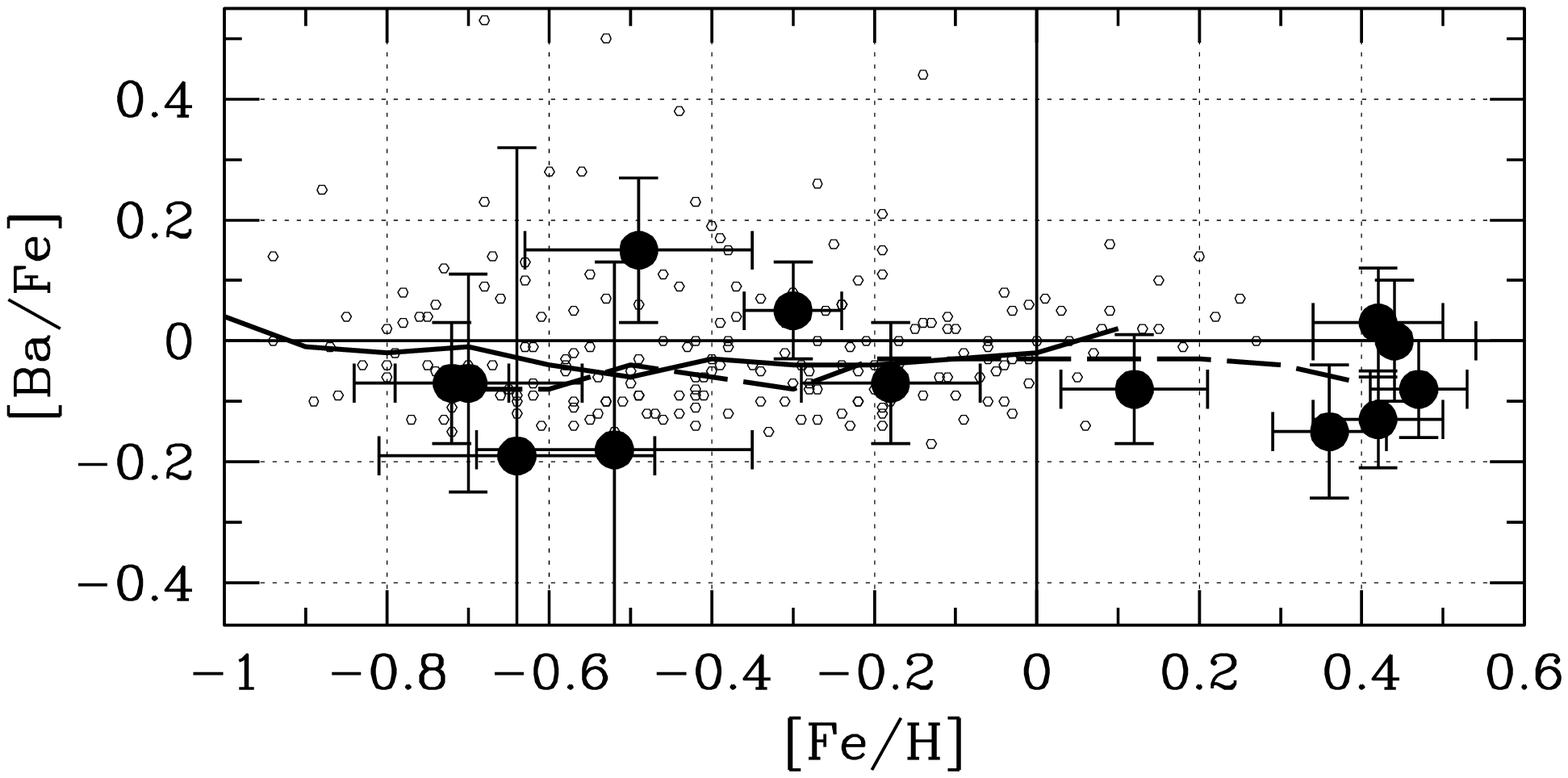}}
\caption{Abundance results for 13 microlensed dwarf and 
subgiant stars in the Bulge (marked by filled bigger circles).
Thick disc stars from Bensby et al.~(2010, in prep.) 
are shown as small circles, and the solid line is the running median
of the shown thick disc sample, and the dashed line the running
median of the (not shown) thin disc sample from 
Bensby et al.~(2010, in prep.).
The error bars represent the total uncertainty in the abundance ratios
(see Sect.~\ref{sec:errors} and Table~\ref{tab:abundances2}).
\label{fig:abundances}}
\end{figure*}

In spite of the small sample, it is notable that we only see the 
low ages for the metal-rich stars while all stars with sub-solar 
[Fe/H] are consistent with the classical view of the Bulge as an 
old population \citep[e.g.,][]{holtzman1993,feltzing2000b,zoccali2003}.
As the stars with sub-solar [Fe/H] also have enhanced levels of 
$\alpha$-elements (see Fig.~\ref{fig:abundances}), these stars all 
appear to adhere to the classical picture of the Bulge as a stellar 
population that formed rapidly early in the history
of the Galaxy (see, e.g., models and discussions
in \citealt{matteucci2001}). 

Overall, the evidence for young stars in the Bulge is scarce.
For instance, the extremely deep CMD of $\sim$\,180\,000 field stars in 
the Bulge by \cite{sahu2006} show no traces of a 
young population. It is therefore surprising to find 
three (MOA-2008-BLG-310S, MOA-2008-BLG-311S, and MOA-2006-BLG-099S)
out of 14 stars to have young ages. At this point we can only speculate
on their origin. One interpretation would be that the older 
stars are all bona fide Bulge stars while the young, metal-rich stars 
are disc interlopers. In the Galactic disc a young age and a 
high metallicity is common 
\citep[e.g.,][]{twarog1980b,feltzing2001,nordstrom2004}. 
Also,  the innermost Galactic disc is expected to be more
metal-rich than the Solar neighbourhood \citep[e.g.][]{colavitti2009}.
However, we still do not now if it is supposed to be young too.

It should furthermore be noted that these young Bulge stars are not 
brighter than the main old turnoff, they are just 
too blue to fall on old isochrones (see Fig.~\ref{fig:ages}).
Also, there are some theoretical limitations of the isochrone fitting
method. First, isochrones at $\rm [Fe/H]>+0.3$ have very few 
calibrators, and, second, the colour of the main sequence is strongly
affected by the Y (helium) content, which for the Bulge, or any
population with such high metallicity, is poorly known.
However, we find a whole range of ages at high metallicities, so 
we don't see a bias in our ages.

\section{Abundance trends in the Bulge}

\begin{figure*}
\resizebox{\hsize}{!}{
\includegraphics[bb=18 160 570 610,clip]{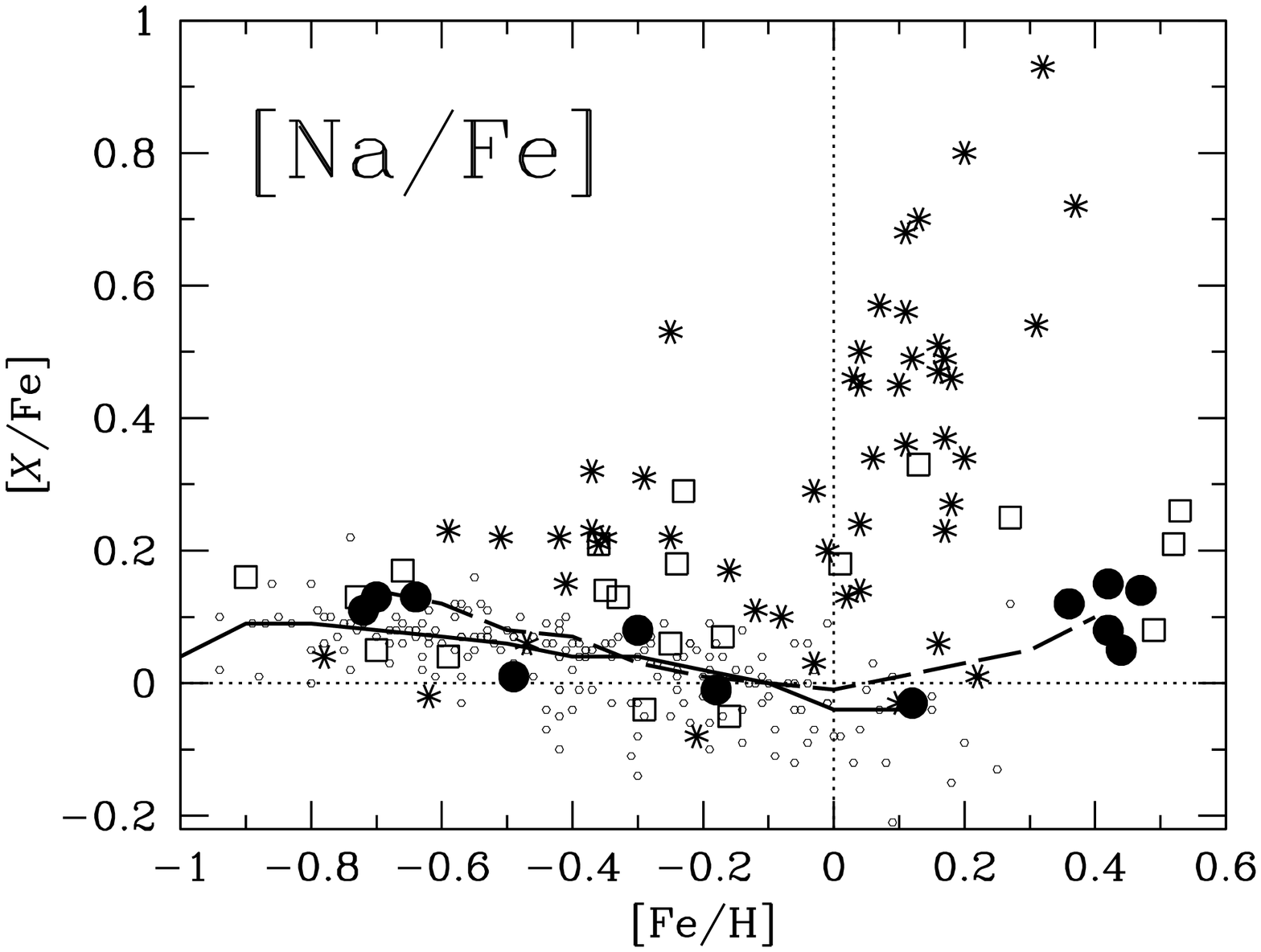}
\includegraphics[bb=95 160 570 610,clip]{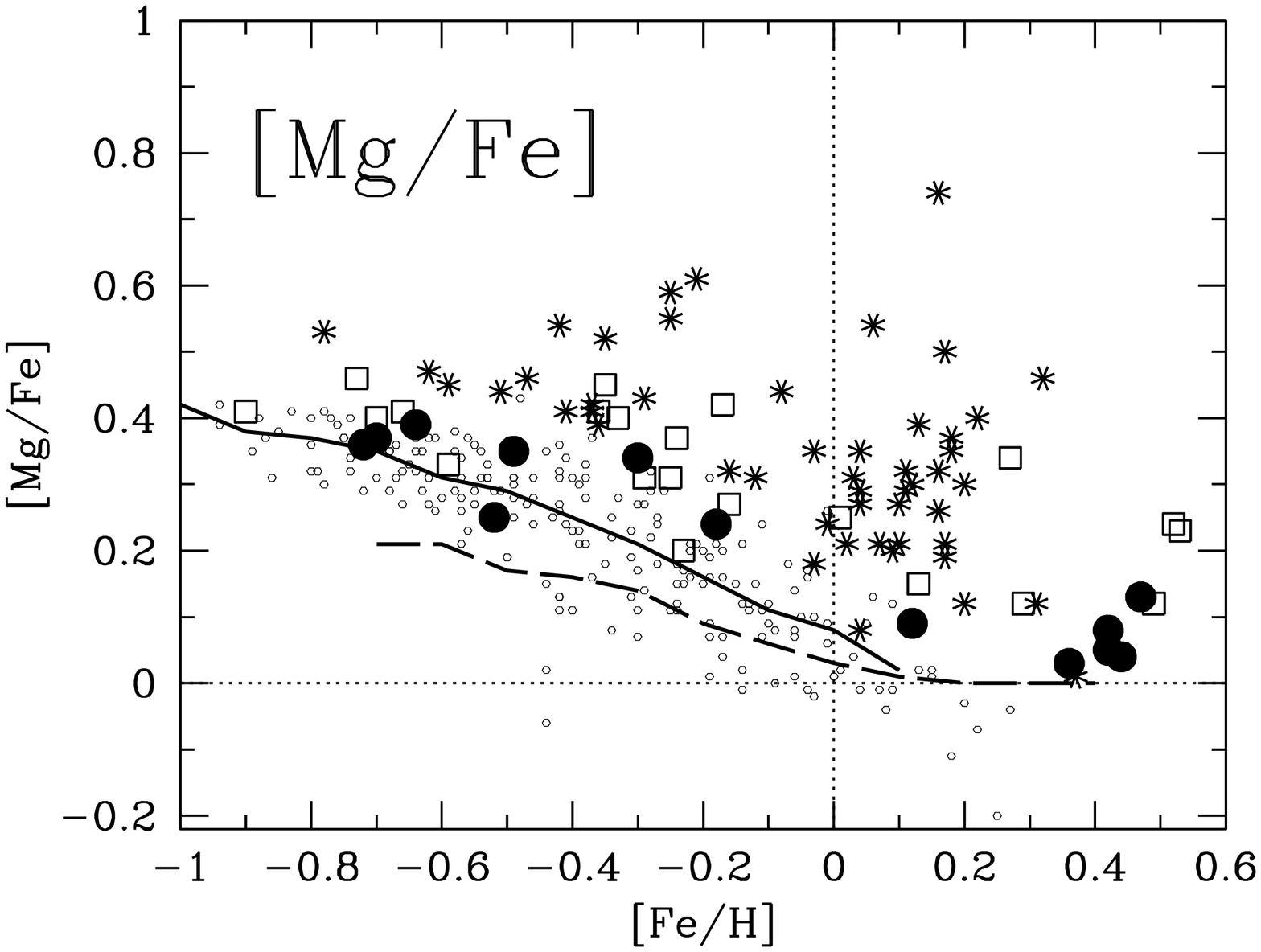}
\includegraphics[bb=95 160 592 610,clip]{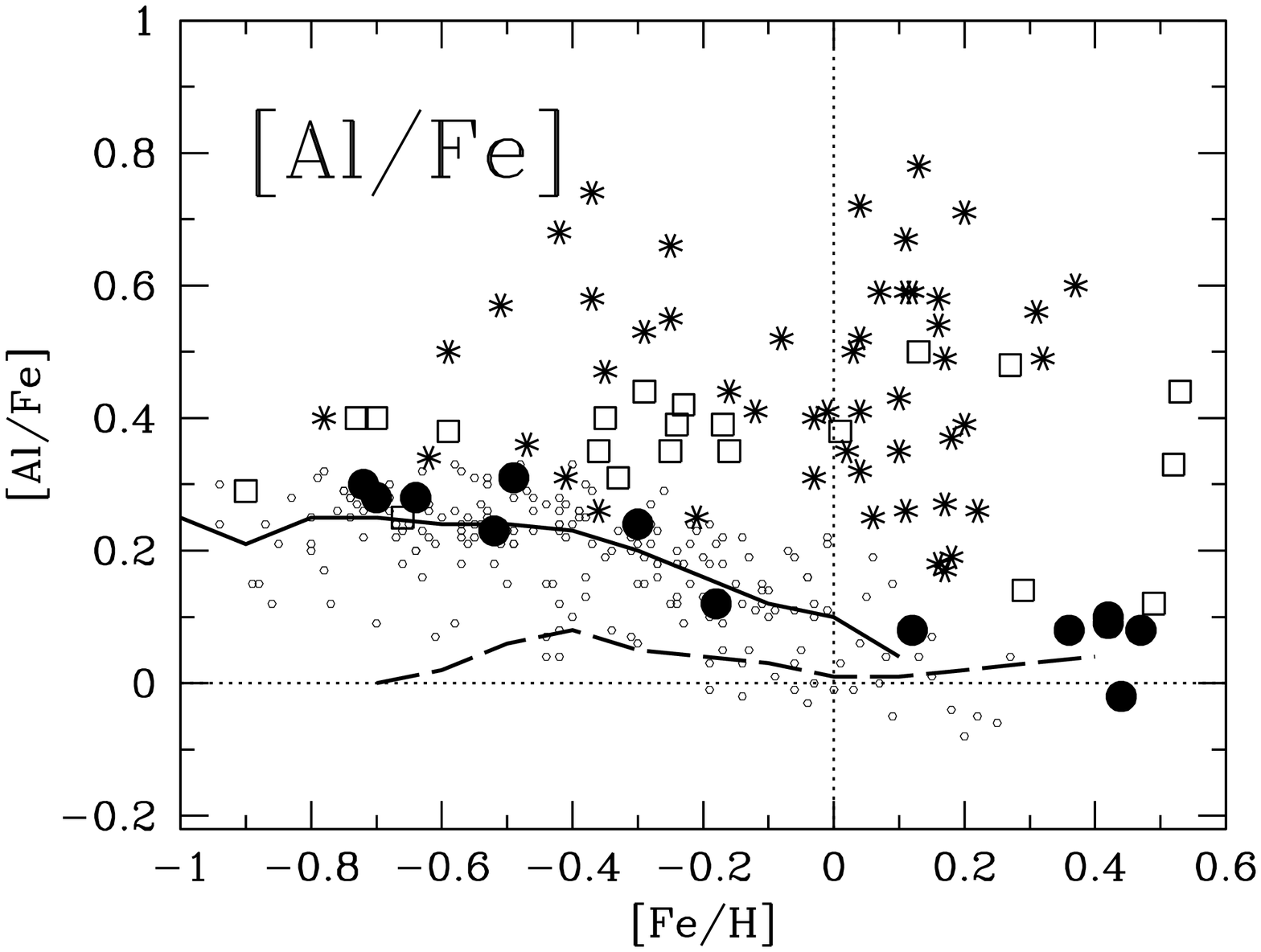}}
\caption{Comparisons of [Na/Fe], [Mg/Fe], and [Al/Fe] between
our microlensed dwarf stars (filled circles), giants from
\cite{fulbright2007} (empty squares), giants from
\cite{lecureur2007} (asterisks), nearby thick disc stars from
Bensby et al.~(in prep.) (small open circles). 
Solid and dashed lines represent the
running median of the thick and thin disc stars, respectively 
(same as in Fig.~\ref{fig:abundances}).
\label{fig:abund_lec}}
\end{figure*}

\subsection{General appearance}

Figure~\ref{fig:abundances} shows the abundance results 
for 13 of the 15 the microlensed dwarf and subgiant stars in the 
Bulge (MOA-2009-BLG-259S were excluded, see Sect.~\ref{sec:errorsum}, 
and for the \cite{epstein2009} star OGLE-2007-BLG-514S we only determined stellar parameters and 
Fe abundances.). 

Regarding the $\alpha$-elements (Mg, Si, Ca, and Ti), the Bulge 
dwarfs show enhanced $\rm [\alpha/Fe]$ ratios at sub-solar [Fe/H], 
that decline when approaching solar metallicities. At higher
metallicities the [$\alpha$/Fe] ratios are around or slightly
higher than solar. The oxygen trend is similar to the $\alpha$-element
trends at sub-solar metallicities, but differs at super-solar
[Fe/H] where it continues to decrease. The oxygen abundances
that are based on the infrared triplet lines at 773\,nm
have been NLTE corrected the according to the empirical formula 
given in \cite{bensby2004}. 

Generally, abundance trends of the 
dwarf stars in the Bulge are very well-defined. The scatter in the 
[Ti/Fe]-[Fe/H] plot for instance is remarkably low. The canonical 
interpretation of the plateau of high $\alpha$-element abundances 
relative to iron at low metallicities is due to early and rapid 
chemical enrichment of the Bulge by massive stars. When these stars 
die, they explode as core-collapse supernovae, producing a lot of 
$\alpha$-elements relative to iron. At some  point low-mass stars 
start to contribute to the chemical enrichment, and since these 
produce less of the $\alpha$-elements the $\rm [\alpha/Fe]$ ratio 
will start to decline.

The enhanced $\rm [Na/Fe]\approx+0.1$ for metal-rich disc stars
was already noticed by \cite{edvardsson1993}, \cite{feltzing1998},
\cite{shi2004}, and also in \cite{bensby2005} an upturn in [Na/Fe] 
can be seen. Our microlensed dwarf stars are in full agreement with
these disc results.

The [Ba/Fe] trend is flat and slightly under-abundant compared
to the Sun for all [Fe/H].

\subsection{Comparisons to the Galactic thick disc}

In Fig.~\ref{fig:abundances} we also show the thick disc abundance 
trends based on dwarf stars in the Solar neighbourhood 
\citep[taken from][and Bensby et al., in prep.]{bensby2003,bensby2005}. 
These thick disc stars have been analysed using the exact same methods 
(spectral line lists, atomic data, model stellar atmospheres, etc.) that
we use for the microlensed dwarf stars. Hence, any differences between 
the stars from the two stellar populations should be real, and not due 
to unknown systematic effects.

The first thing that can be taken away from Fig.~\ref{fig:abundances}
is that, at sub-solar metallicities, the abundance trends for the
Bulge dwarf stars are very similar to those of the thick disc stars. 
The solid lines shown in the figures indicate the median abundance 
ratio as a function of metallicity for thick disc stars. 
Even though the appearance of the abundance trends
at sub-solar [Fe/H] are very similar between the Bulge and the
thick disc, it is also evident that for many elements the Bulge 
stars appear to be slightly more enhanced than the median thick disc. 
The Bulge stars seem to occupy the upper envelope of the thick 
disc abundance trends. Part of this apparent shift in the 
abundance trends between the two populations could be due to that 
the thick disc sample is kinematically selected, and hence will 
unavoidably be mixed, to some degree, with kinematically hot thin 
disc stars \citep[see][]{bensby2007letter2}. The median thick disc
line that we show in Fig.~\ref{fig:abundances} will then be slightly
too low. However, this effect should only be important when approaching
solar metallicities, and we do see a shift between the Bulge and the
thick disc at lower [Fe/H] as well, where the kinematic confusion 
between the thin and thick discs should be negligible.
However, the shift is not for all elements, and it is small,
on the order of 0.05\,dex or less. More microlensing
events will help us to clarify if this shift is real or not.

A possible link between the Bulge and the Galactic thick disc 
based on similarities of abundances was first suggested by 
\cite{prochaska2000}, pointing out the 
``excellent agreement" between the abundance ratios in their sample
of ten thick disc stars to those of the Bulge giants of 
\cite{mcwilliam1994}. As the thick disc sample of \cite{prochaska2000} 
only reached $\rm [Fe/H]\approx -0.4$, the downturn in 
$\rm [\alpha/Fe]$ that we now see in the thick disc at 
$\rm [Fe/H]\approx -0.35$ \citep[e.g.][]{feltzing2003,bensby2007letter2} 
was at that time not known. Hence, when later studies of giant stars 
in the Bulge showed that the $\rm [\alpha/Fe]$ remained high even at 
super-solar metallicities \citep[e.g.][]{fulbright2007}, in contrast
to the declining thick disc trends, the possible connection between 
the Bulge and the thick disc became less clear. It should also be noted
that \cite{prochaska2000} did not analyse both the Bulge and thick disc 
samples. Also, recently, \cite{melendez2008} presented a consistent
analysis of giant stars in both the Bulge and the thick disc that 
found a similarity between them for C, N, O (recently confirmed by 
\citealt{ryde2009b}). The agreement is extended
to other $\alpha$-elements in the upcoming study by 
\cite{alvesbrito2010}. 
However, the first confirmation that the Bulge and the thick disc have 
similar abundance patterns  based on dwarf stars came from 
\cite{bensby2009,bensby2009letter}, and with this study it now 
appears well established that the Bulge and the thick disc have had, 
to some degree, similar histories.

Before interpreting the apparent agreement between the
Bulge and thick disc abundance trends, it is important to recognise
that the MDFs for the thick disc and the Bulge differ in the average
as well as width of their MDFs.
The metallicity distribution for the thick disc peaks at
$\rm [Fe/H]\approx -0.6$ \citep[e.g.,][]{carollo2009},
and its metal-rich tail likely reach solar metallicities 
\citep{bensby2007letter2}. Stars with $\rm [Fe/H]>0$ that can 
be kinematically classified as thick disc stars, are heavily biased
to belong to the high-velocity tail of the thin disc
\citep{bensby2007letter2}.
Even though the average metallicity of the Bulge and the shape of 
its MDF is under debate (see Sect.~\ref{sec:mdf}) it is clear that it
spans the full range of metallicities of the thick disc, and in addition
reaches very high, super-solar [Fe/H].
Hence, the agreement, as regards the abundance trends, between the 
Bulge and the thick disc
can currently only be established for sub-solar 
metallicities. The comparison that can be made at super-solar 
[Fe/H] is between the Bulge stars and thin disc stars 
(dashed lines in Fig.~\ref{fig:abundances}) that reach similar high
metallicities. As the abundance
trends in Fig.~\ref{fig:abundances} show, the agreement 
between the Bulge and the thin disc at $\rm [Fe/H]>0$
is good. Looking at, e.g., the [Na/Fe], [Ni/Fe], and [Ba/Fe]
trends, which are not simply flat at the highest [Fe/H], the Bulge
stars nicely extend the trends that are seen for the nearby 
thin disc stars. However, \cite{epstein2009} found
a [Ba/Fe] value noticeably below the disc for OGLE-2009-BLG-514S. The
\ion{Ba}{ii} lines are strong and very sensitive to the
microturbulence parameter, and therefore measuring the abundances 
of another heavy element, such as La, would be a useful check.

With this study where we compare Bulge {\it dwarf}
stars with disc {\it dwarf} stars, and with the recent studies
by \cite{melendez2008} and \cite{alvesbrito2010}, all 
using internally consistent methods, it appears clear that 
the Bulge and thick disc abundance trends are similar. 
The agreement between the Bulge and the thick disc
means that conclusions from recent theoretical works,
developed under the assumptions that the abundance trends in 
the Bulge and thick disc are different, may not be valid.

We caution that the comparisons that we do and that have been done by 
others are between the Bulge stars close to the Galactic centre
and thin and thick disc stars in the vicinity of the Sun, i.e., at a 
distance of approximately 8\,kpc from the Galactic centre. In order 
verify a possible connection between the Bulge and the Galactic thick 
(and thin) disc(s) we need disc stellar samples much closer to the 
Bulge, say at 4\,kpc from the Galactic centre. At that distance the 
contamination of Bulge stars in a disc stellar sample should be small, 
allowing us to directly compare the populations. 
If, for instance, the star formation has been more rapid in the
inner disc than in the Solar neighbourhood, or in the outer disc,
we should expect the ``knee" in the $\alpha$-element
abundance trends at higher [Fe/H] than what we see in the
Solar neighbourhood. Then that could explain the apparent
slight shift in the abundance trends between the Bulge and
the thick disc in the Solar neighbourhood that we see. However, 
no such inner disc sample is currently available.

\subsection{Comparisons to Bulge giants}

It should be noted that some recent studies of Bulge giant stars 
claim that the Bulge is enhanced in the $\alpha$-elements with 
respect to the stars of the Galactic disc. For instance, 
\cite{rich2005} compare their Bulge giants to nearby thin disc giants 
and find the Bulge ones to be more enhanced;
\cite{zoccali2006,fulbright2007} and \cite{lecureur2007} find 
their Bulge giants to be more enhanced than comparison samples of 
nearby thin and thick disc dwarf stars. In contrast, the recent 
study by \cite{melendez2008}, and the upcoming study by 
\cite{alvesbrito2010}, compare Bulge giants with nearby thick disc 
giants, and find the Bulge and thick disc abundance trends to be 
similar.

\begin{figure*}
\resizebox{\hsize}{!}{
\includegraphics[bb=18 160 592 710,clip]{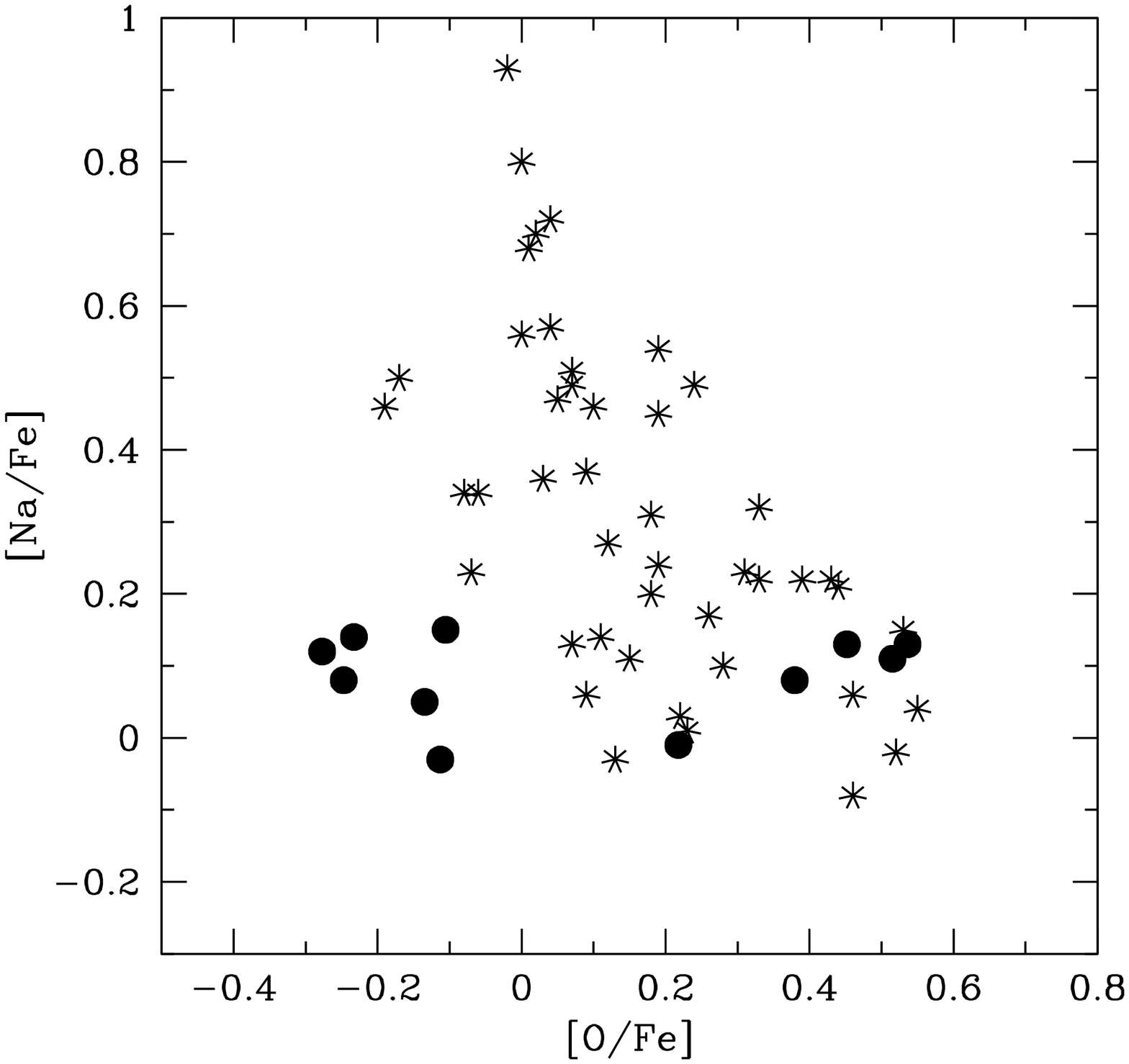}
\includegraphics[bb=18 160 592 710,clip]{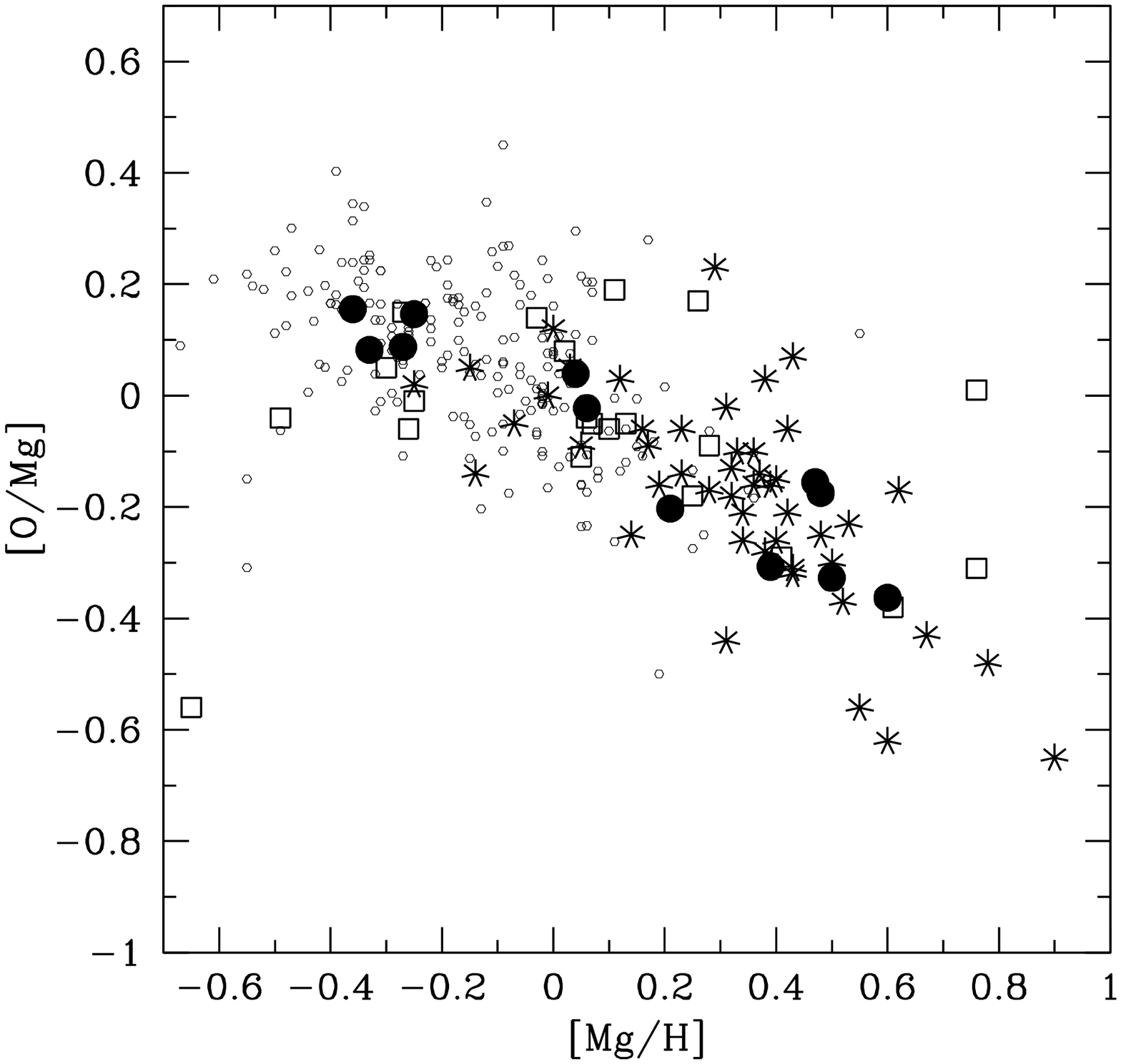}}
\caption{[Na/Fe] versus [O/Fe] ({\sl Left panel}) and  
[O/Mg] versus [Mg/H] ({\sl Right panel}) for our 
microlensed dwarf stars (filled circles), the giants from 
\cite{fulbright2007} (open squares), the giants from 
\cite{lecureur2007} (asterisks), and thick disc dwarf stars 
in the Solar neighbourhood
\citep[][and in prep.]{bensby2003,bensby2005} (small open circles).
\label{fig:ona_anticorr}}
\end{figure*}

In Fig.~\ref{fig:abund_lec} we compare [Na/Fe], [Mg/Fe] and [Al/Fe] 
between our microlensed Bulge dwarfs, the giants from 
\cite{fulbright2007} and \cite{lecureur2007}, and the Solar 
neighbourhood thin and thick disc dwarfs from 
Bensby et al.~(2003, 2005, and in prep.). It is evident that the 
giants from \cite{lecureur2007} are much more enhanced in Na, Mg, 
and Al than any of the other comparison samples,
with our dwarf stars having lower abundances, the \cite{fulbright2007}
giants having intermediate abundances, and the \cite{lecureur2007} 
giants having the highest abundances. The higher abundances of the 
giant stars might be due to the way these abundances have been 
normalised to the Sun. For dwarf stars with spectral types similar 
to the Sun, the normalisation is straightforward, it is just to 
analyse the Sun in the same way and then subtract the solar 
abundance. However, for giants, having very different $\log g$:s 
and $\teff$:s than the Sun, the normalisation is usually done to 
another standard star, such as, e.g., $\mu$\,Leo. This could partly 
explain the levels of the abundances, progressing from our dwarf 
stars, to the \cite{fulbright2007} giants, and to the 
\cite{lecureur2007} giants.

Looking closer at the \cite{lecureur2007} data it appears as if 
there is a very well defined lower envelope just above the thick 
disc trend, and on top of that a large scatter of stars with 
higher abundance ratios, spreading upwards. We suspect that this 
might be due to line blending, and possibly lack of accurate 
continuum points, in the extremely crowded spectra that 
metal-rich giants have. The  lower envelope that is seen in the 
\cite{lecureur2007} is what could be expected if blending is 
the case. A number of stars that have less (or very few) blends 
forming the well-behaved lower envelope of the trends, while others 
are more affected, leading to randomly increased equivalent widths, 
and hence randomly increased abundances.

\subsection{No Na-O anti-correlation}

The Na-O anti-correlation that has seen in all globular 
clusters studied so far is not present for field stars  
\citep[e.g.,][]{carretta2009}. Therefore, one of the most
striking results from \cite{lecureur2007}
was the Na-O anti-correlation that they found for their
Bulge giants, and they claim it is probably an effect
of the Bulge chemical evolution. However, Fig.~\ref{fig:ona_anticorr} 
({\sl left panel}) shows the Na-O plot for our microlensed dwarf stars 
and there is no Na-O anticorrelation present. Instead [Na/Fe] versus 
[O/Fe] is flat. Given the large spreads that \cite{lecureur2007} see 
for Na, Mg, and Al (see Fig.~\ref{fig:abund_lec}) it appears likely 
that their results are affected by systematic errors (blending 
presumably).

\subsection{Metallicity dependent oxygen yields}

In Fig.~\ref{fig:ona_anticorr} ({\sl right panel}) the evolution of 
[O/Mg] vs [Mg/H] is shown, which reveals a very tight correlation. 
Our results from microlensed dwarfs have significantly less scatter 
than the corresponding values for Bulge red giants 
\citep[e.g.][]{lecureur2007,fulbright2007,alvesbrito2010}
presumably due to the lesser influence of blends and better
determined stellar parameters. The dwarf-based slope is also slightly
shallower than for the giants. A declining trend in [O/Mg]
towards higher metallicity is expected with traditional
metallicity-dependent supernovae yields such as those of
\cite{woosley1995} but our slope is steeper, which
may signal an metallicity-dependence introduced for example
by mass-loss in massive stars \citep[e.g.][]{maeder1992,meynet2002}
as discussed by \cite{mcwilliam2008} and \cite{cescutti2009}.
These findings are based on the increased Wolf-Rayet stellar
wind efficiency at higher metallicity that removes a larger fraction of
He and C before they are converted to O and thus decreasing
the O production but leaving the Mg-yield largely unaltered.
The similarity between the Bulge and disc results implies that
these metallicity-dependent nucleosynthesis yields are a
general feature but also argues against substantial differences
in the initial mass function (IMF) between the two populations.

\section{The origin of the Galactic bulge}

To summarise our observations, we find that the
stars with high [$\alpha$/Fe] and low-metallicity in the
Bulge are old, but the stars with high metallicities and
solar (or subsolar) [$\alpha$/Fe] ratios span a range of ages
from 10 Gyr to 3 Gyr. What does this mean for the origin of
the Bulge?

The old age, high [$\alpha$/Fe] population can be
explained through the standard chemical pattern in systems
dominated by Type II SNe that is common to all stellar systems.
That the high [$\alpha$/Fe] ratio persist
to $\rm [Fe/H]>-0.6$ indicates that star formation proceeded very
efficiently in the event(s) that created the Bulge. Such
events could be early mergers of subhalos which drive efficient
star formation as well as contributing their own high [$\alpha$/Fe]
and low [Fe/H] stars \citep{rahimi2009}, or early fragmentation
of the disc into clumps of stars and gas which then rapidly merge
to form the Bulge \citep{immeli2004}. These results may also
be consistent with the secular evolution of the disc, depending
on the age-metallicity-[$\alpha$/Fe] relation present in the
inner disc. In the models of \cite{schonrich2009a}, the
inner disc is composed of stars that have old ages and higher 
[$\alpha$/Fe] at the same [Fe/H] than stars formed in the thin 
disc in the Solar neighbourhood. Some of these inner disc stars 
then migrate outwards to form the local thick disc. Thus it would 
not be surprising that the Bulge (=puffed up inner disc) stars 
should be chemically similar to the local thick disc (=migrated 
inner disc) stars, which is exactly what we see.
The model by \cite{schonrich2009a} requires
a number of assumptions/approximations/parameterisation of the
migration and heating processes, and is tuned to explain the thick 
disc in the Solar neighbourhood, so observational evidence 
(still missing) of the nature of the inner disc is, once again, 
important. Our results are not in agreement with the models of 
\cite{immeli2004} where the gas cools less efficiently and the 
instability that forms the bar sets in at later times, because those 
models predict a decline in [$\alpha$/Fe] starting at much lower 
metallicities (see their Figure 10).

The solar [$\alpha$/Fe] and high metallicity
stars that span a range of ages are more of a puzzle.
The old stars imply that part of the Bulge
got a head start on its chemical evolution, so that some 10 Gyr
old stars were formed from a population that was already producing
Type Ia SNe, while other 10 Gyr stars were forming out of
gas that had just been enriched with Type II SNe. Old, low
[$\alpha$/Fe] stars are found in the dwarf spheroid galaxies, but
accreting a Sagittarius-like object would not explain the pattern,
because the low [$\alpha$/Fe] stars in Sagittarius have too low 
metallicities \citep[e.g.,][]{venn2004}. In the simulations of
\cite{rahimi2009}, the bulges experience a series of mergers over
a period of $\sim$5 Gyr, leading to populations of "old", "intermediate"
and "young" stars. The distributions of [Mg/Fe] for these
populations do show some old stars with low [Mg/Fe], (as well the
expected shift to low [Mg/Fe] for the younger stars) but they caution
that their code suppresses mixing among gas particles, leading to
artificially high abundance ratio dispersions. Nonetheless, this is
what we see in our data, so perhaps the Galaxy found a way to suppress
mixing as well, maybe with a merger history different than the
two cases in \cite{rahimi2009}, where two of the subclumps that merged
had different starting times for star formation relative to today.

Finally, the younger, low [$\alpha$/Fe] and high metallicity stars
show that star formation persisted in the components that
created the Bulge. These stars are seen in the disc fragmentation
models of \cite{immeli2004}, where, for the cold gas model that
agrees with the turnover in the [$\alpha$/Fe] vs. [Fe/H] diagram, 
30\,\% of the baryons are not rapidly converted to stars. Instead, 
low level star formation occurs for several Gyr
and produces a peak in the
[Mg/Fe] histogram of the Bulge at $\rm [Mg/Fe]=-0.2$. This kind of stars
are also seen in the models of \cite{rahimi2009} where new
stars formed after mergers at later times are polluted with
Type Ia ejecta as well. Finally, depending on the star formation of the
thick disc, these stars may be present in that component and then
used to make the Bulge.

To distinguish further among these models, the inner disc of the
Galaxy needs to be better characterised observationally. In addition,
more Bulge dwarfs with accurate ages, metallicities and abundance ratios
would help clarify whether the age spread is confined to the
higher metallicities and whether the oldest low [$\alpha$/Fe] stars
are as old as the lower metallicity high [$\alpha$/Fe] stars. Finally,
the elements produced in Type Ia SNe are not the only chemical
evolution "clock" available,
and measuring elements produced in AGB stars (C, N and s-process), for
example, would test whether chemical evolution really began earlier
for some stars now in the Bulge than for others.

\section{Summary}

With this study we have doubled the number statistics on the
data for microlensed dwarf and subgiant stars in the Bulge.
All stars have been observed with high-resolution spectrographs
and from a detailed elemental abundance analysis we present results 
for O, Na, Mg, Al, Si, Ca, Ti, Cr, Fe, Ni, Zn, Y, and Ba. The method we
utilise is identical to the method used for a large sample
of 702 F and G dwarf stars in the thin and thick discs
in the Solar neighbourhood. Therefore, any differences between the 
Bulge stars and the disc stars should be genuine, and not due to
unknown (systematic) uncertainties. We have also determined stellar
ages for the stars, a task that is impossible to do with giant stars.

The main results and conclusions that can be drawn from our sample
of 15 microlensed stars in the Bulge are:

\begin{enumerate}
 \item The stars span a wide range of metallicities between 
 $\rm [Fe/H]=-0.72$ up to super-solar metallicity of 
 $\rm [Fe/H]\approx+0.54$.
 \item The mean metallicity of the 14 microlensed dwarf 
 and subgiant stars is 
 $\rm \langle[Fe/H]\rangle=-0.08\pm0.47$ in good agreement with
 the 204 giant stars in Baade's window from \cite{zoccali2008}
 that have an average metallicity of $-0.04$\,dex. 
 However, a two-sided KS-test gives only a low 30\,\% probability that
 microlensed dwarf stars and giant stars in BaadeÕs window have the
 same MDFs. The low probability is due to the skewed and uneven
 metallicity distribution of the dwarf stars, with excesses at both low
 and high metallicities.
 More observations of microlensed dwarf stars will certainly
 refine the comparison. It is clear though that the extremely metal-rich
 MDF for the Bulge that \cite{cohen2009} propose is not borne out
 by the larger sample presented here.
 \item The abundance trends that the microlensed dwarf stars show are 
 surprisingly well-defined. At sub-solar [Fe/H] they are more or 
 less coincident with the abundance trends of the Galactic thick disc 
 as traced by nearby dwarf stars 
 \citep{bensby2003,bensby2005,bensby2007letter2}.
 At super-solar [Fe/H] they follow the trends we see for nearby 
 thin disc dwarf stars. However, due to the
 high ages that some of the Bulge stars possess at super-solar [Fe/H], 
 and due to the lack of Bulge stars at sub-solar [Fe/H] with thin disc
 abundance ratios we see no obvious connection between the Bulge and 
 the thin disc.
 \item 
 All stars with sub-solar [Fe/H] are old (around 10\,Gyr) 
 and have high $\rm [\alpha/Fe]$
 ratios, consistent with fast enrichment by core-collapse supernovae
 during the early stages of the formation of the Galaxy.
 At super-solar [Fe/H] we have a few old stars but also three stars
 with ages lower than 5\,Gyr. This is inconsistent with, e.g., recent
 CMDs of field stars in the Bulge based on deep imaging with HST/ACS, 
 that show no evidence for a young stellar component in the Bulge.
 The average age for our sample of microlensed dwarf stars is 
 $8.4\pm3.3$\,Gyr.
 \item 
 Additionally, our results indicate that the red clump 
 stars in the Bulge have $(V-I)_{0} = 1.08$.
\end{enumerate}

Based on these results and conclusions we speculate on the origin
of the Bulge and we must conclude that it is still poorly constrained.

\begin{acknowledgement}

 S.F. is a Royal Swedish Academy of Sciences Research Fellow supported 
 by a grant from the Knut and Alice Wallenberg Foundation. 
 Work by A.G. was supported by NSF Grant AST-0757888. 
 A.G.-Y. is supported by the Israeli Science Foundation, an
 EU Seventh Framework Programme Marie Curie IRG fellowship and the
 Benoziyo Center for Astrophysics, a research grant from the Peter 
 and Patricia Gruber Awards, and the William Z. and Eda Bess Novick 
 New Scientists Fund at the Weizmann Institute.
 A.U. acknowledges  support by the Polish MNiSW grant N20303032/4275.
 S.L. research was partially supported by the DFG cluster of excellence
`Origin and Structure of the Universe'. 
 J.M. is supported by a Ci\^encia 2007 contract, funded by FCT/MCTES
 (Portugal) and POPH/FSE (EC) and he acknowledges financial support from
 FCT project PTDC/CTE-AST/098528/2008.
 T.S. acknowledges support from grant JSPS20740104.
 D.A. thanks David Bolin at the Centre for Mathematical
Sciences (Lund University) for help with statistics. 
 We would like to thank Bengt Gustafsson, Bengt 
 Edvardsson, and Kjell Eriksson for usage of the MARCS model atmosphere 
 program and their suite of stellar abundance (EQWIDTH) programs. 
 We also thank Judy Cohen and Courtney Epstein for providing reduced
 spectra of their microlensing events.
 This research has also made use of the Keck Observatory Archive (KOA), 
 which is operated by the W. M. Keck Observatory and the NASA 
 Exoplanet Science Institute (NExScI), under contract with the 
 National Aeronautics and Space Administration. 
  
\end{acknowledgement}

\bibliographystyle{aa}
\bibliography{referenser}

\end{document}